\documentclass[structabstract]{aa}  
 \usepackage{epsfig}
\usepackage{epstopdf}
\usepackage{graphicx}
\usepackage{natbib}

\usepackage{txfonts}
\newcommand{\Msun}{M$_\odot$}
\newcommand{\Mj}{M$_\mathrm{J}$}
\newcommand{\asec}{$\!''$}
\newcommand{\fmax}{$f_{\mathrm{max}}~$}
\newcommand{\acut}{$a_\mathrm{cutoff}~$}

\begin{document}
   \title{A survey of young, nearby, and dusty stars to understand the formation of wide-orbit giant planets}

   \subtitle{VLT/NaCo adaptive optics thermal and angular differential imaging\thanks{Based on observations collected at the European Organization for Astronomical Research in the Southern Hemisphere, Chile, ESO : runs 084.C-0396A, 085.C-0675A, 085.C-0277B, 087.C-0292A, 087.C-0450B, 088.C-0085A, 089.C-0149A.}
       }

   \author{J. Rameau  \inst{1}
          \and
          G. Chauvin \inst{1}
          \and
          A.-M. Lagrange \inst{1}
          \and
          H. Klahr \inst{2}
          \and
          M. Bonnefoy \inst{2}
           \and
          C. Mordasini \inst{2}     
           \and
          M. Bonavita \inst{3}
         \and
          S. Desidera \inst{4}
           \and
          C. Dumas \inst{5}
            \and
          J. H. Girard \inst{5}
              }

   \institute{UJF-Grenoble 1 / CNRS-INSU, Institut de Plan\'etologie et d'Astrophysique de Grenoble (IPAG) UMR 5274, Grenoble, F-38041, France \\
              \email{julien.rameau@obs.ujf-grenoble.fr}
                  \and
             Max Planck Institute f\"ur Astronomy, K\"onigsthul 17,
             D-69117 Heidelberg, Germany
             \and
             Department of Astronomy and Astrophysics, University of Toronto,
              50 St. George Street, Toronto, Ontario, Canada M5S 3H4
                \and
              INAF - Osservatorio Astronomico di Padova, Vicolo dell' Osservatorio 5, 35122, Padova, Italy 
                \and
         	   European Southern Observatory, Alonso de Cordova 3107, Vitacura, Santiago, Chile
            }

   \date{Received December 21st, 2012; accepted February, 21st, 2013.}


  \abstract
{Over the past decade, direct imaging has confirmed the
  existence of substellar companions on wide orbits from their parent stars. To
  understand the formation and evolution mechanisms of these companions, their
  individual, as well as the full population properties, must be
  characterized.}
  {We aim at detecting giant planet and/or brown dwarf companions around young, nearby, and dusty stars. Our goal is also to provide statistics on the population of giant planets at wide-orbits and discuss planet formation models.}
  {We report the results of a deep survey of $59$ stars, members of young stellar associations. The observations were conducted with
  the ground-based adaptive optics system VLT/NaCo at
  $L~\!'$-band ($3.8\mu m$). We used angular differential imaging to reach
  optimal detection performances down to the the planetary mass
  regime. A statistical analysis of about $60~$\% of the young and southern A-F stars closer than $65~$ pc allows us to
  derive the fraction of giant planets on wide
  orbits. We use gravitational instability models and planet population synthesis models following the core-accretion scenario to discuss the occurrence of these companions.}
  {We resolve and characterize new visual binaries and do not detect any new substellar companion. The survey's median detection performance reaches contrasts of
  $10$~mag at $0.5~\!''$ and $11.5$~mag at $1.0~\!''$. We find the occurrence of planets to be between $10.8$ and $24.8~$\% at $68~$\% confidence level assuming a uniform distribution of planets in the interval $[1,13]~$\Mj~and $[1,1000]~$AU. Considering the predictions of planetary formation models, we set important constraints on the occurrence of massive planets and brown dwarf companions that would have formed by gravitational instability. We show that this mechanism favors the formation of rather massive clump ($M_{\rm{clump}}>30~$\Mj) at wide ($a>40$~AU) orbits which might evolve dynamically and/or fragment. For the population of close-in giant planets that would have formed by core accretion (without considering any planet - planet scattering), our survey marginally explore physical separations ($\le 20~$AU) and cannot constrain this population. We will have to wait for the next generation of planet finders to start exploring that population and even for the extremely large telescopes for a more complete overlap with other planet hunting techniques.}
   {}
  

   \maketitle
%
\section{Introduction}

\begin{table*}[th]
\caption{Frequency of giant planets reported by various surveys around the full spectral type range.}             
\label{tab:stat}      
\centering          
\begin{tabular}{lllllll}     
\noalign{\smallskip}
\noalign{\smallskip}\hline
\noalign{\smallskip}\hline  \noalign{\smallskip}
Sample & Technique & Sep. range & Mass range & Frequency & Distribution  & Reference \\
& & (AU) & (\Mj) & (\%) &(AU) &  \\
\noalign{\smallskip}\hline                  \noalign{\smallskip}
  $102$ M & RV & $\lesssim 1$ & $\lesssim 3$ & $1-5$ & observed & \cite{bonfils13} \\
    $822$ FGK & RV & $\le 5$ & $\ge 0.3 $ & $14$ & observed  & \citet{mayor11}\\
 $31$ old-A & RV & $0.1-3$ & $0.5-14$ & $9-14$ & observed  & \citet{johnson10} \\
  585$$ F-M & RV & $\le 3$ & $0.3-10$ & $10.5$ & observed & \citet{cumming08} \\
  \noalign{\smallskip}\hline \noalign{\smallskip}
  $42$ AF & AO & $5-320$ & $3-14$ & $5.9-18.8$ & flat / Cu08\tablefootmark{a} & \citet{vigan12} \\
    $85$ F-M & AO & $5-500$ & $\le 100$ & $\le 10$ & flat + GI\tablefootmark{b}  & \citet{janson12} \\
    $15$ BA & AO & $\le 300$ & $\le 100$ & $\le 32$ & flat + GI  & \citet{janson11}\\
  $118$ F-M & AO & $25-856$ &  $\ge 4$ & $\le 20$ & flat / Cu08  & \citet{nielsen10}\tablefootmark{c}\\
  $88$ B-M & AO & $\ge 40 $ & $5-13$ & $\le 10$ & Cu08 & \citet{chauvin10}\\
    $88$ FGKM & AO &  $50-250$ & $0.5-13$ &$\le 9.3$ & power laws in $m$ and $a$\tablefootmark{d} &  \citet{lafreniere07} \\
  $22$ GKM & AO & $\ge 30$ & $\ge 3$ & $\le 5$ & power laws in $m$ and $a$\tablefootmark{d} & \citet{kasper07} \\
  \noalign{\smallskip}\hline  \noalign{\smallskip}
\end{tabular}
\tablefoot{Results on the frequency of giant planets are reported at $68~$\% confidence level, except the ones from \citet{janson11,janson12} which are stated at $99~$\% confidence level.
\tablefoottext{a}{\citet{cumming08}}
\tablefoottext{b}{They infer the planet population from boundaries in a planet mass-semi major axis grid considering disk instability model.}
\tablefoottext{c}{They performed their analysis using results from surveys of \citet{masciadri05,biller07,lafreniere07}.}
\tablefoottext{d}{They infer the planet population from power laws distributions with different coefficients to $m$ and $a$ as the ones in \citet{cumming08}. Please be referred to the publication for details.}
}
\end{table*}

  Most of the giant planets have been discovered so far thanks
    to indirect techniques (radial velocity and transit) at short
   orbits ($\le5$~AU). Almost 20 years of systematic
    search lead to numerous surveys around solar-type, lower/higher mass
    \citep{endl06,bonfils13,lagrange09a}, or even evolved
    stars \citep{johnson07,lovis07}. The sample of detected and characterized planets thus becomes
    large enough to perform robust statistical analysis of the
    population and test planetary formation theories. In that sense,
    the planet occurrence frequency has been determined for giant and
    telluric planets.
    \citet{mayor11} find that 50~\% of solar-type stars harbor at least one planet of any mass
    and with period up to 100 days. This occurrence decreases to $14~$\% when considering giant planets larger than $0.3~$\Mj and varies if we consider giant planets around lower/higher mass stars \citep{cumming08,johnson10,mayor11} (see Table \ref{tab:stat}). These rates thus confirm that planet formation is not rare.
    
    Observational evidences regarding close-in planets lead to favor a formation by the core-accretion mechanism (hereafter CA, e.g. \citealt{pollack96}). \citet{sousa11} show that the presence of close-in giant planets is correlated with the metallicity of their host stars. This correlation is also related to, if planets orbit within $3$~AU, their host-star mass \citet{lovis07,johnson10,bowler10}. Another correlation but between the content in heavy elements of the planets and the metallicity of their parent star has also been found \citep[e.g.][]{guillot06,miller11}. All are
    insights that CA is operating at
    short orbits. According to this scenario, the first steps of the growth of giant gaseous planets are identical to those of rocky planets.
    The dust settling towards the mid plane of the protoplanetary disk leads to formation of larger and larger aggregates through coagulation up to meter-sized planetesimals. 
    These cores grow up then through collisions with other bodies until they reach a critical mass of $10~M_\oplus$ \citep{mizuno80}. 
    Their gravitational potentials being high enough, they trigger runaway gas accretion and become giant planets. 
    However, such scenario requires high surface density of solids into the disk to provide enough material to form the planet core and a large amount of gas. 
    Large gaseous planets are not expected to form in situ below the ice line. 
    \citet{lin96,alibert04,mordasini09} refine the model with inward migration to explain the large amount of giant planets orbiting very close to their parent stars.    
    
       At wider ($\ge30$~AU) orbits, the situation is very different
    since this core accretion mechanism has difficulties to form giant
    planets \citep{boley09,dodson09}. The
    timescales to form massive cores become longer than the gas
    dispersal ones and the disk surface density too low. Additional
    outward migration mechanisms must be invoked (corotation torque in
    radiative disks, \citealt{kley12} or planet-planet scattering, 
    \citealt{crida09}). Alternatively, cloud fragmentation can form objects down to the planetary
    mass regime \citep{whitworth07} and is a solid alternative to explain the existence of
    very wide orbits substellar companions. Finally, disk
    fragmentation also called gravitational instability (GI, \citealt{cameron78,stamatellos09}) remains an attractive mechanism for the formation of
    massive giant planets beyond 10 to 20~AU. According to this scenario, a protoplanetary disk becomes unstable if cool enough leading to the excitation of global instability modes, i.e. spiral arms. 
    Due to their self-gravity, these arms can break up into clumps of gas and dust which are the precursor for giant planets. 
    
    Understanding how efficient are these different mechanisms as
    a function of the stellar mass, the semi-major axis, and the disk
    properties are the key points to fully understand the formation of giant
    planets. Understanding how giant planets form and interacts with
    their environment is crucial as they will ultimately shape the
    planetary system's architecture, drive the telluric planet's
    formation, and the possible existence of conditions favorable to Life.

The presence of massive dusty disks around young stars, like HR\,8799 and $\beta$
  Pictoris, might be a good indicator of the presence of exoplanetary
  systems recently formed \citep{rhee07}. Observations at several wavelengths
revealed asymmetry structures, ringlike sometimes, or even warps which could
arise from gravitational perturbations imposed by one or more giant
planet \citep[e.g.][]{mouillet97,augereau01,kalas05}. Thanks to improvement of
direct imaging (DI) technique with ground-based adaptive optics systems
(AO) or space telescopes, a few planetary mass objects and low mass
brown dwarfs have been detected since the first one by
\citet{chauvin04}. One also has to refer to the breakthrough
discoveries of giant planets between $8$ and $68~$AU around young,
nearby, and dusty early-type stars
\citep{kalas08,lagrange09b,marois08,marois10,carson12}. Direct imaging is the
only viable technique to probe for planets at large separations but
detecting planets need to overcome the difficulties due to the angular
proximity and the high contrast involved. 

Nevertheless, numerous large
direct imaging surveys to detect giant planet companions have reported
null detection
\citep{masciadri05,biller07,lafreniere07,ehrenreich10,chauvin10,janson11,delorme12}, nevertheless this allowed to set upper limits to the occurrence of giant planets. Table \ref{tab:stat} reports the statistical results of several direct imaging surveys, as a function of the sample, separation and mass ranges, and planet distribution. All surveys previous to the one of \citet{vigan12} derive upper limits to the occurrence of giant planets, usually more massive than $1-3$\Mj~between few to hundreds of AUs. They find that less than $10-20~\%$ of any star harbor at least one giant planet if the distribution is flat or similar to the RV one, taking into account all the
assumptions beyond this results. \citet{janson11,janson12} include in the planet distribution limitations if giant planets form via GI. They show that the occurrence of planets might be higher for high mass stars than for solar-type stars but GI is still a rare formation channel. On the other hand, \citet{vigan12} takes into account two planetary system detections among a volume-limited set of 42 A-type stars to derive lower limits for the first time. It comes out that the frequency of jovian and massive
giant planets is higher than $5.9~$\% around A-F stars. However, all these surveys suggest a
decreasing distribution of planets with increasing separations, which
counterbalances the RV trend.

In this paper we report the results of a deep direct imaging survey of
$59$ young, nearby, and dusty stars aimed at detecting giant planets
on wide orbits performed between 2009 and 2012. The selection of the target sample and the observations are detailed in Section \ref{sec:strategy}. In
Section \ref{sec:datared}, we describe the data reduction and analysis
to derive the relative astrometry and photometry of companion
candidates, and the detection limits in terms of contrast. 
Section \ref{sec:results} is then dedicated to the main results of the
survey, including the discovery of new visual binaries, the
characterization of known substellar companions, and the detection
performances. Finally, we present in Section \ref{sec:statistics} the
statistical analysis over two special samples: A-F type stars and A-F dusty stars, from which we constrain the
frequency of planets based on different 
formation mechanisms or planet population hypotheses.
			 

\section{Target sample and observations}
\label{sec:strategy}

\subsection{Target selection}

\begin{table*}
\caption{Sample of young, nearby, and dusty stars observed during our VLT/NaCo thermal and angular differential imaging survey.}
\label{tab:target}
\begin{center}
\small
\begin{tabular}{llllllllllcll}     
\noalign{\smallskip}
\noalign{\smallskip}\hline
\noalign{\smallskip}\hline  \noalign{\smallskip}
\multicolumn{2}{c}{Name}&
   $\alpha$               &   
   $\delta$    &
    b &
      $\mu_\alpha\cos(\delta)$      &  
       $\mu_\delta$    &   
       d    & 
         SpT         &
             K     &  
             excess ? & 
              age   &
              Ref. \\
       HIP  & 
         	    HD    &
          (J2000)              & 
           (J2000)    & 
             (deg) &
              (mas.yr$^{-1}$)   &
                (mas.yr$^{-1}$)    & 
                      (pc)      &   
                            & 
                            (mag)          &
                            &
                              (Myr)&
                                \\
\noalign{\smallskip}\hline                  \noalign{\smallskip}
\multicolumn{12}{c}{AB Doradus} \\
\noalign{\smallskip}\hline                  \noalign{\smallskip}
6276	 &-&01 20 32	&-11 28 03&-72.9&	110.69&	-138.85&	35.06	&G9V	&6.55 &y	&70 & 1\\
$\star$18859	& 25457 &04 02 37&	-00 16 08	&-36.9&149.04	&-253.02&	18.83&	F7V	&4.18&y	&70&1\\
30314&45270&06 22 31	&-60 13 08	&-26.8&-11.22	&64.17	&23.49	&G1V	&5.04&y	&70&1\\
$\star$93580  & 177178 &19 03 32 & 01 49 08 & -1.86 &  23.71 & -68.65 & 55.19 &  A4V	& 5.32	&n	& 70& 1\\
 95347  & 181869 &19 23 53 &-40 36 56 & -23.09 &  32.67 & -120.81& 52.08&  B8V	& 4.20&n &	70 & 1\\
  109268 &209952 &22 08 13 &-46 57 38 & -52.47 &127.6  & -147.91 &31.09 & B6V	 &2.02&n	&	70 & 1 \\
  $\star$115738 &220825&23 26 55 &01 15 21& -55.08&  85.6 &  -94.43 & 49.7  &A0	&	4.90	&	y&70	& 1\\
  $\star$117452 &223352&23 48 55 &-28 07 48 & -76.13 &100.03&  -104.04 &43.99& A0V	 &4.53&y	&	70 & 1\\
\noalign{\smallskip}\hline                  \noalign{\smallskip}
\multicolumn{12}{c}{$\beta$ Pictoris} \\
\noalign{\smallskip}\hline                  \noalign{\smallskip}
$\star$11360	&15115&02 26 16	&+06 17 34&-49.5&86.09	&-50.13	&44.78	&F2	&5.86&	y&12&10\\
$\star$21547	&29391&04 37 36	&-02 28 24&-30.7&43.32	&-64.23	&29.76	&F0V	&4.54&n	&12&2\\
$\star$25486	 &35850&05 27 05&	-11 54 03	&-24.0&17.55	&-50.23	&27.04&	F7V	&4.93&y	&12&2\\
$\star$27321	& 39060 & 05 47 17 & -51 03 59 & -30.6 & 4.65 & 83.1 & 19.4 & A6V & 3.53 &y& 12 & 2 \\
$\star$27288	&38678&05 46 57	&-14 49 19&	-20.8&14.84&	-1.18	&21.52	&A2IV/V	&3.29&y	&12&13\\ 
$\star$79881	&146624&16 18 18	&-28 36 50&+15.4	&-33.79	&-100.59	&43.05	&A0V	&4.74&	n&12&2\\
$\star$88399	&164249&18 03 03	&-51 38 03&-14.0&74.02	&-86.46	&48.14&	F5V	&5.91&y	&12&2\\
$\star$92024	&172555&18 45 27	&-64 52 15&-23.8&	32.67	&-148.72	&29.23	&A7V	&4.30&y	&12&2\\
$\star$95261	&181296&19 22 51	&-54 25 26	&-26.2&25.57&	-82.71&	48.22	&A0Vn	&5.01&y	&12&2\\
$\star$95270	&181327&19 22 59	&-54 32 16&-26.2	&23.84	&-81.77	&50.58	&F6V	&5.91&	y&12&2\\
102409&197481&	20 45 09&	-31 20 24	&-36.8&280.37	&-360.09	&9.94	&M1V	&4.53&	y&12&2\\
\noalign{\smallskip}\hline                  \noalign{\smallskip}
\multicolumn{12}{c}{Tucana-Horologium / Columba} \\
\noalign{\smallskip}\hline                  \noalign{\smallskip}
$\star$1134  &984 &00 14 10 &-07 11 56& -66.36 &  102.84&  -66.51  &46.17 & F7V	&6.07&	n&	30& 1\\
$\star$2578	&3003&00 32 44	&-63 01 53&-53.9&	86.15&	-49.85&	46.47	&A0V	&4.99&y	&30&1\\
7805 &10472&01 40 24	&-60 59 57& -55.1   &61.94	&-10.56 & 67.25	&F2IV/V	&6.63&y	&30&5\\
$\star$9685 &12894&02 04 35	&-54 52 54&-59.2	&	75.74	&-25.05&	47.76&	F4V	&5.45&n	&30&1\\
10602	&14228&02 16 31	&-51 30 44&-22.2&	90.75&	-21.9	 &47.48	&B0V	&4.13&n	&30&1\\
12394	&16978&02 39 35	&-68 16 01&-45.8	&87.4	&0.56	&47.01	&B9	&4.25&n	&30&1\\
16449	&21997&03 31 54	&-25 36 51&-54.1&	53.46&	-14.98&	73.80	&A3IV/V	&6.10&	y&30&1\\
$\star$22295	&32195&04 48 05&	-80 46 45	&-31.5&46.66&	41.3	&61.01	&F7V	&6.87&	y&30&1\\
$\star$26453	 &37484&05 37 40&	-28 37 35	&-27.8&24.29	&-4.06	&56.79	&F3V	&6.28&y	&30&1\\
26966	&38206&05 43 22	&-18 33 27&-23.1	&18.45&	-13.2	6 &69.20	&A0V	&6.92&y	&30&1\\
$\star$30030	&43989&06 19 08	&-03 26 20	&-8.8&10.65	&-42.47	&49.75	&F9V	&6.55&y	&30&1\\
30034	&44627&06 19 13	&-58 03 16&-26.9	&14.13	&45.21	&45.52	&K1V	&6.98&y	&30&1\\
$\star$107947&207575&	21 52 10&	-62 03 08&-44.3&	43.57&	-91.84&	45.09	&F6V&	6.03&y	&30&1\\
$\star$114189	&218396&23 07 29	&+21 08 03&-35.6	&107.93	&-49.63	&39.40&F0V	&5.24&y	&30&1\\
$\star$118121&224392&23 57 35	&-64 17 53&-51.8	&	78.86&	-61.1	4 &8.71	&A1V	&4.82&n	&30&1\\
\noalign{\smallskip}\hline                  \noalign{\smallskip}
\multicolumn{12}{c}{Argus} \\
\noalign{\smallskip}\hline                  \noalign{\smallskip}
- 	&	$\star$67945&08 09 39&	-20 13 50&+7.0&-38.6	&25.8&	63.98	&F0V	&7.15&n	&40&3\\ 
$\star$57632	&102647&	11 49 04&	+14 34 19&+70.8&-497.68&	-114.67	&11.00&	A3V	&1.88&y	&40&3\\ 
\noalign{\smallskip}\hline                  \noalign{\smallskip}
\multicolumn{12}{c}{Hercules-Lyra} \\
\noalign{\smallskip}\hline                  \noalign{\smallskip}
544	&166&00 06 37	&+29 01 19	&-32.8 &379.94	&-178.34	&13.70	&K0V	&4.31& y	&200&4\\
7576	&10008&01 37 35	&-06 45 37	&-66.9&170.99	&-97.73	&23.61	&G5V	&5.70&y	&200&4\\
\noalign{\smallskip}\hline                  \noalign{\smallskip}
\multicolumn{12}{c}{Upper Centaurus-Lupus} \\
\noalign{\smallskip}\hline                  \noalign{\smallskip}
78092	&142527&15 56 42	&-42 19 01	&&-11.19	&-24.46	&145.	&F6IIIe	&4.98&y	&5&6\\ 
\noalign{\smallskip}\hline                  \noalign{\smallskip}
\multicolumn{12}{c}{Other} \\
\noalign{\smallskip}\hline                  \noalign{\smallskip}
682 &377&	00 08 26&	+06 37 00 	&+20.6&88.02	&-1.31	&39.08&	G2V	&6.12&y	&30&7\\ 
$\star$7345&9692	&01 34 38	&-15 40 35	&-74.8&94.84& -3.14	&59.4&	A1V	&5.46&	y&20&5\\
7978	&10647&01 42 29	&-53 44 26&-61.7	&166.97	&-106.71	&17.35	&F9V	&4.30&y	&300&5\\ 
$\star$13141	&17848&02 49 01	&-62 48 24&-49.5&	94.53&	29.02&	50.68	&A2V	&5.97&y	&100&5\\ 
18437	&24966& 03 56 29	&-38 57 44	&-49.9&29.46	&0.1	&105.82	&A0V	&6.86&y	&10&5\\
22226	&30447&04 46 50	&-26 18 09&-37.9&	34.34&	-4.63	 &78.125	&F3V	&6.89&	y&100&5\\
$\star$22845	&31295&04 54 54&	+10 09 03	&-20.3&41.49	&-128.73	&35.66	&A0V	&4.41&y	&100&5\\
34276	&54341&07 06 21	&-43 36 39 &-15.8&	5.8	&13.2	&102.35	&A0V	&6.48&y	&10&5\\
38160	&64185&07 49 13	&-60 17 03	&-16.6&-37.41	&140.08	&34.94	&F4V	&4.74&n	&200&8\\ 
$\star$41307	&71155&08 25 40	&-03 54 23&+18.9	&-66.43&	-23.41	&37.51&	A1V	&4.08&y	&100&5\\ 
\end{tabular}
\end{center}
\end{table*}

\begin{table*}[th!]
\addtocounter{table}{-1}
\caption{Continued}
\begin{center}
\small
\begin{tabular}{llllllllllcll}     
\noalign{\smallskip}
\noalign{\smallskip}\hline
\noalign{\smallskip}\hline  \noalign{\smallskip}
\multicolumn{2}{c}{Name}&
   $\alpha$               &   
   $\delta$    &
    b &
      $\mu_\alpha\cos(\delta)$      &  
       $\mu_\delta$    &   
       d    & 
         SpT         &
             K     &   
             excess ? &
              age &
              Ref.   \\
       HIP  & 
         	HD &
          (J2000)              & 
           (J2000)    & 
             (deg) &
              (mas.yr$^{-1}$)   &
                (mas.yr$^{-1}$)    & 
                      (pc)      &   
                            & 
                            (mag)          &
                            &
                              (Myr) &
                               \\
\noalign{\smallskip}\hline                  \noalign{\smallskip}
53524&95086	&10 57 03	&-68 40 02	&-8.1&-41.41&	12.47	&90.42&	A8III	&6.79&	y&50&5\\
59315&105690	&12 10 07	&-49 10 50&+13.1	&-149.21&	-61.81	&37.84	&G5V	&6.05&n	&100&9\\ 
76736&138965	&15 40 12	&-70 13 40	&-11.9&-40.63	&-55.31	&78.49	&A3V	&6.27&y	&20&5\\
$\star$86305&159492	& 17 38 06&	-54 30 02&-12.0&-51.04	&-149.89	&44.56	&A7IV	&4.78&y	&50&11\\
$\star$99273&191089	&20 09 05	&-26 13 27	&-27.8&39.17&	-68.25&	52.22	&F5V	&6.08&y	&30&5\\
$\star$101800&	196544&20 37 49&	+11 22 40	&-17.5&39.15	&-8.26	&7.94&	A1IV	&5.30&y	&30&5\\
108809	&209253&	22 02 33&	-32 08 00&-53.2&-19.41	&23.88	&30.13	&F6.5V	&5.38&y	&200&5\\
114046&217987	&23 05 47&	-35 51 23	&-66.0&6767.26	&1326.66	&3.29	&M2V	&3.46&n	&100&12\\
- & 219498	&23 16 05	&+22 10 02&-35.6	&79.7	&-29.4	&150.0	&G5	&7.38&	y&300&7\\
$\star$116431&221853&	23 35 36&	+08 22 57	&-50.0&65.37	&-40.79	&68.45	&F0	&6.40&y	&100&5\\ 
\noalign{\smallskip}\hline  \noalign{\smallskip}
\end{tabular}
\end{center}
\tablefoot{ Stars with the $\star$ symbol are used for the statistical analysis. Star parameters ($\alpha$, $\delta$, b, $\mu_\alpha\cos(\delta)$, $\mu_\delta$ and d) are extracted from the Hipparcos catalog \citep{leeuwen07}. Unit of right ascension are hours, minutes, and seconds ; units of declination are degrees, arcminutes, and arcseconds. For HD 219498, the distance is extracted from \citet{rocca09}. The K magnitudes are extracted from the 2MASS catalog \citep{cutri03}. The IR excess at $24$ and/or $70~\mu m$ are extracted from \citet{zuckerman11,kains11,morales11, rhee07}. The age of HIP 93580 is still debated in \citet{zuckerman11} due to discordant kinematics. The binarity (if physical) may have some impact on both proper motions and RVs and thus on the membership to AB Dor.\\
The age references are the following :}
\tablebib{(1) \citet{zuckerman11};
(2) \citet{zuckerman01} ;
(3) \citet{torres08} ;
(4) \citet{lopez06} ;
(5) \citet{rhee07} ;
(6) See discusion in \citet{rameau12} ;
(7) \citet{hillenbrand08} ;
(8) \citet{zuckerman06} ;
(9) See for instance \citet{chauvin10} ;
(10) \citet{schlieder12} ;
(11) \citet{song01} ;
(12) See discusion in \citet{delorme12} ;
(13) \citet{nakajima12}
}
\end{table*}

\begin{figure*}[th]
\centering
\includegraphics[width=7cm]{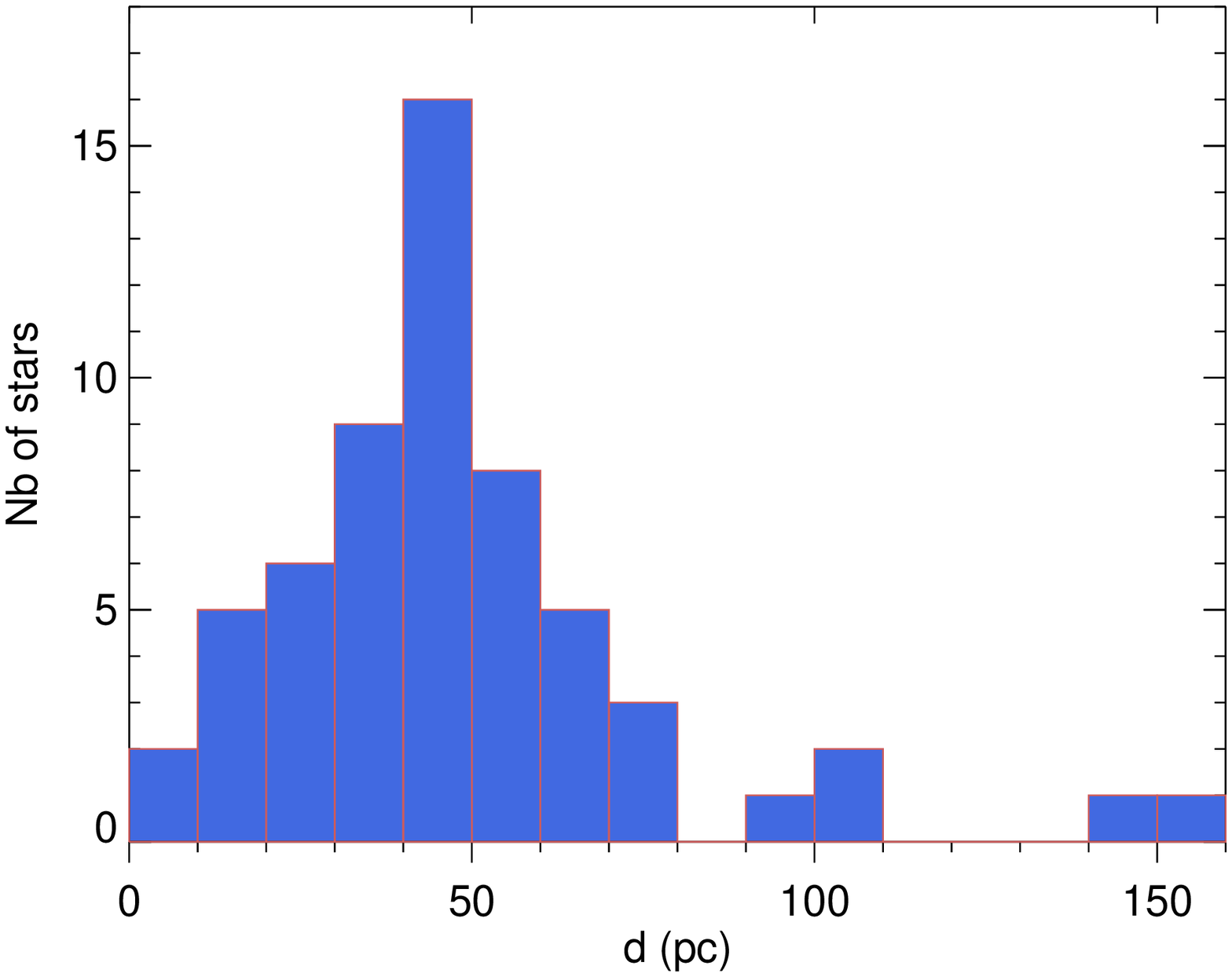}
\includegraphics[width=7cm]{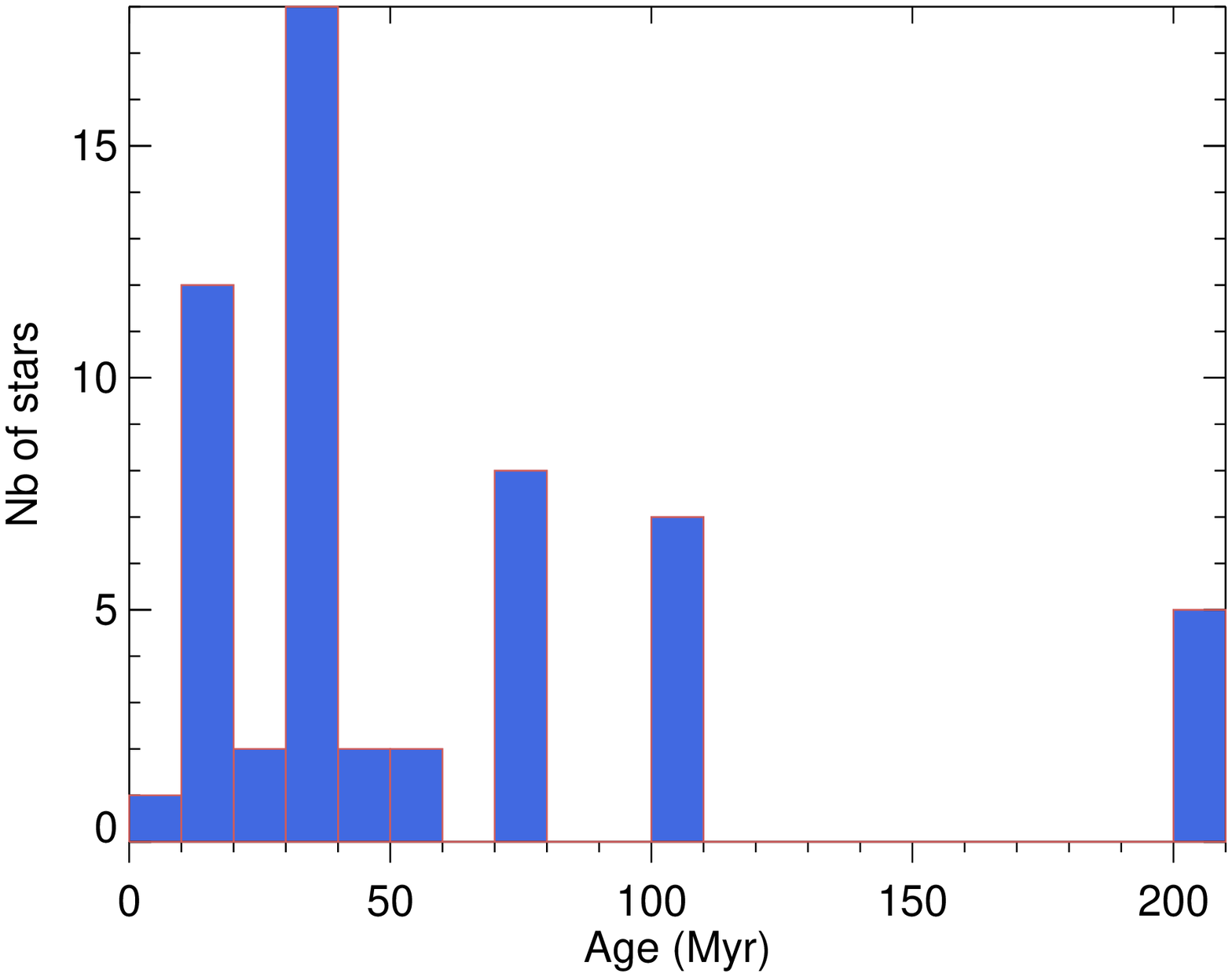}\\
\includegraphics[width=7cm]{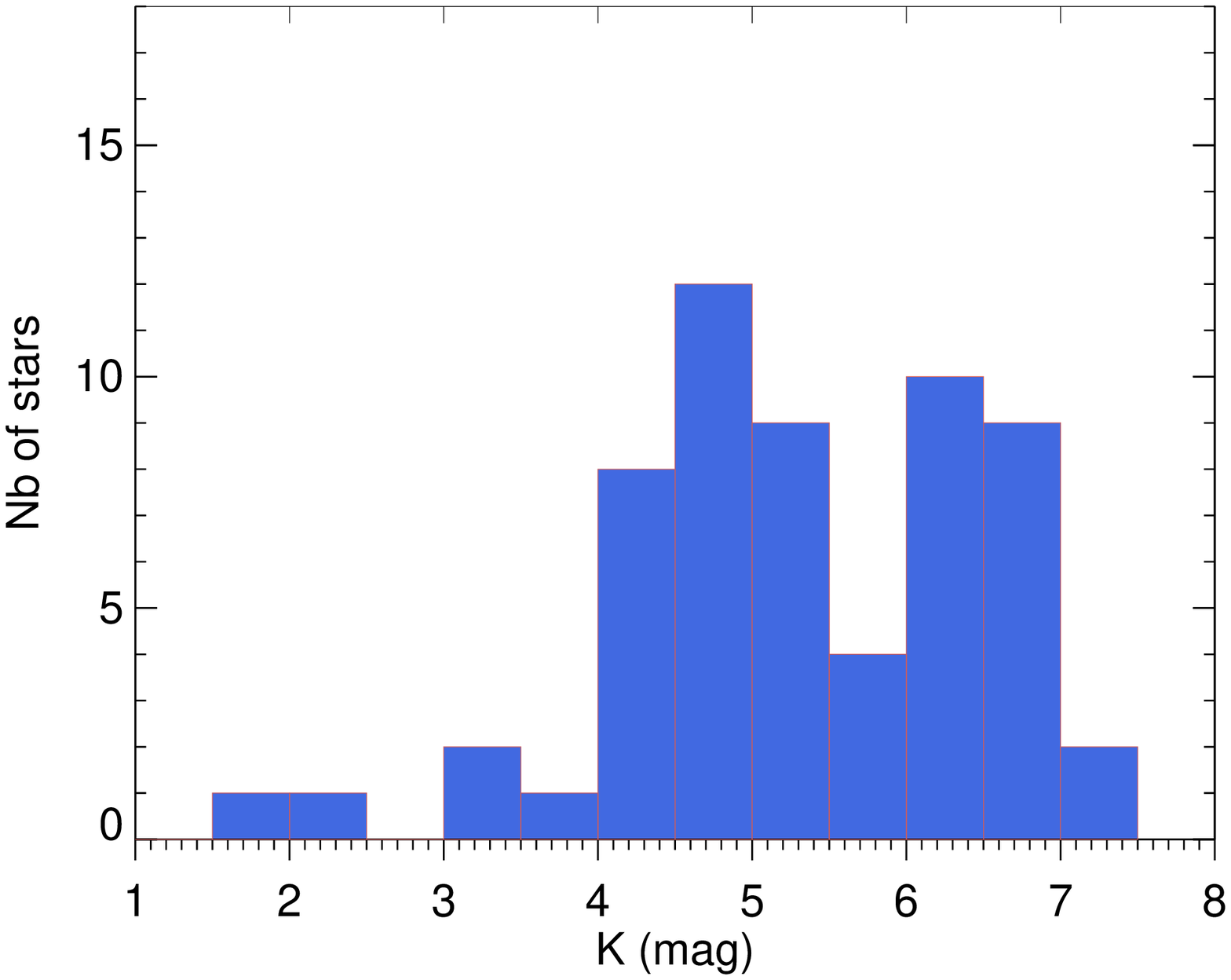}
\includegraphics[width=7cm]{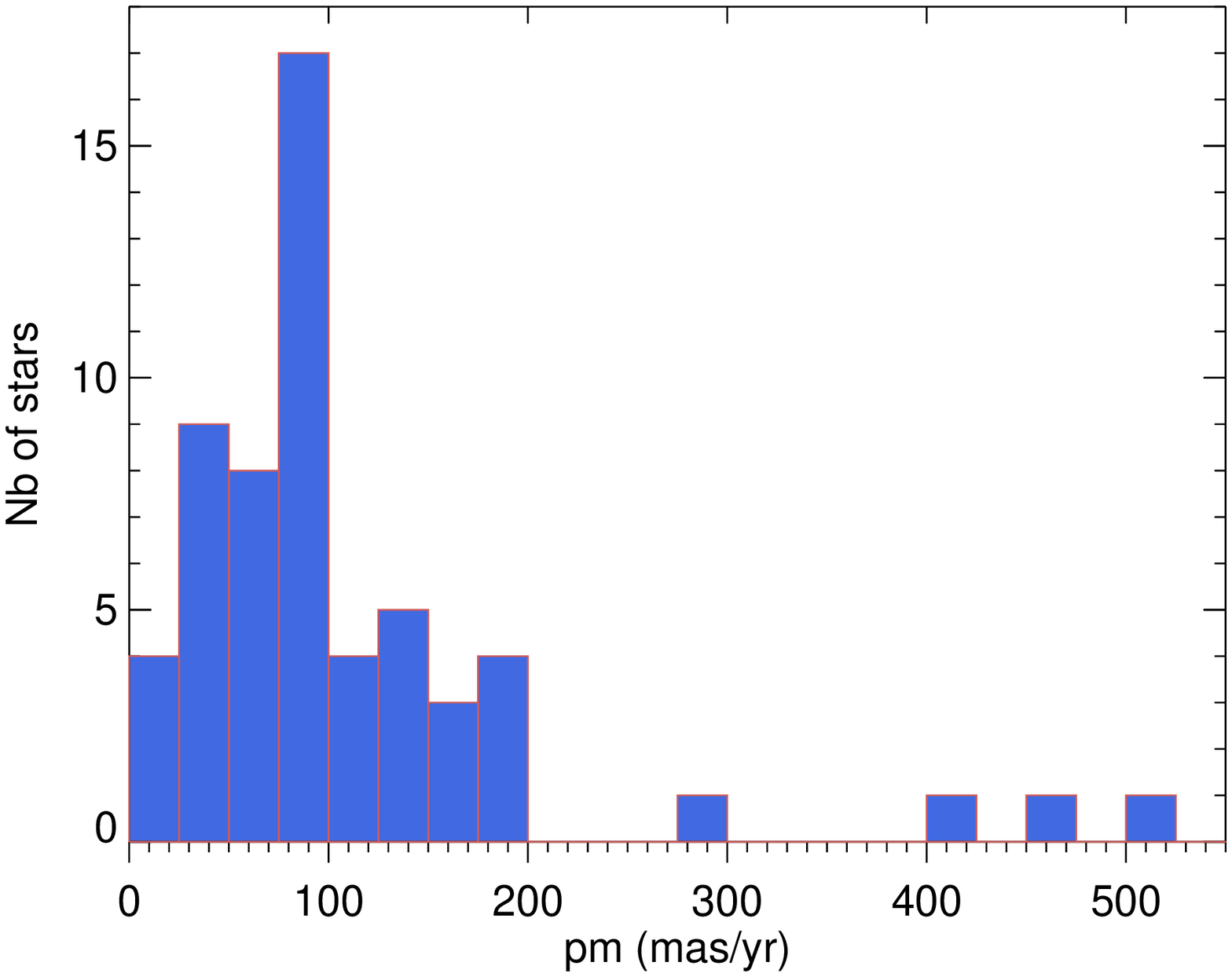}\\
\includegraphics[width=7cm]{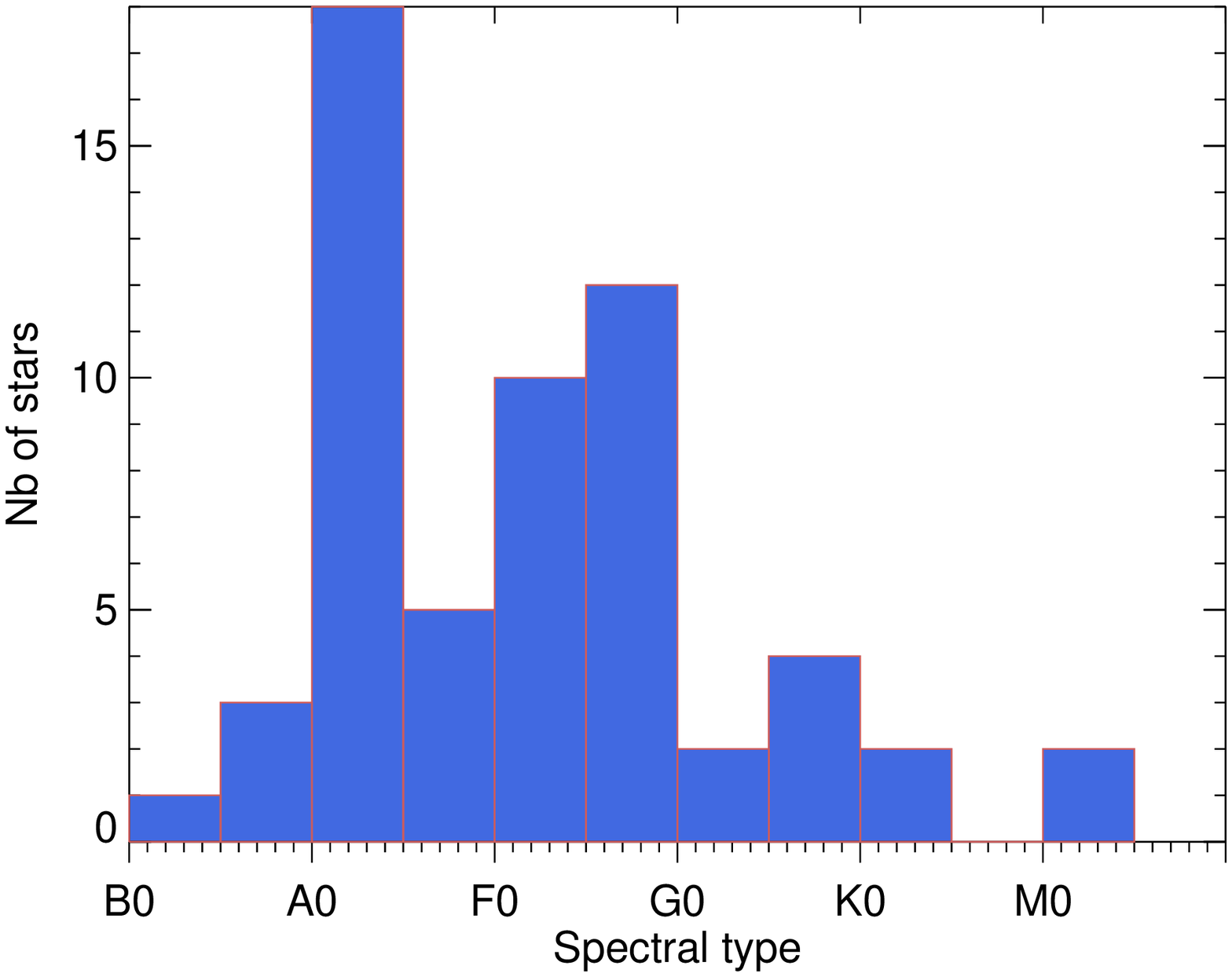}
\caption{Graphical summaries of the main properties on the target stars. \textbf{Top raw : Left : } Histogram of distances with $10~$pc bin. \textbf{Right :} Histogram of ages of known members of considered moving groups (AB Dor - $70~$Myr, $\beta$ Pictoris - $12~$ Myr, Tuc/Hor \& Col - $30~$Myr, Her/Lyr - $200~$Myr, Argus - $40~$Myr, Upper Cen/Lup - $5/10~$Myr) and additional stars with $10~$Myr bin. \textbf{Central raw : Left :} Histogram of K-band magnitude with $0.5~$mag bin. \textbf{Right :} Histogram of apparent proper motions with $25~$mas/yr bin. \textbf{Bottom raw :} Histogram of spectral types. }
\label{fig:targets}
\end{figure*}

The target stars were selected to optimize our chance of
  planet detection according to :
  \begin{itemize}
  \item Distance : with a given angular resolution limited by the
  telescope's diffraction limit, the star's proximity enables to
  access closer physical separations and fainter giant planets when
  background-limited. We therefore limited the volume of our sample to
  stars closer than $150$~pc, even closer than $100$
  pc for $94~\%$ of them (see Figure \ref{fig:targets}, top left
  panel). 
  
  \item Observability and magnitude : stars were selected according to : 1) their declination ($\delta\le25~$deg) for being
observable from the southern hemisphere, 2) their K-band brightness (K
$\lesssim7~$mag) to ensure optimal AO corrections, 3) for being single
to avoid degradation of the AO performances and 4) for being never
observed in deep imaging (see section \ref{sec:strategy}). 

   \item Age : evolutionary models \citep{baraffe03,marley07} predict that
  giant planets are intrinsically more luminous at young ages and
  become fainter with time. Therefore, for a given detection
  threshold, observing younger stars is sensitive to lower mass planets. Our sample selection is based on
recent publications on associations (AB Dor,
$\beta$ Pic, Her/Lyr, Argus, Tuc/Hor, Columba, Upper Cen/Lupus) from
\citet{zuckerman11}, \citet{torres08} and \citet{rhee07}. Indeed,
stars belonging to these moving groups share common kinematic
properties and ages. These parameters are measured from spectroscopy,
astrometry, and photometry (optical and X-rays). 
$64~$\% of the selected stars belongs to young and nearby moving groups.
$88\%$ of the targets are younger
  than $200~$Myr and even $62~\%$ younger than $70~$Myr (see Figure
  \ref{fig:targets}, top right panel). 
  
  \item Spectral Type : recent imaged giant planets have been detected
    around the intermediate-mass HR\,8799, Fomalhaut, and $\beta$
    Pictoris stars with separations from $8$ to $110~$AU. More massive stars
    imply more massive disks, which potentially allow the formation of more
    massive planets. We therefore have biased $79~$\% of our sample towards
    spectral types A and F (see Figure
\ref{fig:targets}, bottom panel).

  \item Dust : dusty debris disks around pre-
    and main-sequence stars might be signposts for the existence of
    planetesimals and exoplanets \citep[see a review
      in][]{krikov10}. $76~$\% of our sample are star with large infrared excess 
      at $24$ and/or $70~\mu m$ (IRAS, ISO and \textit{Spitzer}/MIPS), indicative of the emission of cold dust. 
      The remaining stars have no reported excess in the literature.
\end{itemize}

The name, coordinates, galactic latitude (b), proper motions
($\mu$), spectral type (SpT), distance (d), K magnitude, and age of
the target stars are listed in Table \ref{tab:target} together with
the reference for the age determination and the moving group they
belong to if they do\footnote{We attempt to derive the age in a homogeneous
way. If the star belongs to a moving group, the age of that group is
adopted for this star. If the star does not belong to a known
association, then we ensured that the age determination was done on
similar way than for the membership identification, i.e. the galactic
space motions $UVW$, the Li $\lambda 6708~\AA$ line equivalent width
or the X-ray emission.}. Figure \ref{fig:targets} summarizes the main properties of the target
stars. Briefly, the sample consists of $59$ B- to M-type stars whose
the median star would be a F-type at distance of $40~$pc with an age
of $30~$Myr, a K-magnitude of $5.5$, and an apparent proper motion of
$85~$mas/yr.

\begin{table*}[th!]
\caption{Additional young, nearby and southern A-F stars for the statistical analysis}
\label{tab:add_target}
\begin{center}
\small
\begin{tabular}{llllllllllcll}     
\noalign{\smallskip}
\noalign{\smallskip}\hline
\noalign{\smallskip}\hline  \noalign{\smallskip}
\multicolumn{2}{c}{Name}&
   $\alpha$               &   
   $\delta$    &
    b &
      $\mu_\alpha\cos(\delta)$      &  
       $\mu_\delta$    &   
       d    & 
         SpT         &
             K     &   
             excess ? &
              age &
              Ref.   \\
       HIP  & 
         	HD &
          (J2000)              & 
           (J2000)    & 
             (deg) &
              (mas.yr$^{-1}$)   &
                (mas.yr$^{-1}$)    & 
                      (pc)      &   
                            & 
                            (mag)          &
                            &
                              (Myr) &
                               \\
\noalign{\smallskip}\hline                  \noalign{\smallskip}
$\star$12413 & 16754 & 02 39 48 & -42 53 30 & -63.0 & 88.20 &-17.82 & 39.8 & A1V & 4.46 & n & 30 & 1 \\ 
$\star$14551 & 19545 & 03 07 51 & -27 49 52 & -59.8 & 66.26 & -19.09 & 54.6 & A5V & 5.77 & n & 30 & 1 \\
$\star$26309 & 37286 & 05 36 10 & -28 42 29 & -28.1 &  25.80& -3.04& 56.6 & A2III & 5.86 & y & 30 & 1 \\
$\star$61468 & 109536 & 12 35 45 & -41 01 19 & 21.8 & -107.09 &0.63& 35.5 & A7V &4.57 &n & 100 & 1 \\
\noalign{\smallskip}\hline  \noalign{\smallskip}
\end{tabular}
\end{center}
\tablefoot{The reference corresponds to the publication the star is extracted from : (1) \citet{vigan12}.}
\end{table*}

To analyze an homogeneous and volume-limited sample, we perform the statistical
 study on stars which have 1) d $\le 65~$pc,
2) age $\le 100~$Myr 3) dec $\le25~$deg, and 4) Spectral type = A or F. 
We get set of $68$ young, nearby, and southern A-F stars from the literature.
$33$ stars in our survey fulfill these criteria (flagged with a $\star$ symbol in Table \ref{tab:target}), i.e. $48~$\% of completeness.
To increase this rate, we add $4$ being observed with VLT/NaCo from a previous survey 
\citep{vigan12}\footnote{These additional stars have been also observed using 
ADI techniques but with the Ks filter on VLT/NaCo.} 
(see Table \ref{tab:add_target}), thus reaching a representative rate of $56~$\%.
We will then refer to this sample of stars as the A-F sample.

Moreover, we also aim at constraining the formation mechanism and
 rate of giant planets around $\beta$ Pictoris analogs. 
 A sub-set of stars is extracted from the A-F homogeneous sample by considering an IR excess at
 $24$ and/or $70~\mu m$ (from the same references as for our survey plus \citealt{mizusawa12,rebull08,hillenbrand08,su06}). Among the full set of $68$ stars, $39$ are dusty. Our survey
 plus $1$ stars from \citet{vigan12}, i.e. $28$ stars, reach a complete level of $72~$\%. 
 We will then refer to this sample as the A-F dusty sample.\newline
 
 \begin{table}[t!]
\caption{Sample definitions}
\label{tab:sample}
\begin{center}
\small
\begin{tabular}{lclc}     
\noalign{\smallskip}
\noalign{\smallskip}\hline
\noalign{\smallskip}\hline  \noalign{\smallskip}
Name&
   Number             &   
   SpT    &
    Representative rate                  \\
        & 
           & 
           & 
              (\%)                             \\
\noalign{\smallskip}\hline                  \noalign{\smallskip}
 Survey & 59 & B-M & --\\
 A-F sample & 37 & A-F & 56 \\
 A-F dusty sample & 29 & A-F & 72 \\ 
\noalign{\smallskip}\hline  \noalign{\smallskip}
\end{tabular}
\end{center}
\end{table}

Table \ref{tab:sample} summarizes all different samples.


\subsection{Observing strategy}

The survey was conducted between 2009 and 2012 with the NAOS adaptive
optics instrument \citep{rousset03} combined to the CONICA
near-infrared camera \citep{lenzen03}. NaCo is mounted at a Nasmyth
focus of one of the $8.2~m$ ESO Very Large Telescopes. It provides
diffraction-limited images on a $1024\times1024$ pixel Aladdin 3 InSb
array. Data were acquired using the L27 camera, which provides a
spatial sampling of $\simeq27.1~$mas/pixel and a field of view (FoV
hereafter) of $28~\!''\times28~\!''$. In order to maximize our chance
of detection, we used thermal-infrared imaging with the broadband
$L~\!'$ filter ($\lambda_c=3.8~\mu m$, $\Delta\lambda=0.62~\mu m$) since
it is optimal to detect young and warm massive planets with a peak of
emission around $3-4~\mu m$.

NaCo was used in pupil-tracking mode to reduce instrumental
speckles that limit the detection performances at inner angles,
typically between $0.1~\!''$ and $2.0~\!''$. This mode provides
rotation of the FoV to use angular differential imaging (ADI,
\citealt{marois06}). The pupil stabilization is a key element for the
second generation instruments GPI and VLT/SPHERE. Note that NaCo suffered from a drift of the star with time (few pixels/hours
depending on the elevation) associated to the pupil-tracking mode
until october 2011 \citep{girard12}. Higher performance was obtained after the correction of the drift. To
optimize the image selection and data post-processing, we recorded short individual exposures coupled to the windowing mode of $512\times514$
  pixels (reduced FoV of $\approx 14~\!''\times14~\!''$). The use of the dithering pattern
combined to the cube mode also ensure accurate sky and instrumental
background removal. The detector integration time (DIT) was set to
$0.2~$s to limit the background contribution to the science images.

\begin{table}[t]
\caption{Mean platescale and true north orientation measured on $\theta_1$ Ori C field with NaCo and $L~\!'$/L27 set-up.}             
\label{tab:astro-calib}      
\centering          
\begin{tabular}{llll}     
\noalign{\smallskip}
\noalign{\smallskip}\hline
\noalign{\smallskip}\hline  \noalign{\smallskip}
ESO program & UT-date & Platescale & True north\\
& & (mas) & (deg)\\ 
\noalign{\smallskip}\hline                  \noalign{\smallskip}
  084.C-0396A & 11/24/2009 & $27.09\pm0.02$ & $-0.08\pm0.01$\\
  085.C-0675A & 07/27/2010 & $27.12\pm0.03$ & $-0.36\pm0.05$\\
        085.C-0277B & 09/28/2010 & $27.11\pm0.04$ & $-0.36\pm0.11$\\ 
    087.C-0292A & 12/18/2011 & $27.10\pm0.03$ & $-0.60\pm0.01$\\
     087.C-0450B & 12/08/2011 &$27.16\pm0.08$ & $-0.52\pm0.07$\\ 
  088.C-0885A & 02/19/2011 & $27.10\pm0.03$ & $-0.38\pm0.03$\\
  089.C-0149A & 08/24/2012 & $27.11\pm0.02$ & $-0.41\pm0.07$\\
\noalign{\smallskip}\hline  \noalign{\smallskip}
\end{tabular}
\end{table}

Each observing sequence lasted around $90~$minutes, including telescope
pointing and overheads. It started with a sequence of short
unsaturated exposures at five dither positions with the neutral
density filter $ND_{long}$ (transmission of $1.17~\%$). This allowed
the estimation of the stellar point spread function (PSF) and served
as photometric calibrator. Then, saturated science images were
acquired with a four dithering pattern every two DIT$\times$NDIT
exposures with NDIT= $100$ stored into a datacube and this was
repeated over more than $100~$times to provide sufficient FoV rotation
for a given star. The PSF core was saturating the detector over a
$\simeq5~$pixel-wide area. Twilight flat-fields were also acquired. For some
target stars, second epoch data on NaCo were acquired with the same
observing strategy. Finally, $\theta_1$ Ori C field was observed as
astrometric calibrator for each observing run. The same set of stars
originally observed with HST by \citet{mccaughrean94} (TYC058, 057,
054, 034 an 026) were imaged with the same set-up ($L~\!'$ with the L27
camera). The mean platescale and true North orientation were measured
and reported on Table \ref{tab:astro-calib}.

 \subsection{Observing conditions}
 
\begin{figure*}[th]
\centering
\includegraphics[width=8cm]{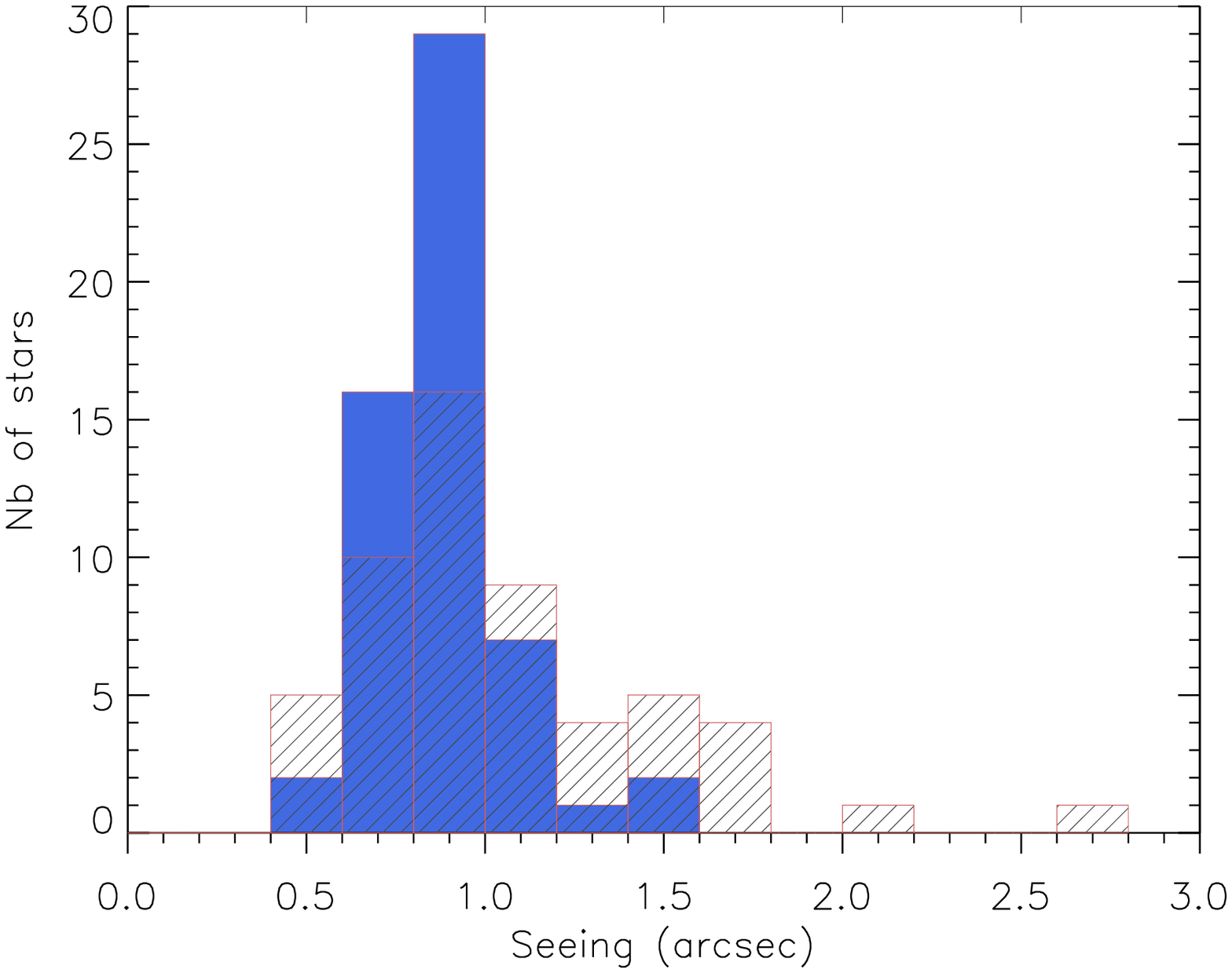}
\includegraphics[width=8cm]{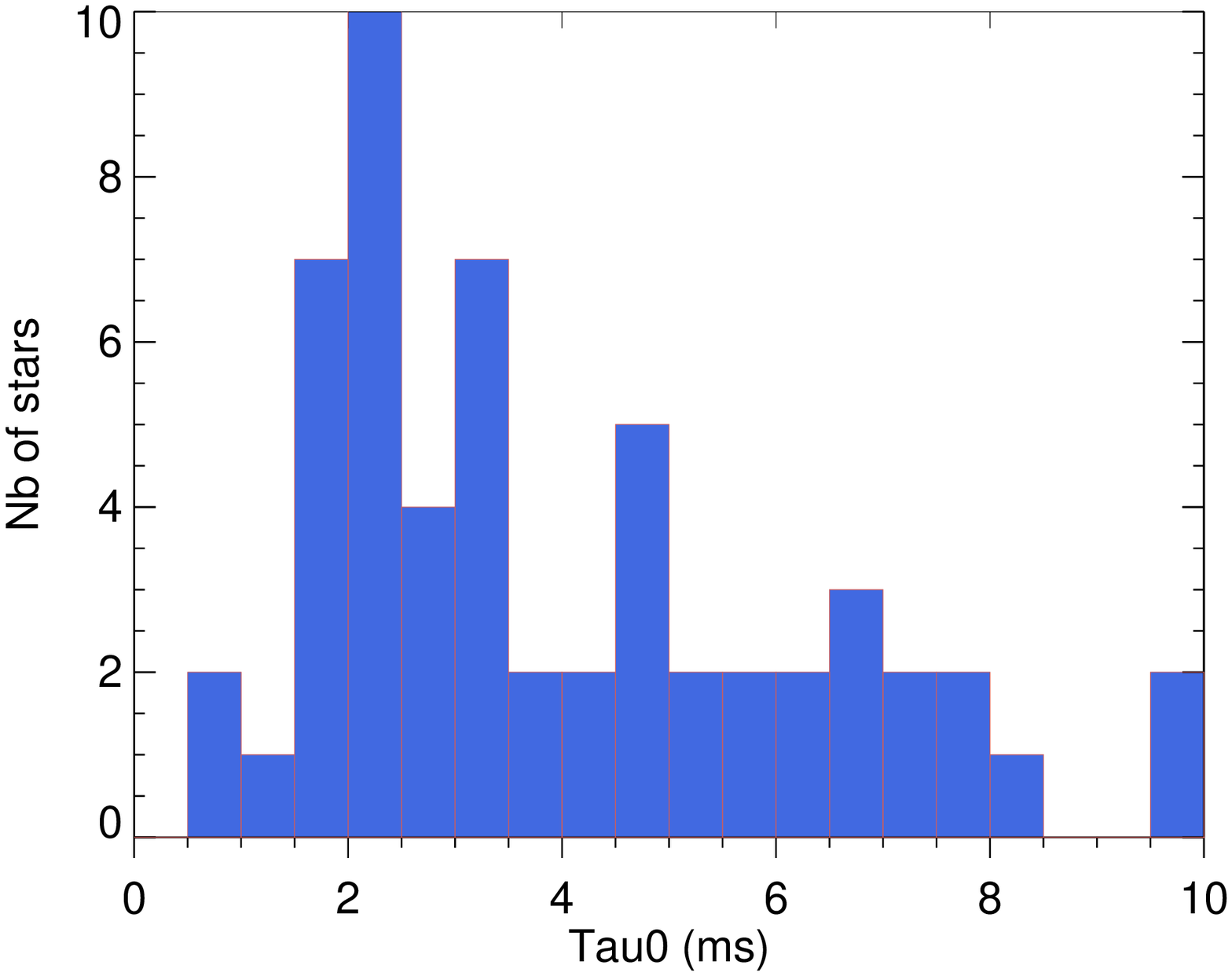}\\
\includegraphics[width=8cm]{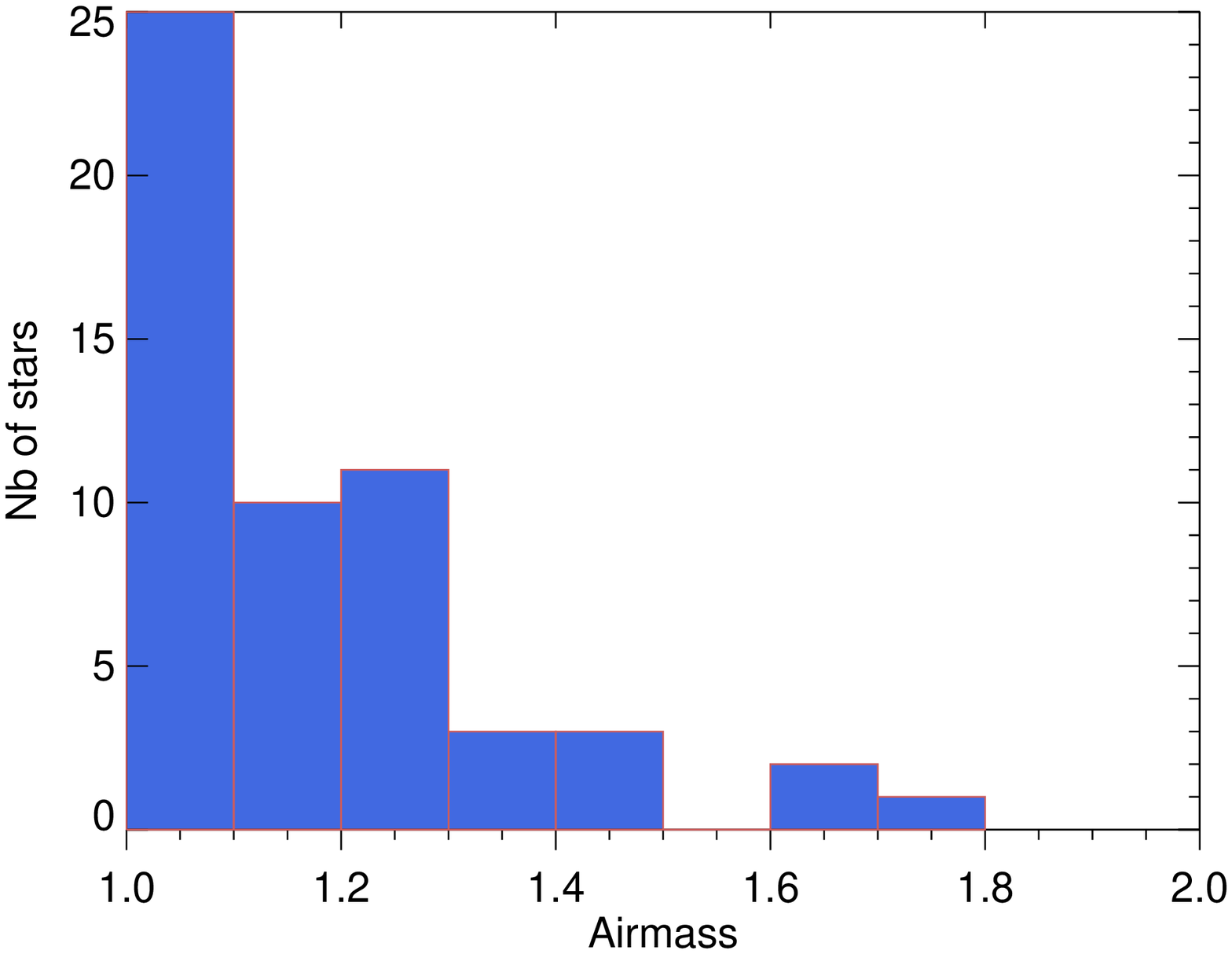}
\includegraphics[width=8cm]{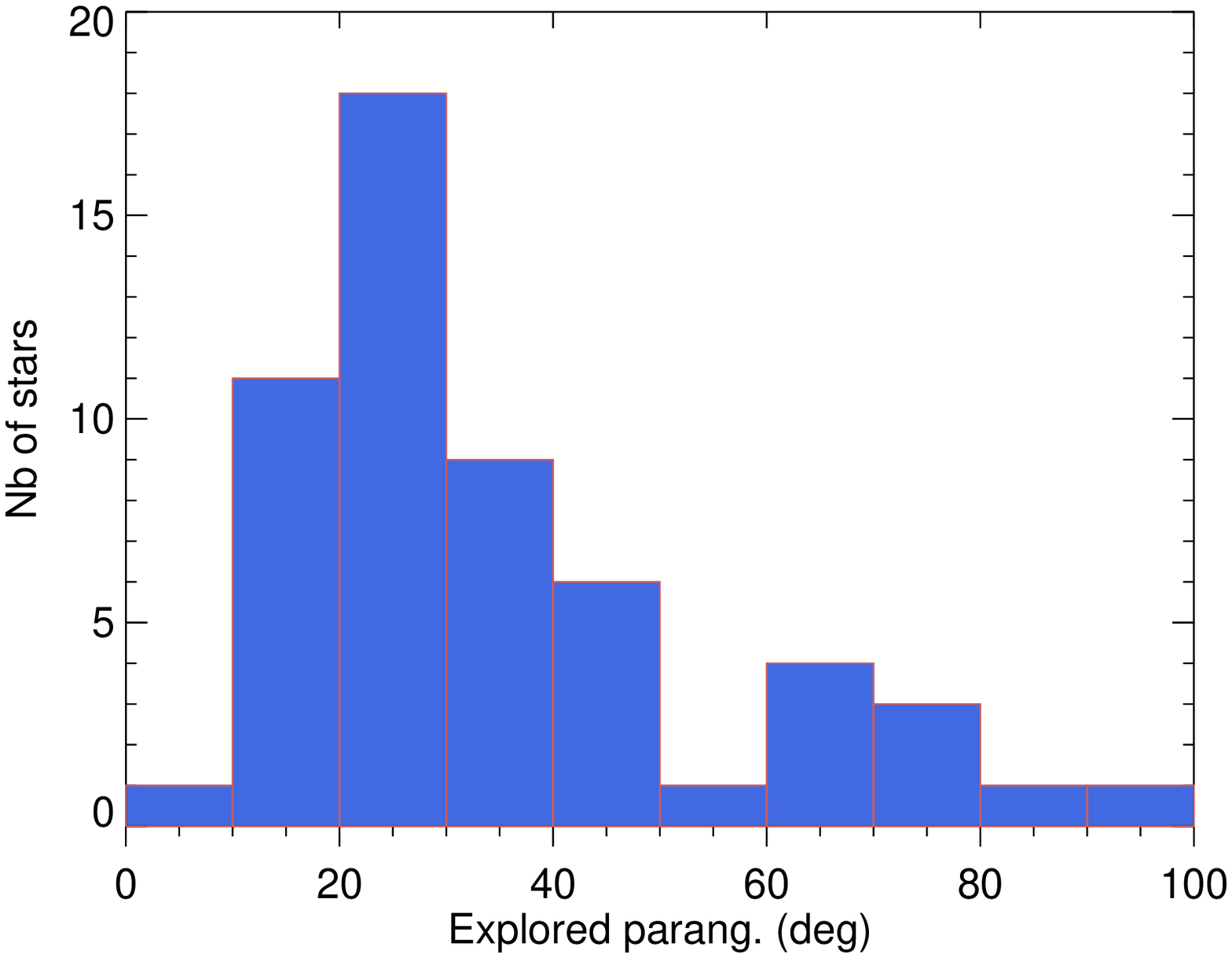}
\caption{Graphical summaries of the log of the observations of the target young, nearby dusty stars observed with VLT/NaCo between 2009 and 2012. \textbf{Top left :} Histogram of median seeing (image quality seen by the active optics sensor) with $0.2~\!''$ bin. The DIMM seeing has been over plotted with dashed columns. \textbf{Top left :} Histogram of median $\tau_0$  with $0.5~$ms bin as calculated by NAOS. \textbf{Bottom left :} Histogram of median airmass with $0.1~$ bin. \textbf{Bottom right :} Histogram of explored parallactic angles with $10~$deg bin.}
\label{fig:histo-log}
\end{figure*}

Observations in period 84 and 85 were done in visitor mode as
  the pupil stabilization mode was not offered in service mode. Observations in period 87, 88 and 89 were then completed in
  service mode to benefit from optimal atmospheric conditions. A
  summary of the observing conditions is reported on Figure
\ref{fig:histo-log} showing histograms of explored parallactic angle
ranges, airmass, as well as the atmospheric conditions : seeing and
coherence time $\tau_0$. Note that NAOS corrects the atmospheric
turbulences for bright stars when $\tau_0$ remains longer than $2~$ms
($63~$\% of the observations). When $\tau_0$ decreases, the image
quality and precision for astrometric and photometric measurements are
degraded. The observations were however conducted under good
conditions since the median seeing is $0.9~\!''$, the median $\tau_0$
is $3.2~$ms and the median airmass is $1.15$. Finally, $72\%$ of the stars were
observed with a parallactic angle exploration larger than $20~$deg. 
 
\section{Data reduction and analysis}
\label{sec:datared}

\subsection{Unsaturated images}

The unsaturated dithered exposures of each star were processed with
the \textit{Eclipse} software developed by \citet{devillar97}: bad
pixels removal, sky subtraction constructed as the median of the
images followed by flat-fielding were applied to data; the final PSF
image was then obtained by shifting and median combining the images.

\subsection{Saturated angular differential images}

The reduction of the ADI saturated dithered datacubes was performed
with the dedicated pipeline developed at the Institut de
Plan\'etologie et d'Astrophysique de Grenoble (IPAG). This pipeline
has been intensively used and gave probing results :
\citet{lagrange10, bonnefoy11, chauvin12, delorme12}; Lagrange et
al. (2012); Rameau et al. (2012). We describe the main steps in the
following.

Getting twilight flats allowed us to achieve optimal flat-fielding and
bad pixel identification. We used the \textit{Eclipse} software to
extract those calibrations frames. The raw data were then divided by
the flat-field and removed for the bad and hot pixels through
interpolation of the closest neighbor pixels. Sky estimation was
performed by taking the median of the $400~$s closest in time dithered
exposures within a cube and then subtracted to each frame. Frames with
low quality were removed from the cubes following a selection based on
cube statistics such as the flux maximum, the total flux, and the
encircled energy in each frame in an annulus outside the saturated
pixels. Poor quality frames due to
degraded atmospheric conditions were rejected (typically less than
10\% of the complete observing sequence, see \citealt{girard10} for the cube advantages). The good-quality frames were
recentered to a common central position using the \textit{Eclipse}
software for the shift and Moffat profile fitting \citep{moffat69} on
the PSF wings for the registration of the central star position. We
ended up with good-quality cleaned and recentered images within a
single master cube associated with their parallactic angle
values. A visual inspection
was done to check the quality of the final frames.

Subsequent steps were the estimation and subtraction of the stellar halo for each image then
derotation and stacking of the residuals. The most critical one is the estimation of the stellar
halo which drives the level of the residuals. We applied
different ADI algorithms to optimize the detection performances and to
identify associated biases. Since the quasi-static speckles limit the
performances on the inner part of the FoV, we performed the ADI
reduction onto reduced frames, typically $200\times200$ pixels. We
recall here the difference between the four ADI procedures :
\begin{itemize}

\item in classic ADI
  (cADI, \citealt{marois06}), the stellar halo is estimated as the median of all individual
  reduced images and then subtracted to each frame. The residuals are then median-combined
  after the derotation;

\item in smart ADI (sADI, \citealt{lagrange10}), the PSF-reference for
  one image is estimated as the median of the $n$ closest-in-time
  frames for which the FoV has rotated more than $\alpha\times$ \textit{FWHM}
  at a given separation. Each PSF-reference is then subtracted to each
  frame and the residuals are mean-stacked after the derotation; We
  chose a PSF-depth of $n=10$ frames for the PSF-building satisfying
  a separation criteria of $\alpha=1.$~\textit{FWHM} at a radius of $1.3~\!''$;

\item the radial ADI (rADI, \citealt{marois06}) procedure is an
  extension of the sADI where the $n$ frames with a given rotation
  used for the stellar halo building are selected according to each
  separation. The PSF-depth and the $\alpha$ coefficient were chosen
  as for sADI ($n=10$ and $\alpha=1.$~\textit{FWHM}). The radial extent of the
  PSF-building zone is $\Delta r=1.4~$\textit{FWHM} below $1.6~\!''$ and
  $3~$\textit{FWHM} beyond;
  
\item in the LOCI approach \citep{lafreniere07}, the PSF-reference is
  estimated for each frame and each location within this frame. Linear
  combinations of all data are computed so as to minimize the
  residuals into an optimization zone, which is much bigger than the
  subtraction zone to avoid the self-removal of point-like sources. We
  considered here a radial extent of the subtraction zone $\Delta
  r=0.9~$\textit{FWHM} below $1.6~\!''$ and $3$ beyond; a radial to azimuthal
  width ratio was set to $g=1$; a standard surface of the optimization
  zone was $N_A=300~$PSF cores; the separation criteria of
  $N_\delta=1~$\textit{FWHM}.

\end{itemize}

All the target stars were processed in a homogeneous way using
similar set of parameters.  It appears that when the PSF remained very
stable during a sequence (i.e. $\tau_0 \ge 4~$ms), advanced ADI techniques do not strongly
enhance the performance. 

ADI algorithms are not the best performant tools for background-limited regions 
as the PSF-subtraction process add noise. We thus processed the data
within the full window (i.e. $512\times514~$pixels with the dithering
pattern) with what we called the non-ADI (nADI) procedure. It consists
in 1) computing an azimuthal average of each frame within 1-pixel wide
annulus, 2) circularizing the estimated radial profile 3) subtracted
the given profile to each frame and then 4) derotating and
mean-stacking the residuals. nADI by-products can help to distinguish
some ADI artifacts from real features as well. 

For each star, a visual
inspection of the five residual maps was done to look for candidate
companions (CC).

\subsection{Relative photometry and astrometry}

Depending on the separation and the flux of the detected CC,
  different techniques were used to retrieve the relative photometry
  and astrometry with their uncertainties:
\begin{itemize}
\item for bright visual binaries, we used the deconvolution algorithm of
  \citet{veran98};
  
\item for CCs detected in background-limited regions (in nADI final images),
  the relative photometry and astrometry were obtained
 using a 2D moffat fitting and classical aperture
  photometry  \citep{chauvin10}. The main limitation of this technique remains the
  background subtraction which affects the level of residuals;
  
\item for speckle limit objects, fake planets were injected following
  the approach of \citet{bonnefoy11,chauvin12} with the scaled PSF-reference
  at the separation of the CC but at different position angles. The
  injections were done into the cleaned mastercubes which were processed with the same setup. 
  We then measured the position and the flux of the fake planets which minimized the difference with the real CC.
  The related uncertainties associated
  with this method were also estimated using the various set of fake planets injected at
  different position angles. 
\end{itemize}

In both algorithms, the main error for the relative astrometry
  is the actual center position of the saturated PSF (up to
  $0.5~$pixel). Other sources of errors come from the Moffat fitting,
  the self-subtraction, the residual noise, or the PSF shape. The
  reader can refer for more details to the dedicated analysis on
  uncertainties on CC astrometry using VLT/NaCo ADI data \citep{chauvin12}. For a CC observed at several epochs (follow-up or
  archive), we investigated its status (background source or comoving
  object) by determining its probability to be a stationary background
  object, assuming no orbital motion. This approach is the same as in
  \citet{chauvin05} by comparing the relative positions in $\alpha$
  and $\delta$ from the parent-star between the two epochs, from the
  expected evolution of positions of a background object, given the
  proper and parallactic motions and associated error bars.

\subsection{Detection limits}

The detection performances reached by our survey were estimated by
computing 2D detection limit maps, for each target star, at $5\sigma$
in terms of $L~\!'$ contrast with respect to the primary. For each set
of residual maps for each target, we computed the pixel-to-pixel noise
within a sliding box of $1.5\times1.5$ \textit{FWHM}. The second step was
to estimate the flux loss due to self-subtraction by the ADI
processing. We created free-noise-cubes with bright fake planets
($100$ ADU) at three positions, i.e. $0$, $120$ and $240~$deg, each
$20~$pixels from the star, with the same FoV rotation as real
datacubes for each star then processed the ADI algorithms with
the same parameters. Note that for LOCI, we injected the fake planets
in the cleaned and recentered datacubes before applying the
reduction. Then the comparison between the injected flux to the
retrieved one on the final fake planets images was done by aperture
photometry. This allowed to derive the actual attenuation for all
separations from the central star by interpolating between the
points. Finally, the $5\sigma$ detection limits were derived by taking the flux loss and the transmission of the
neutral-density filter into account, and were normalized by the unsaturated PSF flux. 2D contrast maps were therefore available for
each star, with each reduction techniques.

To compare the detection performances between the stars, we built 1D contrast curves. An azimuthaly averaging
within 1-pixel annuli of increasing radius on the noise map was
performed, followed by flux loss correction, and unsaturated PSF flux
scaling. This approach however tends to degrade the performances at
close-in separations due to asymmetric speckle and spider residuals on
NaCo data, or even the presence of bright binary component. To
retrieve the detection performances within the entire FoV, we created
composite maps between ADI processed and nADI processed ones. Indeed,
beyond $2~\!''$ from the central star where the limitations are due to
photon and read-out noises, nADI remains the most adapted reduction
technique. It has been shown that the limiting long-lived (from few
minutes to hours) quasi-static speckles are well correlated for
long-time exposures \citep{marois06} thus leading to a non Gaussian
speckle noise in the residual image. Therefore, the definition and the
estimation of $\sigma$ to provide a detection threshold in the region
limited by the quasi-static speckle noise might not be well
appropriate and overestimated. However, the Gaussian noise
distribution being achieved in the background noise regime, the
$5\sigma$ detection threshold corresponds to the expected confidence
level. Moreover, the conversion from contrast to mass detection limits
is much more affected by the uncertainties on the age of the target
stars and the use of evolutionary models than by uncertainties
on the detection threshold.

\section{Results}
\label{sec:results}

Our survey aims at detecting close-in young and warm giant planets,
even interacting with circumstellar disks in the case of stars with IR
excess. Four stars in the sample have been identified as hosting
substellar companions in previous surveys. The redetection of these companions allowed us to
validate our observing strategy, data reduction, and might give additional
data points for orbital monitoring. We also imaged a transitional disk
at an unprecedented resolution at $3.8~\mu m$ for the first time
around HD\,142527 (HIP\,78092) which was presented in a dedicated
paper \citep{rameau12}. However, we did not detect any new
substellar companions in this study. 

In this section, we describe the properties of the newly
  resolved visual binaries, we review the observed and characterized
  properties of known substellar companions and of the candidates
  identified as background sources. We then report the detection
  performances of this survey in terms of planetary masses explored. Finally, 
  we briefly summarize the results on some previously resolved disks, especially about
  HD\,142527, in the context of a deep search for giant planets in its
  close environment.

\begin{table}[t]
\caption{Relative astrometry and photometry of the new visual binaries resolved with VLT/NaCo L' and ADI imaging mode.}             
\label{tab:cc}      
\centering                          
\begin{tabular}{lllll}        
\noalign{\smallskip}
\noalign{\smallskip}\hline
\noalign{\smallskip}\hline  \noalign{\smallskip}
\tiny{Name} &  \tiny{Date} & \tiny{Sep.} &\tiny{PA} & \tiny{$\Delta~$L'} \\    
 & & \tiny{(arcsec)} &\tiny{(deg)} & \tiny{(mag)} \\ 
\noalign{\smallskip}\hline  \noalign{\smallskip}
    \tiny{HIP9685B  }&  \tiny{11/20/2011 } &  \tiny{$0.303\pm0.013$  }&  \tiny{$242.26\pm1.16$  }&  \tiny{2.7$\pm0.1$} \\      
    \tiny{HIP38160B  }&  \tiny{11/25/2009 } &  \tiny{$0.141\pm0.013$  } & \tiny{$ 117.08\pm2.28$  } &   \tiny{3.1$\pm0.3$}\\
       \tiny{HIP53524B }	&  \tiny{01/11/2012 } & \tiny{$ 4.540\pm0.009$  }& \tiny{ $319.04\pm0.7$  }&  \tiny{6.2$\pm0.1$ }\\ 
    \tiny{HIP59315B  }&  \tiny{07/27/2010 } & \tiny{$0.36\pm0.01 $ } & \tiny{$ 259.2\pm2.5 $ }&  \tiny{5.1$\pm0.3$ }\\
    \tiny{HIP88399B  } &  \tiny{07/29/2011 } &  \tiny{$6.439\pm0.009$  }&  \tiny{$88.8\pm0.7 $ }&  \tiny{4.9$\pm0.1$}\\
    \tiny{HIP93580B  }&  \tiny{07/29/2012  }&  \tiny{$0.242\pm0.013  $ }&  \tiny{$94.4\pm1.3 $ }&  \tiny{3.9$\pm0.3$ }\\
    \tiny{HIP117452B  }&  \tiny{07/12/2012  }&  \tiny{$3.667\pm 0.009 $ }&  \tiny{$237.8\pm0.8$ }&  \tiny{3.8$\pm0.1$ }\\    
        \tiny{HIP117452C  }&  \tiny{07/12/2012  }&  \tiny{$3.402\pm 0.009 $ }&  \tiny{$238.6\pm0.8$ }&  \tiny{4.0$\pm0.1$ }\\    
\noalign{\smallskip}\hline  \noalign{\smallskip}
\end{tabular}
\end{table}

\subsection{Binaries}

\begin{figure}[th]
\centering
\includegraphics[width=9cm]{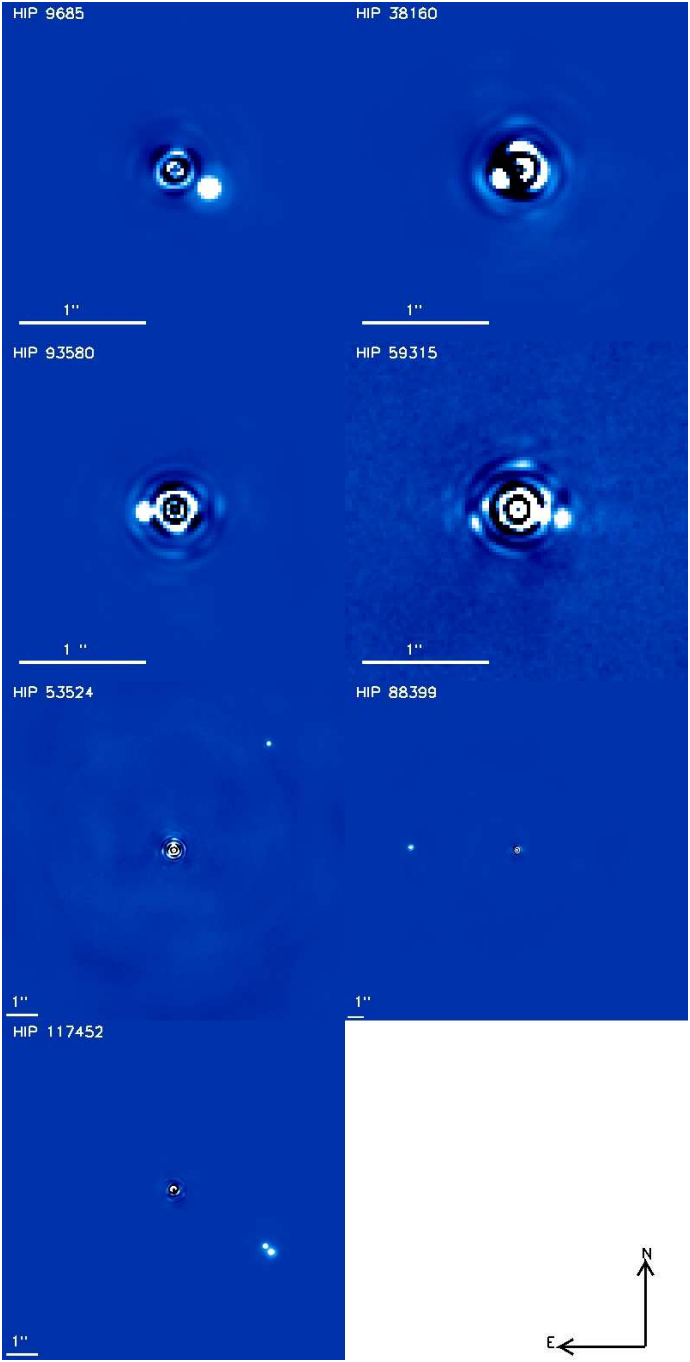}
\caption{Resolved visual binaries with VLT/NaCo in ADI-L' imaging mode. HIP 9685, HIP 53524 and HIP 93580 have not been confirmed as double systems from second-epoch observations or archived data. North is up, east to the left. Counts are displayed in linear scale but differently for each panel.}
\label{fig:cc}
\end{figure}

Despite the rejection of known binaries with $1.-2.~\!''$ separation, 8 visual
multiple systems were resolved (Figure \ref{fig:cc}). Their relative
position and magnitude are reported on Table \ref{tab:cc}. 4 pairs are
very close-in, with separations below $0.4~\!''$ whereas the remaining
ones lie in the range $4-7~\!''$. Only HIP 38160 has been observed at a second epoch and confirms as a
comoving pair. HIP 88399 B and HIP 117452 B and C were
known from literature and HIP 59315 B might be indeed bound to its
host-star based on archive data. 

\textit{HIP 9685 --} HIP 9685 is referenced as a $\Delta~\mu$ astrometric binary \citep{makarov05} and was associated to a ROSAT source by \citet{haakonsen09}. In this work, we report the detection of a close-in binary candidate companion at a projected separation of $14~$AU if we adopt $47.7~$pc of distance. In the 2MASS images in JHK taken in october 1999, a point source is visible toward the North-East direction, at a separation around $12~\!''$ and a position angle of $\simeq15~$deg. From the two relative positions, it came out the $12~\!''$-CC in the 2MASS images is not compatible with a background star at the position of our $0.3~\!''$-CC. It is also very unlikely that the 2MASS $12~\!''$-CC has travels in projection from $540~$AU to $14~$AU in ten years. Instead, the existence of the astrometric acceleration suggests that our NaCo $0.3~\!''$-CC is bound and is responsible for the astrometric signature. The 2MASS PSF being symmetric, our $0.3~\!''$-CC may have been at much smaller separation in 1999 since the orbital motion is significant, which does not contradict the proposed status. If it is true, we derive the mass of our $0.3~\!''$-CC from the measured $M_{L'}=4.1~$mag using the isochrones from \citet{siess00}, assuming a solar metallicity, and an age of $30~$Myr, to be $0.8~$\Msun, matched with a K6 star. 

\textit{HIP 38160 --} A companion with a magnitude of $L'=7.84$ is present at two different epochs (2009-november and 2010-december) $4.8~$AU away ($0.141~\!''$) from HIP 38160 at $35~$pc. The companion shows a common proper motion with the central star between the two epochs. The 2MASS JHK images taken in 2000 also reveal an asymetric PSF which tends to confirm the bound status of this companion. According to \citet{siess00} isochrones for pre- and main-sequence stars, this companion should be of $0.6-0.7$ \Msun, assuming an age of $200~$Myr and a solar-metallicity. Hence, HIP 38160 B could be a late K or ealy M star. HIP 38160 was already catalogued as an astrometric binary \citep{makarov05} and as a double-star system in the Catalog of Component of Double or Multiple stars \citep[CCDM][]{dommanget02}. However, with a separation of $23.3~$arcsec, this additional candidate turns to be only a visual companion (WDS, \citealt{mason2001}).

\textit{HIP 53524 --} HIP 53524 lies at a very low galactic latitude ($b\le 10~$deg). It is therefore very likely that the candidate companion, located at a large separation from the central-star ($\ge4~\!''$), is a background star. Indeed, from \textit{HST}/NICMOS archive data taken in 2007, we measured the relative position of the well seen CC. Even not considering the systematics between the two instruments, the CC turns out to be a background object.

\textit{HIP 59315 --} The star HIP 59315 is not catalogued as being part of a multiple physical system. However, it lies at relative low galactic latitude ($b=13~$deg) but only one point source has been detected with VLT/NaCo in ADI and L' imaging in 2010. \citet{chauvin10} observed this star with VLT/NaCo in H band in coronographic mode and have identified an additional background source more than $5.5~$\asec away with a PA around $100~$deg. If our $0.36~\!''$-CC  is a background contaminant, it would lie in April 2004 at a separation of $1.35~$\asec and a position angle of $254~$deg so that it would have been detected on NaCo H images. The other possibility is that our $0.36~\!''$-CC is indeed bound to HIP 59315 and was occulted by the mask. Therefore, it is likely that it is indeed bound to the star. This would imply for the companion a projected separation of $14~$AU and an absolute magnitude $M_{L'}=8.2$ at $37.8~$pc. The mass derived from the COND model \citep{baraffe03} assuming an age of $100~$Myr is $0.1$\Msun, consistent with a late M dwarf.

\textit{HIP 88399 --} HIP 88399 is referenced as a double star in SIMBAD with a M2 star companion (HIP 88399 B) at $6.35~\!''$ from the 2MASS survey. Given the separation in our observation and the L' magnitude, the CC is indeed the M dwarf companion, lying at $310~$AU from the primary.

\textit{HIP 93580 --} The star is 70 Myr old A4V star at 55.19 pc and $b=-68.6~$deg. A point source 3.9 magnitudes fainter than the primary is detected at a projected separation of $13.24~$AU. Neither archive nor second-epoch observations could infer the status of this CC. Comparison to the \citet{siess00} isochrones at 70 Myr with a solar metallicity would place this object as beeing an early M-type dwarf, with a mass of $0.5~$\Msun.

\textit{HIP 117452 --} Already known as a triple system \citep{rosa11} from observations taken in 2009, the two companions are detected from our data in july, 2012. The brightest companion lies at $\approx160~$AU in projection from the primary while the third component is at a separation of $11~$AU from HIP 117452 B. The large error bars on the astrometry in \citet{rosa11} make difficult to infer any orbital motion of both companion in two years. \\
\newline\\
We also have two spectroscopic binaries (HIP 101800 and HIP 25486) in our survey. \citet{pourbaix09} give a period of about $11~$d, an eccentricity of $0.23$, and a velocity amplitude of the primary of $K_1=26~$km/s for HIP 101800. For HIP25486, \citet{homlberg07} report a standard deviation for the RV signal of $4.4~$km/s, a SB2 nature with an estimated mass ratio of $0.715$. However, we do not detect any source with a contrast from $5~$mag at $100~$mas up to $12~$mag farther out of $1.5~\!"$ around both stars. Both companions are likely too close to their primary for being resolved, or even behind them.

\subsection{Substellar companions}

\begin{table}[t]
\caption{Relative astrometry and photometry of the known substellar companions observed with VLT/NaCo L' and ADI imaging mode.}             
\label{tab:subcc}      
\centering                          
\begin{tabular}{lllll}        
\noalign{\smallskip}
\noalign{\smallskip}\hline
\noalign{\smallskip}\hline  \noalign{\smallskip}
 Name  &  Date  & Sep.  &PA  & $\Delta~$L '  \\    
 & & (arcsec)  &(deg)  & (mag)  \\ 
\noalign{\smallskip}\hline  \noalign{\smallskip}
   \tiny{HR 7329 b} & \tiny{08/13/2011} & \tiny{$4.170\pm0.009$} & \tiny{$167.43\pm0.7$} & \tiny{$6.7\pm0.1$} \\      
   \tiny{AB Pictoris b} & \tiny{11/26/2009} & \tiny{$5.420\pm0.009$}  & \tiny{$175.2\pm0.7$}  &  \tiny{$7.0\pm0.1$}\\
    \tiny{$\beta$ Pictoris b} & \tiny{09/27/2010} & \tiny{$0.383\pm0.11$} & \tiny{$210.28\pm1.73$} & \tiny{$7.8\pm0.3$} \\ 
   \tiny{HR 8799 b} & \tiny{08/07/2011} & \tiny{$1.720\pm0.025$} & \tiny{$62.9\pm1.3$}   & \tiny{$9.9\pm0.1$}\\ 
  \tiny{HR 8799  c} & \tiny{08/07/2011} & \tiny{$0.940\pm0.016$} & \tiny{$321.1\pm1.5$}  & \tiny{$7.1\pm0.2$}\\
    \tiny{HR 8799 d} & \tiny{08/07/2011} & \tiny{$0.649\pm0.016$}  & \tiny{$207.5\pm1.4$}  & \tiny{$7.1\pm0.2$}\\ 
\noalign{\smallskip}\hline  \noalign{\smallskip}
\end{tabular}
\end{table}

\begin{figure}[ht]
\centering
\includegraphics[width=8cm]{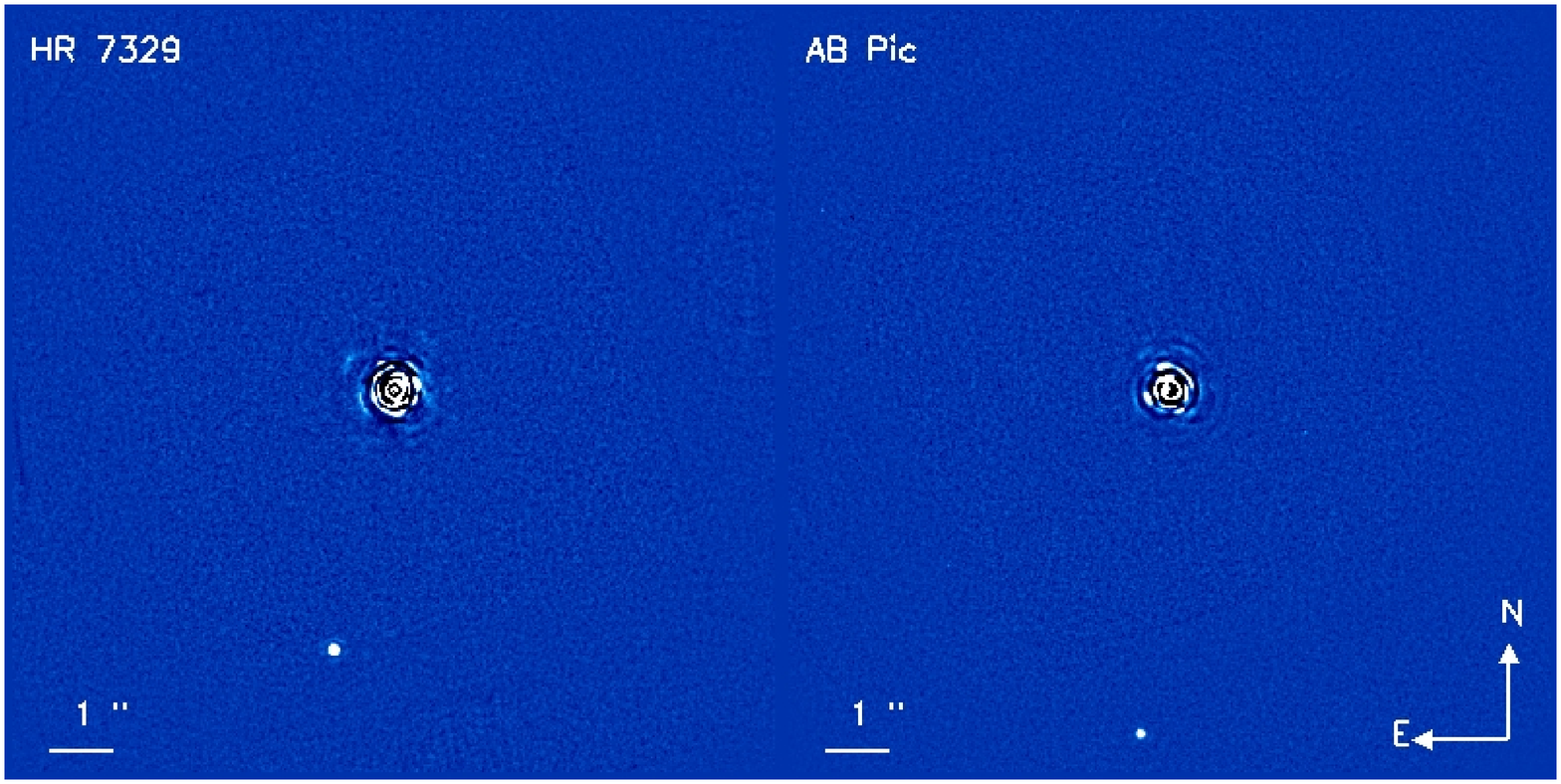}
\includegraphics[width=8cm]{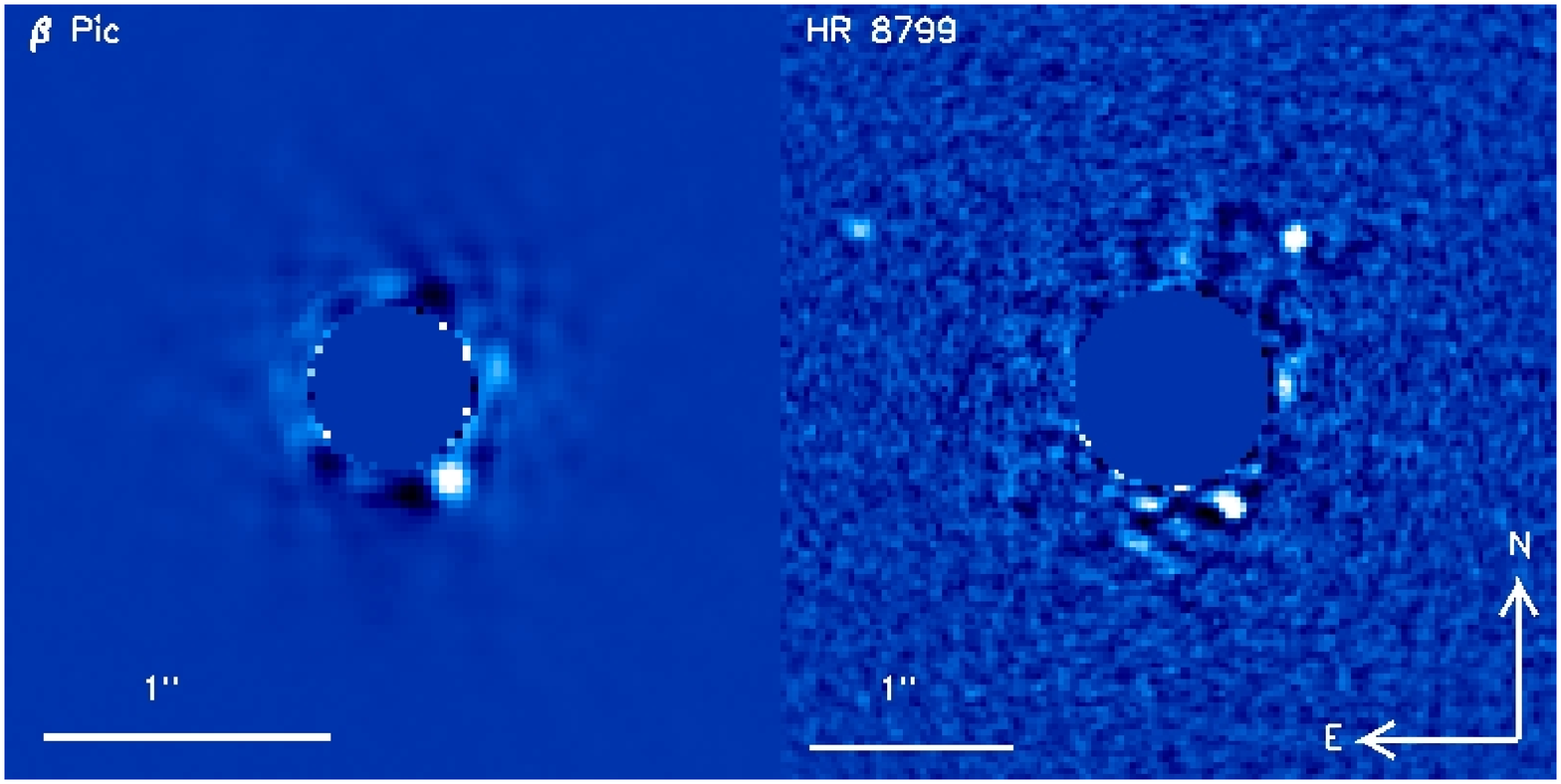}
\caption{Known substellar companions of HR 7329, AB Pic (\textbf{Top left and right}), and $\beta$ Pictoris and HR 8799 (\textbf{Bottom left and right}) observed with VLT/NaCo in ADI-L' modes. North is up, east to the left the count scale is linear. Note that, we could not retrieve with a good SNR HR 8799 e due to the low parallactic angle amplitude ($20.9~$deg).}
\label{fig:subcc}
\end{figure}

Four targets in the sample - HR 7329, AB Pictoris, $\beta$ Pictoris, and HR 8799 - have previously reported to host a brown dwarf and/or planet companions
\citep{lowrance00, chauvin05,lagrange10,marois08, marois10}. Only one identified substellar CC to HIP 79881 has also been stated as background object. We review
below the latest results about these companions since their initial
confirmation. Table \ref{tab:subcc} lists their relative astrometry
and photometry from our observations (see Figure \ref{fig:subcc}). 

\textit{HIP 79881 --} Clearly seen in 2010 july observations in L', the $11.6~$mag-contrast CC to HIP 79881 (separation of $4.528\pm0.008~\!''$ and a position angle of $175.54\pm0.8~$deg) has been also resolved in Keck/NIRC2 images in 2003 and 2005. The relative positions of the CC monitored for 7 years clearly showed that it is a background object.

\textit{HIP 95261 / HR 7329 --} HR 7329 b was discovered by
\citet{lowrance00} with the \textit{Hubble Space Telescope}/NICMOS. It
is separated from its host star, a member of the $\beta$ Pic. moving group
which harbors a debris disk \citep{smith09}, of $4.17\pm0.09~\!''$
($\simeq200~$AU at $48~$pc) a position angle of
$167.4\pm0.7~$deg. The age and the known distance of the star together
with \textit{HST}/STIS spectra and photometry from H to L' bands are
consistent to infer HR 7329 B as a young M7-8 brown dwarf with a mass
between $20$ and $50~$M$_J$. \citet{neuhauser11} conduct a $11~$yr
followed-up to confirm the status of the companion and try to
constraint the orbital properties. Due to the very small orbital motion,
they concluded that HR 7329 B relies near the apastron of a very
inclined - but not edge-one - and eccentric orbit. Our observations
are consistent with the previous ones and exclude the presence of
additional companions down to $4$~M$_J$ beyond $40$~AU.

\textit{HIP 30034 / AB Pic --} This member of the Columba association
hosts a companion at the planet-brown dwarf boundary of
$13\pm2~$M$_J$, discovered by \citet{chauvin05}. Located at
$5.4\pm0.07~\!''$ and $175.2\pm0.7~$deg, AB Pic b has a mass of
$13-14~$M$_J$ deduced from evolutionary models and JHK
photometries. Later on, \citet{bonnefoy10} and recently Bonnefoy et al. (2012, submit.) conduct observations with
the integral field spectrograph VLT/SINFONI to extract
medium-resolution ($R_\lambda=1500-2000)$ spectra over the range
$1.1-2.5~\mu m$. They derive a spectral type of L0-L1, an effective
temperature of $\simeq1700-1800$K, a surface gravity of $log
(g)\simeq4.5~$dex by comparison with synthetic spectra. The relative
astrometry and photometry we measured from our observations are similar to
the previously reported ones. Further investigations are mandatory to derive, if similar, 
similar conclusions as for HR 7329.
Due to our highest spatial resolution
and sensitivity, surely planets more massive than $3$~M$_J$ can be
excluded with a semi-major axis greater than $80$~AU.

\textit{HIP 27321 / $\beta$ Pictoris --}$\beta$ Pic b
\citep{lagrange10} remains up to now the most promising case of imaged
planet probably formed by core accretion. Recent results by
\citep{chauvin12}, including measurements from this survey, refined
the orbital parameters with a semi-major axis of $8-9~$AU and an
eccentricity lower than $e\le0.17$. In addition, \citet{lagrange12b}
could accurately show that the planet is located into the
second-warped component of the debris disk surrounding the star, which
confirms previous studies \citep{mouillet97,augereau01} suggesting that the
planet plays a key role in the morphology of the disk. More recently,
\citet{lagrange12a} directly constrain the mass of the planet
through eight years high-precision radial velocity data, offering thus
rare perspective for the calibration of mass-luminosity relation of
young massive giant planets. Finally, \citet{bonnefoy13} build for the first time the infrared spectral energy distribution of the planet. They derive temperature ($1600-1800~$K), log g ($3.5-4.5$), and luminosity ($log (L/L_{\odot} = -3.87\pm0.08$) for $\beta$ Pic b from the set of new and already published photometric measurements. They also derive its mass ($6-15.5$\Mj) combining predictions from the latest evolutionary models (``warm-start", "hot-start") and dynamical constraints.

\textit{HIP 114189 / HR 8799 --} HR 8799 is a well-known $\lambda$
Boo, $\gamma$ Dor star, surrounded by a debris disk \citep{patience11}
and belonging to the $30~$Myr-old Columba association \citep{zuckerman11}. It hosts four
planetary-mass companions between $14$ and $68~$AU
\citep{marois08,marois10} which awards this multiple planet system
being the first imaged so far. Spectra and photometry studies
\citep[e.g.][]{bowler10,janson10} inferred those planets to rely
between $5$ and $7~$M$_J$. \citet{soummer11} monitor the motion of
the planets b, c and d thanks to HST/NICMOS archive giving $10~$yr
amplitude to constrain the orbits of these planets. Invoking
mean-motion resonances and other assumptions for the outer planets,
they derive the inclination of the system to $\sim28~$deg. \citet{esposito12} consider also the planet e for the dynamical analysis of the system. They show that the coplanar and circular system cannot be dynamically stable with the adopted planet masses, but can be consistent when they are about $2~$\Mj~lighter. 
  In our images, HR 8799 b, c and d are clearly detected. 
  The measured contrasts between the host star and each planet are very similar with the previously reported ones.
  No new orbital motion for planet b, d, and d is found from our observations compared to the latest reported astrometric measurements by then end of 2010. 
  Finally, the e component could not be retrieved with high signal to noise ratio due to the short amplitude of parallactic angle excursion ($20.9~$deg).

\subsection{Detection performances}
\label{sec:det_perf}

\begin{figure}[t]
\centering
\includegraphics[width=9cm]{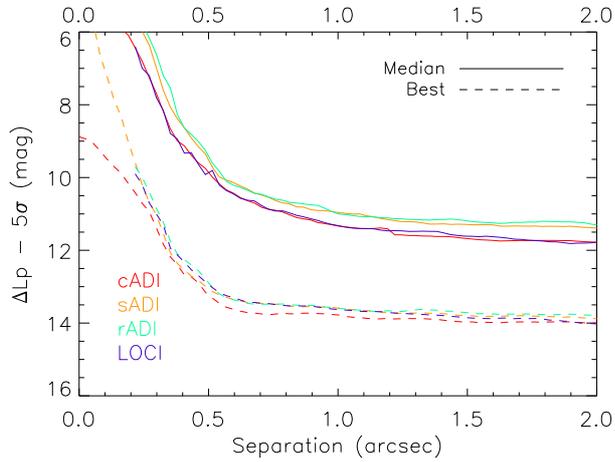}
\caption{Survey detection limits in L'-band contrast between the central star and any point source vs the angular separation, at the $5\sigma$ level, using VLT/NaCo in ADI mode. Solid lines are representative of median performances whereas the bottom dashed ones are for the best performance reached by our survey. cADI (red) and LOCI (blue) are very similar whereas sADI (orange) and rADI (aqua) remain slightly above. Note that the LOCI curves stop at $0.15~$\asec due to the central exclusion area.}
\label{fig:detlim}
\end{figure}

\begin{figure*}[t]
\centering
\includegraphics[width=9cm]{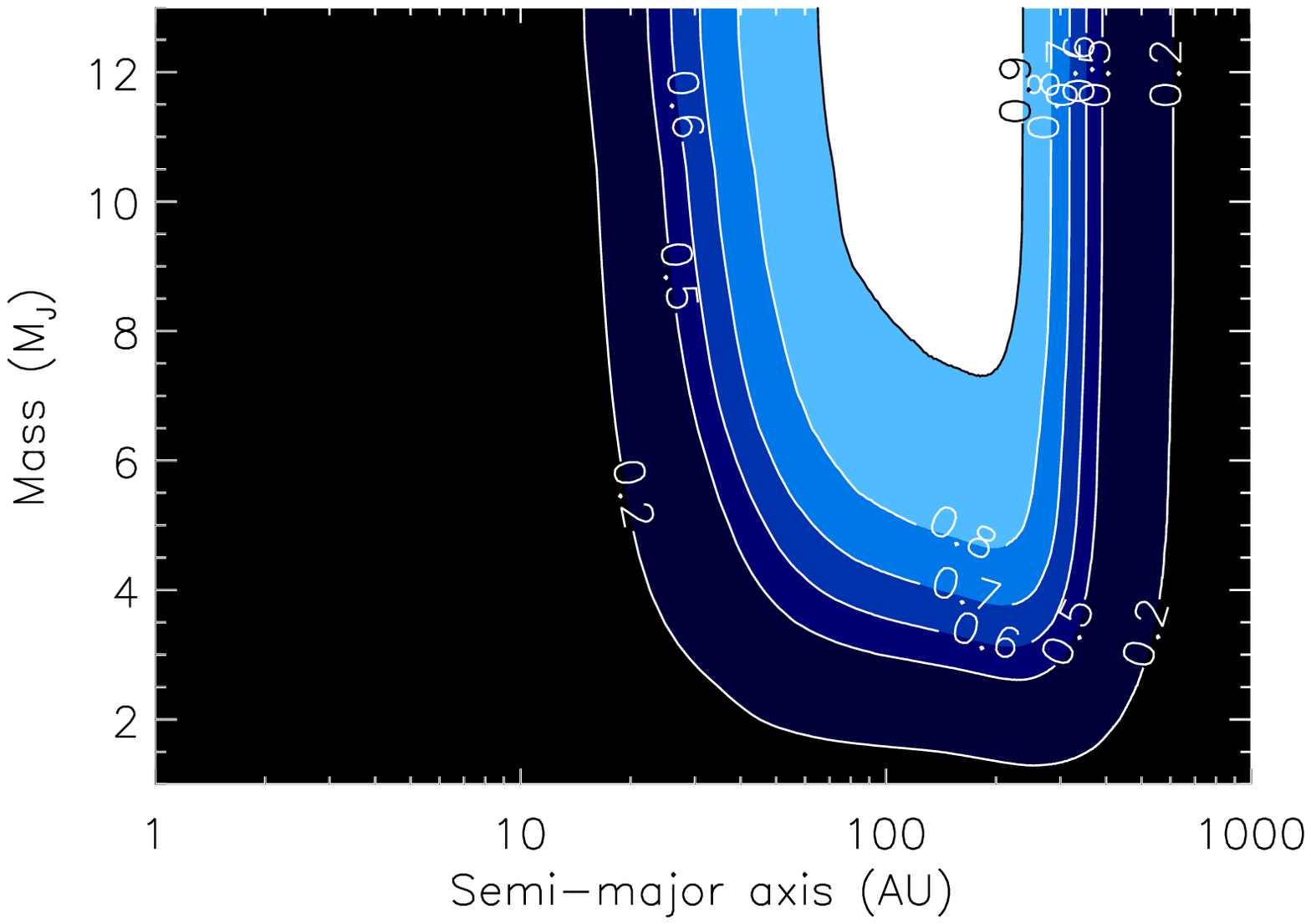}
\includegraphics[width=9cm]{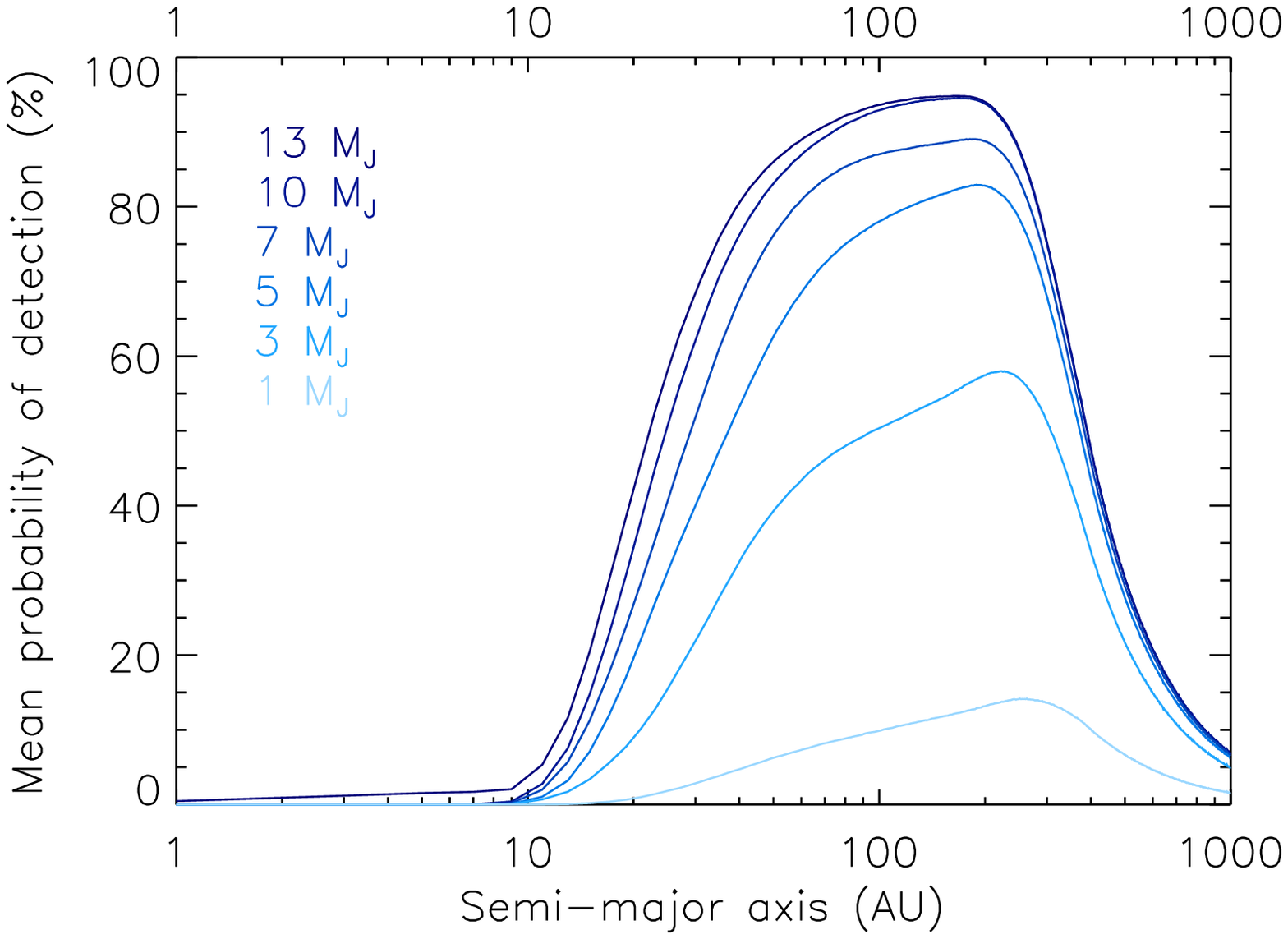}
\caption{Mean detection probability map of our survey as a function of the mass and semi-major axis  (\textbf{left}) and probability curves for different mass bin as function of the semi-major axis (\textbf{right}). The mean is obtained over all targets of the survey. The detectability of the simulated planets was compared to the detection maps from LOCI and nADI algorithms and using COND03 \citep{baraffe03} evolutionary models.}
\label{fig:meanprob}
\end{figure*}

Typical contrasts reached by our survey using ADI algorithms and
nADI algorithms are presented on Figure \ref{fig:detlim}. Note that
using cADI/sADI/rADI or LOCI, the performances are very similar,
except within the exclusion area of LOCI. The median azimuthally
averaged L' contrast vs the angular separation is plotted together
with the best curves. The typical range of detection
performances at all separations beyond $0.5~$\asec is about $2~$mag
with a median contrast of $10~$mag at $0.5~$\asec, $11.5~$mag at
$1~$\asec ,and slightly below $12~$mag at $2~$\asec. Our best
performances even reach very deep contrast, up to $13.5~$\asec at
$0.5~$\asec around HIP 118121.

The detection limits (2D-maps and 1D-curves) were converted to
absolute L' magnitudes using the target properties and to predicted
masses using the COND03 \citep{baraffe03} evolutionary models for the NaCo passbands.

The overall sensitivity of our survey can be estimated using
Monte-Carlo simulations. We use an optimized version of the MESS code \citep{bonavita12} to
generate large populations of planets with random physical and orbital
parameters and check their detectability by comparing with the
deep detection limits of our survey. We performed simulations with a
uniform grid of mass and semi-major axis in the
interval $[1,20]~$\Mj~and $[1,1000]~$AU with a sampling of
$0.5~$\Mj~and $2~$AU. For each point in the
grid, $10^4$ orbits were generated. These orbits are randomly oriented
in space from uniform distributions in $\sin(i)$, $\omega$, $\Omega$,
$e\le0.8$, and $T_p$\footnote{They correspond to respectively the inclination, 
the argument of the periastron with respect to the line of nodes, the longitude of the ascending node, 
the eccentricity, and the time of passage at periastron.}. The on-sky projected position (separation and
position angle) at the time of the observation is then computed for
each orbit. Using 2D informations, one can take into account
projection effects and constrain the semi-major instead of the
projected separation of the companion. We ran these
simulations for each target to compute a completeness map with
  no a priori information on the companion population and therefore
  considering a uniform distribution in mass and semi-major axis. In this case,
the mean detection probability map of the survey is derived by
averaging the $59$ individual maps. The result is illustrated on
Figure \ref{fig:meanprob} with contour lines as function of
sma and masses. Note that the decreasing detection
probability for very large semi-major axis reflects the fact that such
companion could be observed within the FoV only a fraction of their
orbits given favorable parameters. The peak of sensitivity of our
survey occurs at semi-major axis between $40$ and $300~$AU, with the
highest sensitivity around $100~$AU. Our survey's completness
  peaks at $94~$\%, $80~$\%, and $58~$\% for $10~$\Mj~and $5~$\Mj~at $100~$AU, and $3~$\Mj~at
  $220~$AU, respectively. The overall survey's sensitivity at $1~$\Mj
  is very low, with a maximum of $15~$\% at $141~$AU.

\subsection{Giant planets around resolved disks}

Among the targets of our sample, HD 142527 remains the youngest
one. It has been identified by \citet{acke04} as member of the very
young ($\simeq5~$Myr) Upper Centaurus-Lupus association. HD 142527 was
selected to take advantage of the capability offered by VLT/NaCo in
thermal and angular differential imaging to resolve for the first time
at the sub arcsecond level its circumstellar environment. Indeed,
\citet{fukagawa06} detected a complex transitional disk in NIR with a
huge gap up to $100~$ AU. Our observations reported by
\citet{rameau12} confirm some of the previously described structures
but reveal important asymmetries such as several spiral arms and a
non-circular large disk cavity down to at least $30~$AU. The achieved
detection performances enable us to exclude the presence of brown
dwarfs and massive giant planets beyond $50~$AU. In addition, two
sources were detected in the FoV of NaCo. The relative astrometry of
each CC was compared to the one extracted from archived observations
and to the track of the relative position of stationary background
object. Both were identified as background sources. Finally, the
candidate substellar companion of $0.2~$M$_\odot$ at $13~$AU recently
reported by \citet{biller12} may be responsible for the structured
features within the disk. These structures might also be created by type II migration or by planetary formation through GI. Further investigations
of the environment of HD 142527 could set constraints on planetary
formation and disk evolution.

\begin{figure}[t!]
\centering
\includegraphics[width=9cm]{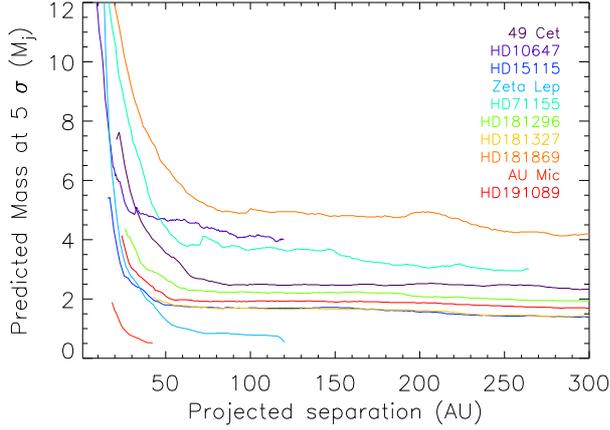}
\caption{Detection limits of giant planets in Jupiter mass vs the projected distance from the central star for stars with known debris disks. The detections are expressed at the $5\sigma$ level, using VLT/NaCo in ADI mode, after LOCI (close-in region) and nADI (background limited region) processing. The COND03 \citep{baraffe03} mass-luminosity relationship is used to convert from L' contrast to Jupiter mass. Note that for AU Mic, the used model does not go under $0.5~$\Mj, so we cut the curve when it reaches this limit.}
\label{fig:lim_disk}
\end{figure}

In our sample, we also have included some stars surrounded by debris disks
that have been resolved in previous observations ( 49 Cet, HD 10647, HD 15115, Zeta Lep, 30 Mon, HD 181296, HD 181327, HD 181869, HD 191089, and AU Mic). Both observation and reduction processes having been designed to search for point sources,
we do not report results about disk properties. However, we investigate the presence of giant planets and plot in the Figure \ref{fig:lim_disk},
the detection limits about these stars in terms of mass vs projected separation. These limits could help to constrain some disk properties which can be created by gravitational perturbation of giant planets. The detection sensitivity around AU Mic reaches the sub Jovian mass regime at very few AUs from the star because AU Mic is a very nearby M dwarf. On the other hand, the detection limits around HD 181869 are not very good due to bad quality data. We remind that these limits are azimuthally average so that there might be affected by the presence of the disk.

\section{Giant planet properties, occurrence and formation mechanisms}
\label{sec:statistics}

The frequency of giant planets $f$ can be derived using known planets and detection limits in case of a null detection. For an arbitrary giant planet population, one can compute  within the mass and semi-major axis ranges probed by the survey.
Numerous deep imaging surveys did not report the detection of at least one substellar or planetary 
mass companion. The authors \citep[e.g.][]{kasper07,lafreniere07,nielsen10,chauvin10} 
nevertheless performed statistical analysis with MC simulations to fully exploit the potential of their data and provided upper limits to the frequency of planets. \citet{vigan12} tooks into account the planets already
identified ($\beta$ Pic, HR 8799) to derive also lower limits to this frequency.

In this section, we derive the rate of wide-orbit giant planets following the statistical approach used in previous works
 \citep{carson06,lafreniere07,vigan12} and described in the appendix. Similarly to \citet{bonavita12}, we take into account the binary status of some stars to exclude semi-major axis values for orbits which would be unstable. The whole section relies on the two previously defined sub-samples : 37 A-F stars and 29 A-F dusty stars (Table \ref{tab:sample}). First, the frequency of wide-orbits planets is derived assuming a uniform distribution. We then use and discuss the extrapolation of RV statistics to wide-orbits in light of DI planets. Planet formation is finally considered, GI and CA, to estimate the impact on the observed occurrence of giant planets.

\subsection{Occurrence of giant planets from a uniform distribution}
\begin{figure}[t]
\centering
\includegraphics[width=9cm]{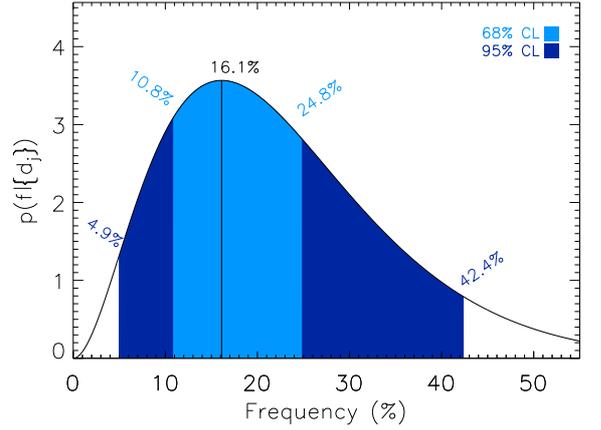}
\caption{A-F sample probability density of the fraction $f$ of stars hosting at least one giant planet at wide orbit, taking into account the detections ($\{d_j\}$) of the two planetary systems
$\beta$ Pictoris and HR 8799 and a linear-flat prior. The interval considered is $[1,1000]~$AU and  $[1,13]~$\Mj. The confidence interval at $68~$\% confident level, labelled as CL, (blue) and at $95~$\% CL (dark blue) are over plotted to the distribution. A uniform planet population has been generated with random orbital parameters.}
\label{fig:proba_density}
\end{figure}

Assuming a uniform distribution of planets in a grid $[1,1000]~$AU and $[1,13]~$\Mj, we use MC 
simulations to measure the detection probability $p_j$ around each star given the detection sensitivity, 
as in section \ref{sec:det_perf}. The probability density function is then derived using equations
 \ref{eq:likelihood} and \ref{eq:proba_density}, assuming a flat prior. Finally, the confidence interval of the true $f$ is computed using equation \ref{eq:tail}. 
  
First, we focus on the relevant A-F statistical sample (see Table \ref{tab:stat}). In this sample, 2 stars harbor at least one giant planet, $\beta$ Pictoris and HR 8799. Since these two stars match our selection criteria, they were originally included in our sample, even if the giant planets have been discovered by other observations. We thus take, in our analysis, at least two planetary detections (since our observations lead to the confirmation of their status). The Figure \ref{fig:proba_density} shows the posterior distribution as function of $f$ in the interval $[1,1000]~$AU and $[1,13]~$\Mj. The observed rate of giant giant planets at wide orbit leads $f$ to be $16.1_{-11.2}^{+26.3}~$\% with a confidence level (CL) of $95~$\%. At $68~$\% CL, this rate becomes $16.1_{-5.3}^{+8.7}~$\%. If one considers the sample of 29 A-F dusty stars, thus with the same planet detections, the giant planet occurrence is $21.4_{-14.9}^{+35.7}~$\% at $95~$\% CL or $21.4_{-7.1}^{+13.8}~$\% at $68~$\% CL. Due to our poor sensitivity to close-in and/or low mass planets, these values are relatively high. If we restrain the interval of interest to $[5,320]~$AU and $[3,14]~$\Mj~as in \citet{vigan12}, then the A-F sample has an occurrence of giant planets of $7.4_{-2.4}^{+3.6}~$\% at $68~$\% CL which matches the results obtained by the authors. 

The same approach can be done to derive the frequency of brown dwarfs. Taking into account the detection of HR 7329 b in the A-F sample, $f$ is $6.5_{-5.0}^{+16.6}~$\% and $8.7_{-6.6}^{+29.4}~$\% in the A-F dusty sample in the interval $[1,1000]~$AU and $[14,75]~$\Mj~at $95~$\% CL. The confidence interval is smaller compared to the statistical results for giant planets due to our high sensitivity to brown dwarfs.

Finally, the full survey of 59 young, nearby, and B- to M-type stars can also give some constraints on the occurrence of planets within a broad sample of stars. Since the companions to AB Pic, HR 7329, HR 8799, and $\beta$ Pictoris were previously detected out of our observations, we consider here a null detection within $55$ stars of our survey. Using equation \ref{eq:CL_null}, an upper limit to the frequency of giant planets can be derived with our detection limits and MC simulations. It comes out that less than $25~$\% among our $55$ stars harbors a giant planet in the range $[40,600]~$AU and $[5,13]~$\Mj~at $95~$\% CL. Note that this upper limit sharply increases towards smaller mass planets and also to a wider semi-major range due to our poor sensitivity. We also remind that this sample is statistically less relevant than the previous ones since it is more heterogenous in terms of stellar mass, distance, and spectral type.

\subsection{Giant planet population extrapolating radial velocity results}
\label{sec:RV}

Radial velocity results provided a lot of statistical results on the giant planet properties but also on the distribution of the population with respect to the mass and/or semi-major axis. However, such results are intrinsically limited so far to close-in planets (typically $3-5~$AU). Numerous publications present statistical analysis on giant planet detected by deep imaging using extrapolation of the RV frequencies and distributions to planets on larger orbits \citep{lafreniere07,kasper07,nielsen10,vigan12}. We briefly present in the following sections the outcomes of our sample based on the same approach, considering the detections around $\beta$ Pic and HR 8799.

\subsubsection{Extrapolation of the radial velocity planets distribution} 

\begin{figure}[t!]
\centering
\includegraphics[width=9cm]{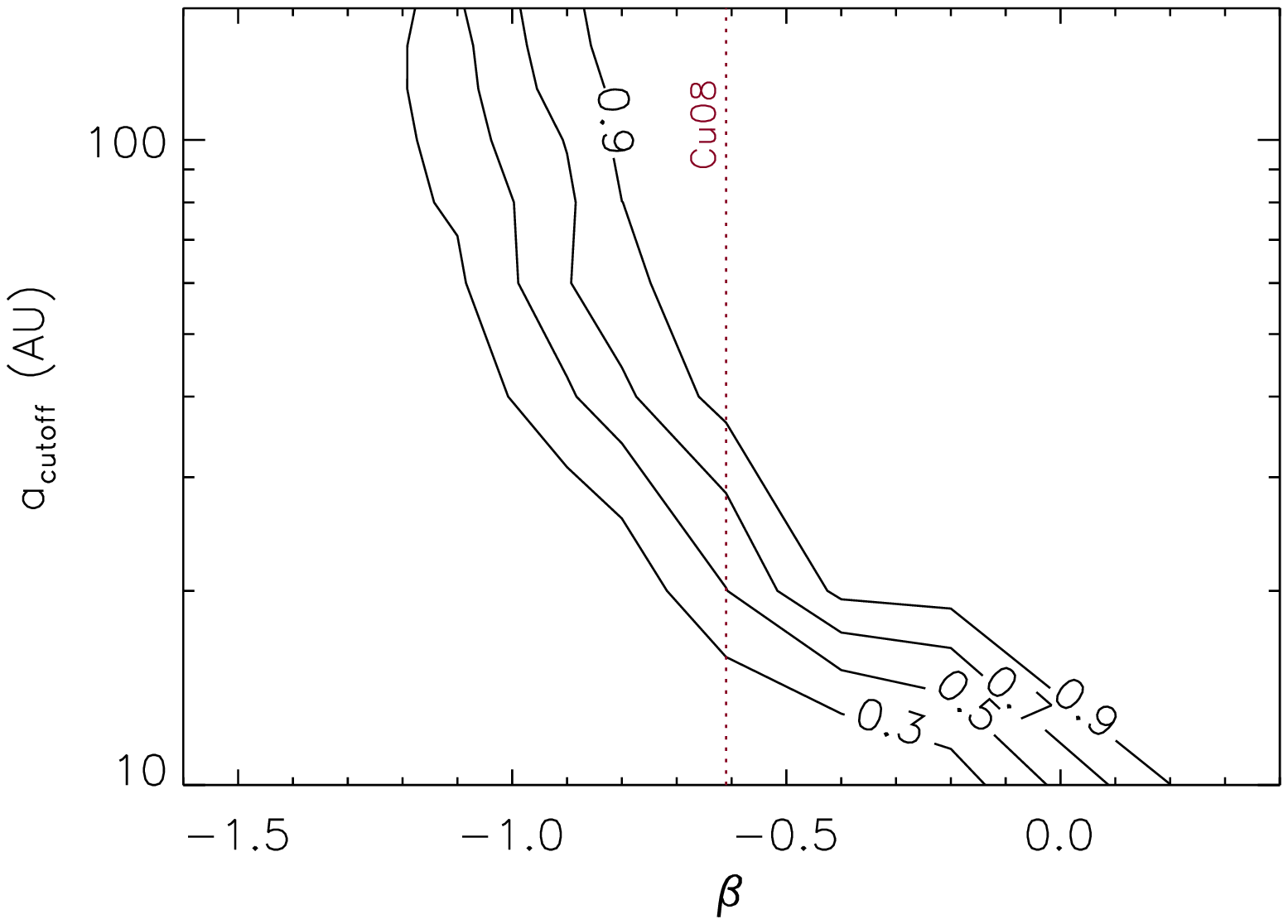}\\
\includegraphics[width=9cm]{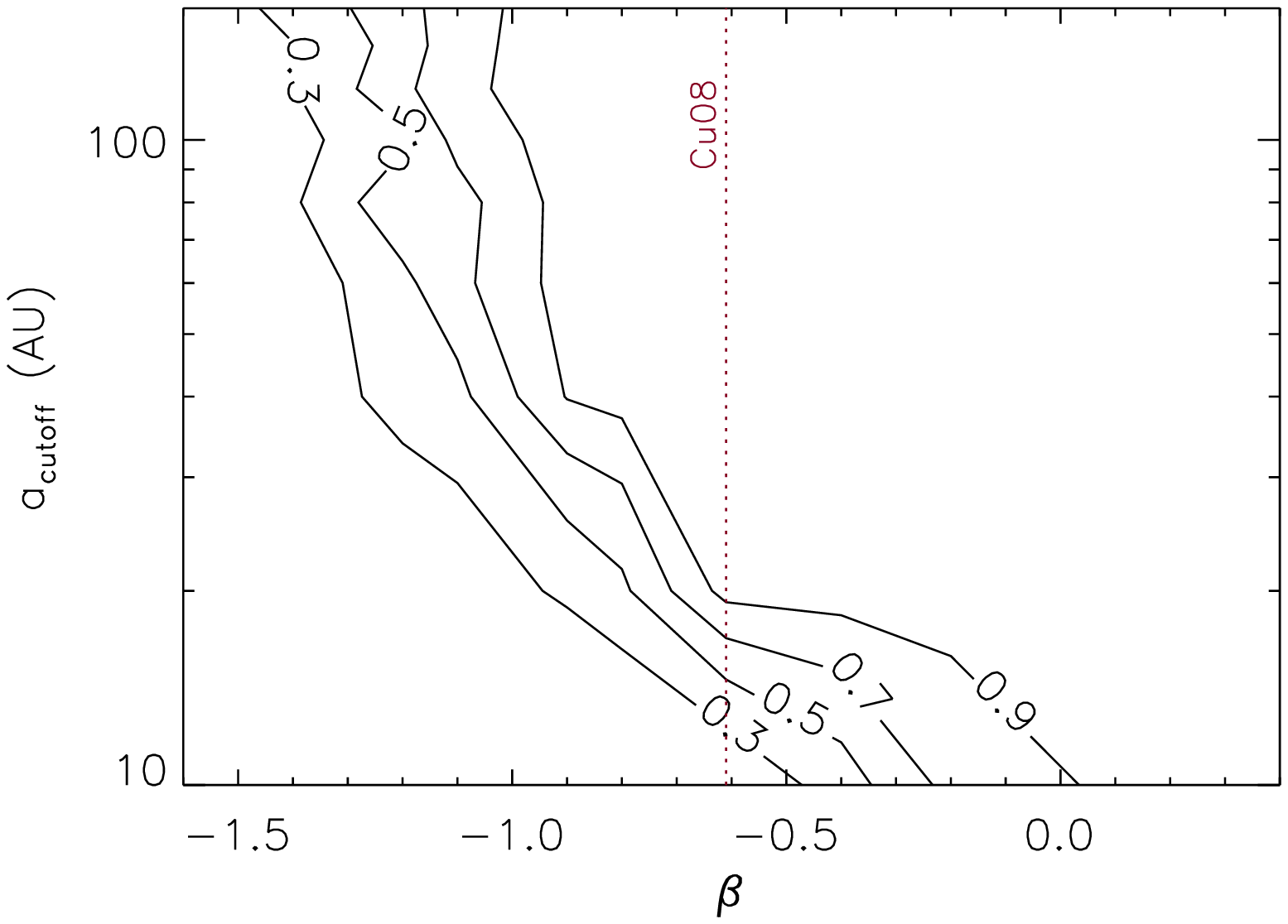}
\caption{Contours showing the confidence level at which we can reject a planet distribution following a mass power law $dN\propto M^\alpha dM$ and a semi-major axis one $dN\propto a^\beta da$ at a given semi-major axis cutoff. \textbf{Top :} Semi-major axis cutoff vs semi-major axis power law index $\beta$ for $\alpha=-1.5$. \textbf{Bottom :} Semi-major axis cutoff vs semi-major axis power law index $\beta$ for $\alpha=1.1$, i.e. with more massive planets. The upper cutoff for the \citet{cumming08} $\beta$ index of$ -0.61$ is over plotted (dotted line). These figures use LOCI and nADI detection limits and the COND03 \citep{baraffe03} for the mass-luminosity relationship.}
\label{fig:f_Cu08_cutoff}
\end{figure}

We assume that the mass and semi-major axis distributions follow the simple parametric laws of index $\alpha$ and $\beta$ : $dN\propto M_p^\alpha dM_p$ with $\alpha=-1.31$ and $dN\propto a^\beta da$ with $\beta=-0.61$\footnote{While they derived the distribution for mass and period  in logarithmic bins using $\alpha$ as the index for $P$, we used the mass and semi-major axis distribution with linear bins using $\alpha$ referring to $a$.}  \citep{cumming08}. Here, we blindly extrapolate the distribution to larger semi major axis while it is formally valid only for planets with semi-major axis below $\simeq3~$AU.

For this calculation, we populate a grid of mass and semi-major axis in the intervals $[1,13]$ and $[1,1000]$ (\Mj~and AU) to the \citet{cumming08} power-laws and run MC simulations to derive the probability density distribution as in section \ref{sec:statistics}. If the giant planet population on wide orbits follows the RV power-laws, then their frequency range from our study is $22.0_{-15.3}^{+37.4}~$\% at $95~$\% CL or $22.0_{-7.3}^{+14.8}~$\% at $68~$\% CL in the A-F sample. This rate becomes $28.3_{-19.7}^{+37.9}~$\% at $95~$\% CL or $28.3_{-9.6}^{+19.6}~$\% at $68~$\% CL in the A-F dusty sample.

However, there are intrinsic limitations on this study and the outputs, eventhough close in values as the ones reported among an uniform distribution, have to be taken with care. The used distribution fits the statistic for solar-type stars up to few AU (coming from RV surveys) and is arbitrarily extrapolated to large separations. There is also no evidence that the few planets detected so far at large orbit separations have similar properties and distributions. 

\subsubsection{Constraining the parametric laws for the giant planet distribution}

This likelihood approach answers the question : 
'How consistent is a given giant planet population with our observing results?' Answering this 
question requires 1/ to know all giant planet population parameters
and 2/ to know the fraction of stars with giant planets according to this given distribution. For each star, we can
 then derive the number of expected detections given the detection sensitivity and compare to our observations. 
 Such comparison allows us to constrain a given distribution of wide orbit giant planets. Likewise, a giant planet population in which $95~$\% of the predicted planets would have lead to detections can be considered as strongly inconsistent with our survey.
Finally, this study relies on the strong assumption that we know the frequency of giant planets in the range where our survey is sensitive to.

In the following, we use a population of planets given by power laws similar to the ones from \citet{cumming08} and we also add an additional parameter which is \acut, the semi-major axis beyond which there are no planets. Our intervals of interest for the simulation are $[1,1000]~$AU and $[1,13]~$\Mj, normalized with $f=10.5~$\% over the range $[0.3,10]~$\Mj, $[2,2000]~$ days in period from \citet{cumming08}. $f_\mathrm{norm}$ is thus set with the ratio of the integrated power laws for a pair ($\alpha$, $\beta$) over $[1,13]~$\Mj~and $[1,1000]~$AU and the same over the RV ranges.

We explored a grid of $\alpha$, $\beta$, and \acut with a sampling of $0.2$ for the power law indices and $20~$AU for the cutoff to derive the expected number of planets for each combination of parameters over the A-F sample (similar results are obtained with the A-F dusty sample). We illustrate the confidence level at which we can reject each model in Figure \ref{fig:f_Cu08_cutoff} as a function of $\beta$ and \acut for the A-F sample for two values of $\alpha$ : $-1.5$ and $1.1$, values corresponding to the extrema of our grid and thus showing the trend of the rejections. All results (ours and previous publications) are consistent with a decreasing number of giant planets ($\beta \le -0.61$) while their mass increase. Considering the \citet{cumming08} distributions, a semi-major axis cutoff around $45-65~$AU at $95~$\% CL is found.

 We remind that mixing power-laws derived from RV and giant planets with possibly different formation processes and evolutions has to be considered with caution.
 
\subsection{Giant planet formation by gravitational instability}
\label{sec:GI}

\begin{figure*}[t!]
\centering
\includegraphics[width=9cm]{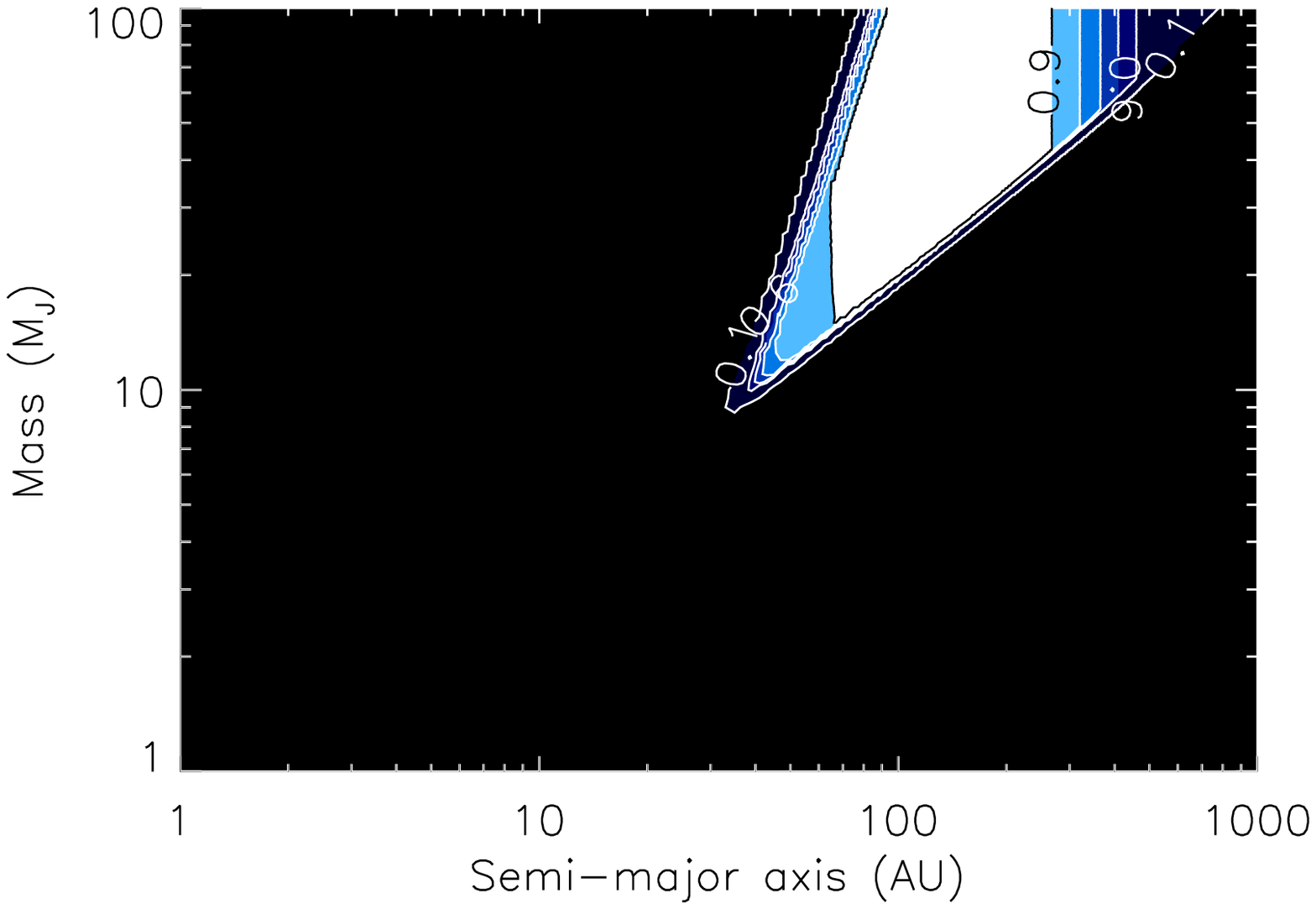}
\includegraphics[width=9cm]{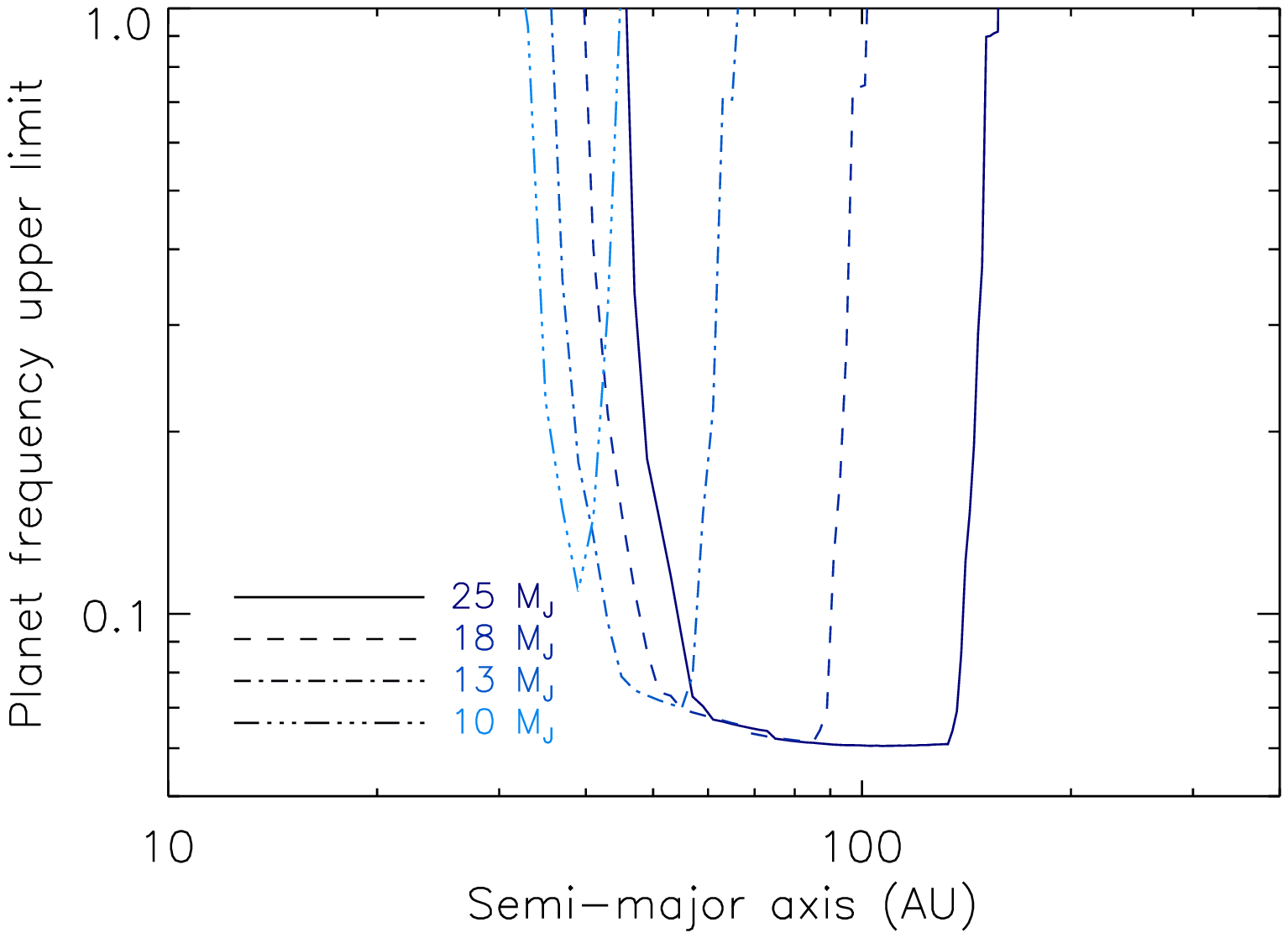}
\caption{\textbf{Left :} Mean detection probability map of the A-F sample as function of the mass and semi-major axis of the substellar companion. A uniform grid distribution has been used to generate the population but the formation limits derived from the gravitational instability models exclude for each star the planets which do not fulfill both criteria. The detection maps from LOCI and nADI algorithms were used with COND03 \citep{baraffe03} evolutionary models to convert from contrast to mass. Contour lines are regular from 0.5 to 0.9 plus one at 0.1. \textbf{Right : }Corresponding estimation of the upper limit, with a confidence level of $95~$\%, on the fraction of stars from the A-F sample, harboring at least one object companion in the same semi-major axis range. The curves for $25~$\Mj, $18~$\Mj, $13~$\Mj~and $10~$\Mj~have been plotted since lower mass planets are not allowed to form via the disk instability mechanism.}
\label{fig:f_GI}
\end{figure*}

Gravitational instability is a competitive scenario to form giant planets, specially at large separations. Such a process becomes more efficient within massive disks, i.e. around massive stars. Since our statistical sample contains A-F and/or dusty stars, i.e. massive stars, we were strongly tempted to test the predictions of GI models with our observing results. We hence adopted the same approach as \citet{janson11}. The reader is refered to \citet{gammie01} for a detail description. The 1D current model of disk instability provides formation criteria, which if fulfilled, create an allowed formation space in the mass-sma diagram. The first one is the well known Toomre parameter \citep{toomre81} which has to be low enough to allow local gravitational instability in a keplerian accretion disk :
$$
Q=\frac{c_s\kappa}{\pi G\Sigma} \le 1
$$
where Q is the Toomre parameter, $c_s$ the sound speed, $\kappa$ the epicyclic frequency and $\Sigma$ the gas surface density. The Toomre parameter is fulfilled at larger radius only when the local mass is high enough. Therefore, fulfilling the Toomre criteria leads to a given value $\Sigma$ which can be converted to mass and thus states a lower limit in the mass-sma diagram :
$$
\pi r^2 \Sigma \ge \frac{H}{r}M_\star
$$ 
where $H=c_s/\Omega$ is the disk scale height. The other parameter which drives the instability is the cooling time, $\tau_c$ which, if higher than a few local keplerian timescale $\Omega^{-1}$, i.e. at small separation, stabilizes the disk through turbulent dissipation \citep{gammie01,rafikov05}. It thus puts an upper boundary in the mass-sma diagram :
$$
M_f=\Sigma (2\pi H)^2
$$
where $2\pi H$ is the wavelength of the most unstable mode. Such boundaries, being global and excluding long term evolution, assume planets formed in-situ with masses of the disk fragments.

\begin{figure}[th]
\centering
\includegraphics[width=9cm]{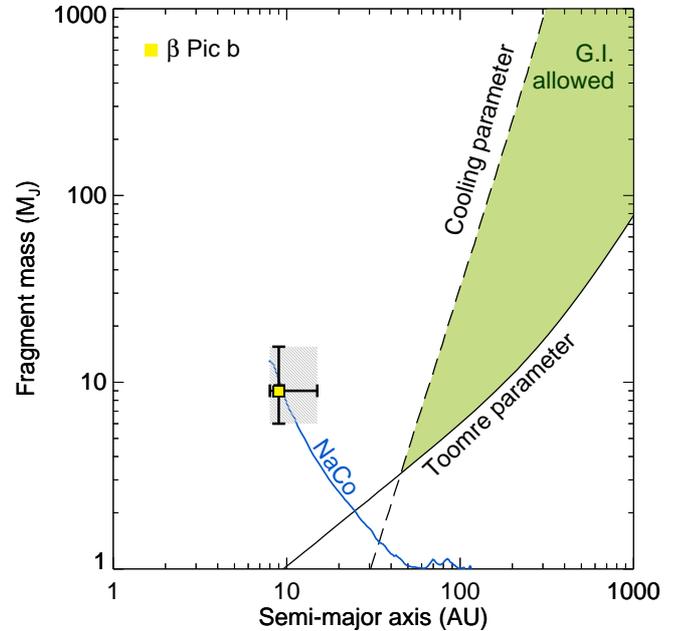}
\caption{Disk instability model predictions for $\beta$ Pictoris. The lower solid curve corresponds to the Toomre criteria which excludes the formation of planets below it. The upper dashed curve corresponds the cooling criteria which only allows disk fragmentation below it. The allowed formation space is in between. The 1D average detection limit curve has been over plotted (blue solid line) but with a projected separation. The location of $\beta$ Pictoris b has been over plotted to the graph with new error bars from \citet{bonnefoy13}.}
\label{fig:boundaries}
\end{figure}

The model computes both boundary curves for each star in the sample, taking into account the stellar mass, luminosity, and metallicity, the later being extracted from \citet{ammons06} or set to the solar one when the information was not available and luminosities derived from isochrones of \citet{siess00} using their absolute K magnitude, spectral type age, and metallicity. The model is very sensitive to the stellar luminosity since strong illumination favors the disk to be gravitationally stable \citep{kenyon87}. Figure \ref{fig:boundaries} shows one example for $\beta$ Pictoris. The Toomre and cooling criteria are fulfilled around $40~$AU for very massive planets ($\ge 3~$\Mj) and this trend rapidly increases with the separation, thus leading to the brown dwarf and stellar regimes. Note that considering a lower mass star would lead to push the boundaries inwards.

We then run MC simulations in a uniform grid of mass and semi-major axis in the interval $[1,100]$ and $[1,1000]~$(\Mj~AU) as in section \ref{sec:statistics}. Points of the grid out of the allowed range for each star are removed according to the formation limits. We remind that these predictions are not normalized due to the absence of knowledge on physical and statistical properties of protoplanetary disks in which GI starts. The mean detection probability over the A-F sample is plotted on Figure \ref{fig:f_GI}, left panel. Only high mass planets and brown dwarfs fulfill the formation criteria and there are almost all detectable. 

$\beta$ Pictoris b, HR 8799 b, c, and d are too light and too close to their stars, so they do not fulfill both GI boundary conditions ($M\le10~$\Mj~below $70~$AU). Therefore, we cannot use these detections to derive the rate of giant planets according to GI mechanism. We instead estimate the upper limit on $f$, \fmax using equation \ref{eq:CL_null}. We derive and plot \fmax (Figure \ref{fig:f_GI}, right panel) for the A-F sample only for the mass regime allowed by this approach, which extends between very few tens of AU. The curves are offset one from another due to the fact that higher mass object can be formed in-situ at larger distance from the central star. It came out that less than $20~$\% ( $\le 23~$\% for the A-F dusty sample) stars harbor at least a $13~$\Mj~planet between $40$ and $60~$AU and less than $25~$\% ($\le27~$\%) a $10~$\Mj~in the range $[32,45]$~AU.

On the other hand, 1/ Figure \ref{fig:f_GI}, left panel, shows our high sensitivity to brown dwarf on wide orbits, and 2/ HR 7329, belonging to the A-F dusty sample so as to the A-F one, hosts a detected brown dwarf companion for which the formation is allowed according to our GI model. We can therefore estimate the rate of formed objects as in section \ref{sec:statistics}. Since GI can form planetary to brown dwarf mass objects, we explore the full range $[1,75]$ and $[1,1000]~$(\Mj~and AU). We found that $f$ equals to $3.2_{-1.0}^{+2.2}~$\% for the A-F sample and $4.3_{-1.3}^{+2.4}~$\% for the A-F dusty one at $68~$\% CL if formed by this mechanism.

It comes out that such GI boundaries prevent the formation of low mass and close-in giant planets but would enhance the presence of brown dwarf and low mass star companions. Since high mass stars would facilitate the GI mechanism by harboring massive disks, one would expect to find a higher occurrence of low mass stars or substellar companions rather than planets and a continuous distribution between the wide orbit giant planets detected so far and higher mass objects \citep{kratter10}.

This approach is a first step towards understand planet formation by GI and the analysis can be improved by taking into account the following steps. First, \citet{meru11} shows that using proper 3D global radiative transfer codes and hydrodynamical simulations, closer-in disk region might become unstable, phenomena which was prevented assuming global simple cooling time law. \citet{kratter11} refines the definition of $Q$ and the cooling time leading GI to be possible at smaller separations. Second, the probability of clump formation towards planets was assumed to be one but long lived clumps require careful considerations about disk dynamics \citep{durisen07}. Then, clump evolution \citep[e.g.][]{galvagni12} and fragmentation might lead to the formation of lower mass planets. HR 8799 seems a good test-case for such hypothesis. Indeed, the three outer planets orbit the star too far away to have form via core accretion. Gravitational instability naturally comes out as alternative scenario. However, each planet, with its mass and separation, does not fulfill the Toomre and cooling time criteria following our models. Considering all three together, even four mass planets ($\simeq30~$\Mj) onto a single disk fragment at a mean separation satisfies our boundaries. One might speculate that this clump would have then broken after collapse leading to individual evolution of the planets. Finally, long term clump evolution was also not taken into account in our study. A self-graviting clump will still accrete a large amount of gas. Even if the disk fragment into an initially planetary mass clump, this fragment will accrete gas, become more massive, and thus might exceed the deuterium burning limit mass \citep{boss11}. However such formation takes about $10^5~$yr, gas accretion is expected to be turned off by disk dissipation by strong UV irradiation of the surrounding high mass stars in the host-star forming region \citep{durisen07} so that one might expect light clump growth to stop before getting too massive. 

\subsection{Giant planet formation by core-accretion}

\begin{figure}[t]
\centering
\includegraphics[width=9cm]{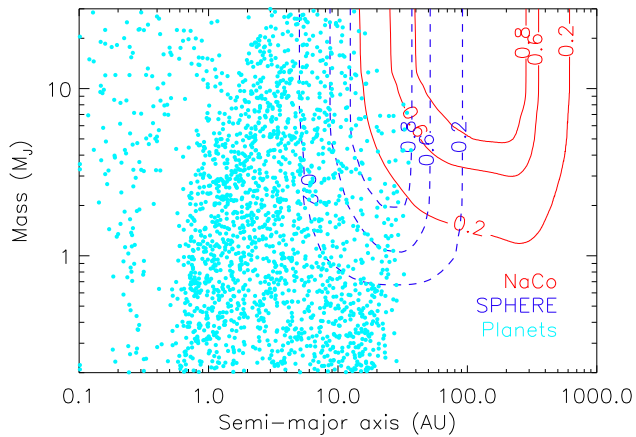}
\caption{Synthetic planetary population predicted assuming the core-accretion scenario similar as in Mordasini et al. (2012) for the case of $2~$\Mj~central stars. The $20~$\%, $60~\%$ and $80~$\% detection probability curves (red) are overplotted to the planet population revealing the poor sensitivity of our survey on such kind of formed planets. The SPHERE/IFS performances (dashed) have also been plotted assuming contrast curves from \citet{mesa11}. The short high-sensivity window is due to the small FoV of the IFS ($1.77~\!''$) which can be overcome with the larger IRDIS FoV. }
\label{fig:f_CA}
\end{figure}

We now investigate the planet formation and evolutionary model of \citet{mordasini12} which predicts the final state of planets following the core accretion scenario and normalized by the frequency of observed disks. The synthetic population is calculated assuming a $2~$\Msun~central star, a mean disk lifetime of $4$~Myr, that gap formation does not reduce gas accretion\footnote{This question is still debated since gap formation might lead to a reduction (e.g. \citealt{lubow99}), but this might be counterbalanced by the effects of eccentric instability \citep{kley06}.}, and considering one embryo per disk-simulation (hence no outward migration of resonant pairs or scattering possible). A comparison with RV data shows that this simulation produces too massive and too close-in giant planets, but this synthetic population can be considerer as a rough approximation (see also in \citealt{alibert11}). We then try to test the predicted expected population at wide orbits with this approach so with direct imaging results. We run the MC simulations ran with planets extracted from this synthetic population. $10^4$ random orbits were generated for each planet and the projected position on the sky was computed as before.

In Figure \ref{fig:f_CA}, we show the extracted planet population (already normalized) as well as the detection probability at $20~$\%, $60~$\%, and $80~$\% for the A-F sample. It comes out that there is no incompatibility between the synthetic population and the results of our survey. Indeed, we are marginally sensitive to the farthest predicted giant planets (the predicted fraction with detectable planets is around $0.06~$\%) so we cannot reject their existence. Notwithstanding, CA is expected to become inefficient to form planets at separations larger than a tens of AU.  Moreover, the domain probed by our detection sensitivity, i.e. beyond $40~$AU, well matched the region where CA inoperates as seen in Figure \ref{fig:f_CA}. The $5-20~$AU gap between deep imaging surveys and those from radial velocity will be at least partly fill in thanks to the forthcoming extreme adaptive optic instruments VLT/SPHERE \citep{beuzit08} and Gemini/GPI \citep{macintosh08} thanks to excellent detection limits and lower inner working angles. Using the same MC simulations, we compute the mean detection probability curves (Figure \ref{fig:f_CA})\footnote{Only a small semi-major axis range is covered by the curves since we considered only the IFS instrument which has a small FoV ($1.77~\!''$). Larger FoV will be provided by the IRDIS focal instrument.}. We show that the improved capabilities of SPHERE will allow indeed to decrease this gap, by detecting a few Jupiter-like planets down to $5~$AU. However, its overall sensibility (in mass and separation) will not allow to entirely probe the predicted giant planet population (the predicted fraction with detectable planets being around $0.6~$\%). Another complementary way to fill this gap is to use both RV and direct imaging on selected young targets, as demonstrated in Lagrange et al. (2012, subm.).


\section{Concluding remarks}

Here, we have reported the observations and analysis of a survey of $59$ stars with VLT/NaCo at $L~\!'$-band ($3.8\mu m$) with the goal to detect and characterize giant planets on wide-orbits. The selected sample favors young, i.e. $\le 70~$Myr, nearby, $\le 100~$pc, dusty, and early-type stars to maximize the range of mass and separation over which the observations are sensitive. The optimized observation strategy with the angular differential imaging in thermal-band and a dedicated data reduction using various algorithms allow us to reach a contrast between the central star and an off-axis point source of $12~$mag at $0.3~\!''$, $13.5~$mag at $0.5~\!''$ up to $14~$mag farther away in the best case. Despite the good sensivity of our survey, we do not detect any new giant planet. New visual binaries have been resolved, HIP 38160 confirmed as a comoving pair and HIP 79881 and HIP 53524 confirmed as background objects. We also report the observations of a perfect laboraty-case!
  for disk evolution with the sub-arcsecond resolved disk surrounding HD 142527 (dedicated publication in \citet{rameau12}.

We used Monte-Carlo simulations to estimate the sensitivity survey performance in terms of planetary mass and semi-major axis. The best detection probability matches the range $40-300~$AU, with maxima at $93~$\% for a $10~$\Mj~planet and $58~$\% for a $3~$\Mj~planet. Brown dwarfs would have been detected with more than $70~$\% probability within the same semi-major axis range.

\begin{table}[t]
\caption{Confidence interval of the frequency of giant planets with a confidence level of $68~$\% reported in this work around young, nearby and dusty A-F stars assuming different planet populations, the detections of $\beta$ Pictoris b and the system around HR 8799 for planets, and the detection of HR 7329 b as brown dwarf.}             
\label{tab:stat_final}      
\centering          
\begin{tabular}{llll}     
\noalign{\smallskip}
\noalign{\smallskip}\hline
\noalign{\smallskip}\hline  \noalign{\smallskip}
 Sep. range & Mass range & Frequency & Distribution \\
 (AU) & (\Mj) & (\%) &   \\
\noalign{\smallskip}\hline                  \noalign{\smallskip}
\multicolumn{4}{c}{A-F sample} \\
\noalign{\smallskip}\hline                  \noalign{\smallskip}
$[1,1000]$ & $[1,13]$ & $10.8-24.8$ & flat \\
$[1,1000]$ & $[1,13]$ & $14.8-36.8$ & Cu08 \\
$[1,1000]$ & $[1,75]$ & $2.2-5.4$ & flat+GI\\
\noalign{\smallskip}\hline                  \noalign{\smallskip}
\multicolumn{4}{c}{A-F dusty sample} \\
\noalign{\smallskip}\hline                  \noalign{\smallskip}
$[1,1000]$ & $[1,13]$ & $14.3-35.2$ & flat \\
$[1,1000]$ & $[1,13]$ & $18.7-47.9$ & Cu08 \\
$[1,1000]$ & $[1,75]$ & $2.9-6.7$ & flat+GI\\
  \noalign{\smallskip}\hline  \noalign{\smallskip}
\end{tabular}
\tablefoot{Results on the frequency of giant planets are reported according to a flat/uniform giant planet distribution or to a power law distribution of giant planets with the mass and semi-major axis \citep{cumming08} or driven by formation boundaries according to the gravitational instability scenario. The first two scenarii consider the detections of two planetary systems whereas the later set with one brown dwarf detection since planet formation through GI remains very low because they are not allowed to form closer-in and become too massive farther out.}
\end{table}

A dedicated statistical analysis was carried out to understand and constrain the formation mechanism of giant planets. From literature and archive data, we focused on two volume-limited samples, representatives of almost $60~$\% to more than $70~$\% of the full set of stars being younger than $100~$Myr, closer than $65~$pc, to the South (dec $\le 25~$deg), A-or F-type, and with/without infrared excess at $24~$ and/or $70\mu m$. We computed the frequency of giant planets at wide orbits, in the interval $[1,13]$ and $[1,1000]~$(\Mj~and AU), summarized in Table \ref{tab:stat_final} :
\begin{itemize}
\item in both A-F and A-F dusty samples, two giant planetary systems have been detected so far : $\beta$ Pictoris and HR 8799, yielding a wide-orbit giant planet occurrence between $4.9~$\% and $42.4~$\% for the A-F sample and between $6.5~$\% and $56.9~$\% for the A-F dusty sample at $95~$\% CL, assuming a uniform distribution. These results are consistent with the upper limit found in the litterature and also with the rate of planets around the volume-limited sample of 42 A-type stars by \citet{vigan12} (see results in Table \ref{tab:stat}).
\item  if the population of giant planets on wide orbits follows the distribution of the ones detected by RV below $5~$AU from \citet{cumming08}, the $95~$\% confidence interval for $f$ is $6.7-59.5~$\% for the A-F sample, and $8.6-66.2~$\% for the A-F dusty sample. We remind that such an assumption is probably incorrect as it implicitly assumes that wide orbit giant planets have similar origins and properties as close-in ones around Sun-like stars.
\item planets formed via gravitational instability within protoplanetary disks are expected to be massive and to orbit far away from their host stars if they remain in situ. We consider such planets to form and remain in situ where such instability could occur. Since $\beta$ Pic b, HR 8799 b,c, and d are not allowed to form via GI according to our model, we only computed upper limit to the frequency of giant planets. We find that less than $25~$\% of stars could form and retain a $10-13~$\Mj~between $30-60~$AU in the A-F sample (so as for the A-F dusty sample). Closer in, disk instabilities are quickly prevented so that no planet can be formed, whereas the disk fragmentation rapidly leads to brown dwarfs and stellar regimes farther away.
\end{itemize}
 
These results would corroborate a correlation between the presence of debris disk and giant planets since the rates tend to be slightly eventhough higher in the A-F dusty sample than in the A-F one. They also point towards a similar occurrence of giant planets on wide (from AO imaging) and close (from RV measurement) separations. They suggest a break up of the positive correlation between the separation, the mass, and the distribution derived from close-in CA planets and the population of wide orbit giant planets. The later is consistent with a decreasing distribution with larger semi-major axis. All previous surveys reported such bimodal behavior of the distribution which would be a signpost of different modes of gas giant formation \citep{boley09}. Upcoming extreme AO surveys will probe the transition region between the two regimes revealing if it is continuous, i.e. same formation process or, not \citep{kratter10}.

Since our survey is very sensitive to high mass objects (i.e $\ge 10~$\Mj), we can derive the rate of brown dwarfs to be $6.2_{-1.9}^{+3.6}~$\% in the A-F sample and $8.7_{-2.7}^{+7.8}~$\% in the A-F dusty one, at $68~$\% CL in the interval $[14,75]$ and $[1,1000]~$(\Mj~AU) assuming a uniform distribution, results which are consistent with the literature. From the GI formation boundaries, these rate become $3.2_{-1.0}^{+2.2}~$\% in the A-F sample and $4.2_{-1.3}^{+3.5}~$\% in the A-F dusty one. 

We finally recall that all the detection limit estimations are based on mass-luminosity relations that are still debated. Also, they strongly rely on age estimates which are much less accurate for early-type stars when they do not belong to a moving group. The long term dynamical evolution of planetary system (migration in e.g. \citealt{papaloizou07}, scattering \citealt{crida09,raymond12}) plays also a key role, and the giant planet distribution at a given age could be different from the one at formation stages.

 \begin{acknowledgements}
 The authors are grateful to the anonymous referee and editor for useful advices and comments that improved the readability of this publication.
 JR thanks ESO staff for conducting the observations and A. Vigan for fruitful discussion about MC simulations and for providing his published detection limits.
      This research has made use of the SIMBAD database and the VizieR catalogue access tool,
operated at CDS, Strasbourg, France. The original description of the VizieR service was published in A\&AS 143, 23. This work makes also use of EURO-VO TOPCAT software. The EURO-VO has been funded by the European Commission through contracts RI031675 (DCA) and 011892 (VO-TECH) under the 6th Framework Programme and contracts 212104 (AIDA) and 261541 (VO-ICE) under the 7th Framework Programme. We also use data products from the Two Micron All Sky Survey, which is a joint project of the University of Massachusetts and the Infrared Processing and Analysis Center/California Institute of Technology, funded by the National Aeronautics and Space Administration and the National Science Foundation.
 JR, GC, and AML acknowledge financial support from the French National
Research Agency (ANR) through project grant ANR10-BLANC0504-01. SD acknowldeges partial support from PRIN INAF 2010 "Planetary systems
at young ages".

\end{acknowledgements}

\nocite*{}
\bibliographystyle{aa}
\bibliography{biblio}

\begin{thebibliography}{133}
\expandafter\ifx\csname natexlab\endcsname\relax\def\natexlab#1{#1}\fi

\bibitem[{{Acke} \& {van den Ancker}(2004)}]{acke04}
{Acke}, B. \& {van den Ancker}, M.~E. 2004, \aap, 426, 151

\bibitem[{{Alibert} {et~al.}(2004){Alibert}, {Mordasini}, \&
  {Benz}}]{alibert04}
{Alibert}, Y., {Mordasini}, C., \& {Benz}, W. 2004, \aap, 417, L25

\bibitem[{{Alibert} {et~al.}(2011){Alibert}, {Mordasini}, \&
  {Benz}}]{alibert11}
{Alibert}, Y., {Mordasini}, C., \& {Benz}, W. 2011, \aap, 526, A63

\bibitem[{{Allard} \& {Homeier}(2012)}]{allard12}
{Allard}, F. \& {Homeier}, D. 2012, ArXiv e-prints

\bibitem[{{Ammons} {et~al.}(2006){Ammons}, {Robinson}, {Strader}, {Laughlin},
  {Fischer}, \& {Wolf}}]{ammons06}
{Ammons}, S.~M., {Robinson}, S.~E., {Strader}, J., {et~al.} 2006, \apj, 638,
  1004

\bibitem[{{Augereau} {et~al.}(2001){Augereau}, {Nelson}, {Lagrange},
  {Papaloizou}, \& {Mouillet}}]{augereau01}
{Augereau}, J.~C., {Nelson}, R.~P., {Lagrange}, A.~M., {Papaloizou}, J.~C.~B.,
  \& {Mouillet}, D. 2001, \aap, 370, 447

\bibitem[{{Baraffe} {et~al.}(2003){Baraffe}, {Chabrier}, {Barman}, {Allard}, \&
  {Hauschildt}}]{baraffe03}
{Baraffe}, I., {Chabrier}, G., {Barman}, T.~S., {Allard}, F., \& {Hauschildt},
  P.~H. 2003, \aap, 402, 701

\bibitem[{{Beuzit} {et~al.}(2008){Beuzit}, {Feldt}, {Dohlen}, {Mouillet},
  {Puget}, {Wildi}, {Abe}, {Antichi}, {Baruffolo}, {Baudoz}, {Boccaletti},
  {Carbillet}, {Charton}, {Claudi}, {Downing}, {Fabron}, {Feautrier},
  {Fedrigo}, {Fusco}, {Gach}, {Gratton}, {Henning}, {Hubin}, {Joos}, {Kasper},
  {Langlois}, {Lenzen}, {Moutou}, {Pavlov}, {Petit}, {Pragt}, {Rabou}, {Rigal},
  {Roelfsema}, {Rousset}, {Saisse}, {Schmid}, {Stadler}, {Thalmann}, {Turatto},
  {Udry}, {Vakili}, \& {Waters}}]{beuzit08}
{Beuzit}, J.-L., {Feldt}, M., {Dohlen}, K., {et~al.} 2008, in Society of
  Photo-Optical Instrumentation Engineers (SPIE) Conference Series, Vol. 7014,
  Society of Photo-Optical Instrumentation Engineers (SPIE) Conference Series

\bibitem[{{Biller} {et~al.}(2012){Biller}, {Lacour}, {Juh{\'a}sz}, {Benisty},
  {Chauvin}, {Olofsson}, {Pott}, {M{\"u}ller}, {Sicilia-Aguilar}, {Bonnefoy},
  {Tuthill}, {Thebault}, {Henning}, \& {Crida}}]{biller12}
{Biller}, B., {Lacour}, S., {Juh{\'a}sz}, A., {et~al.} 2012, ArXiv e-prints

\bibitem[{{Biller} {et~al.}(2007){Biller}, {Close}, {Masciadri}, {Nielsen},
  {Lenzen}, {Brandner}, {McCarthy}, {Hartung}, {Kellner}, {Mamajek}, {Henning},
  {Miller}, {Kenworthy}, \& {Kulesa}}]{biller07}
{Biller}, B.~A., {Close}, L.~M., {Masciadri}, E., {et~al.} 2007, \apjs, 173,
  143

\bibitem[{{Boley} {et~al.}(2009){Boley}, {Lake}, {Read}, \&
  {Teyssier}}]{boley09}
{Boley}, A.~C., {Lake}, G., {Read}, J., \& {Teyssier}, R. 2009, \apjl, 706,
  L192

\bibitem[{{Bonavita} {et~al.}(2012){Bonavita}, {Chauvin}, {Desidera},
  {Gratton}, {Janson}, {Beuzit}, {Kasper}, \& {Mordasini}}]{bonavita12}
{Bonavita}, M., {Chauvin}, G., {Desidera}, S., {et~al.} 2012, \aap, 537, A67

\bibitem[{{Bonfils} {et~al.}(2013){Bonfils}, {Delfosse}, {Udry}, {Forveille},
  {Mayor}, {Perrier}, {Bouchy}, {Gillon}, {Lovis}, {Pepe}, {Queloz}, {Santos},
  {S{\'e}gransan}, \& {Bertaux}}]{bonfils13}
{Bonfils}, X., {Delfosse}, X., {Udry}, S., {et~al.} 2013, \aap, 549, A109

\bibitem[{{Bonnefoy} {et~al.}(2013){Bonnefoy}, {Boccaletti}, {Lagrange},
  {Allard}, {Mordasini}, {Beust}, {Chauvin}, {Girard}, {Homeier}, {Apai},
  {Lacour}, \& {Rouan}}]{bonnefoy13}
{Bonnefoy}, M., {Boccaletti}, A., {Lagrange}, A.-M., {et~al.} 2013, ArXiv
  e-prints

\bibitem[{{Bonnefoy} {et~al.}(2010){Bonnefoy}, {Chauvin}, {Rojo}, {Allard},
  {Lagrange}, {Homeier}, {Dumas}, \& {Beuzit}}]{bonnefoy10}
{Bonnefoy}, M., {Chauvin}, G., {Rojo}, P., {et~al.} 2010, \aap, 512, A52

\bibitem[{{Bonnefoy} {et~al.}(2011){Bonnefoy}, {Lagrange}, {Boccaletti},
  {Chauvin}, {Apai}, {Allard}, {Ehrenreich}, {Girard}, {Mouillet}, {Rouan},
  {Gratadour}, \& {Kasper}}]{bonnefoy11}
{Bonnefoy}, M., {Lagrange}, A.-M., {Boccaletti}, A., {et~al.} 2011, \aap, 528,
  L15+

\bibitem[{{Boss}(2011)}]{boss11}
{Boss}, A.~P. 2011, \apj, 731, 74

\bibitem[{{Bowler} {et~al.}(2010){Bowler}, {Liu}, {Dupuy}, \&
  {Cushing}}]{bowler10}
{Bowler}, B.~P., {Liu}, M.~C., {Dupuy}, T.~J., \& {Cushing}, M.~C. 2010, \apj,
  723, 850

\bibitem[{{Burrows} {et~al.}(2003){Burrows}, {Sudarsky}, \&
  {Lunine}}]{burrows03}
{Burrows}, A., {Sudarsky}, D., \& {Lunine}, J.~I. 2003, \apj, 596, 587

\bibitem[{{Cameron}(1978)}]{cameron78}
{Cameron}, A.~G.~W. 1978, Moon and Planets, 18, 5

\bibitem[{{Carson} {et~al.}(2012){Carson}, {Thalmann}, {Janson}, {Kozakis},
  {Bonnefoy}, {Biller}, {Schlieder}, {Currie}, {McElwain}, {Goto}, {Henning},
  {Brandner}, {Feldt}, {Kandori}, {Kuzuhara}, {Stevens}, {Wong}, {Gainey},
  {Fukagawa}, {Kuwada}, {Brandt}, {Kwon}, {Abe}, {Egner}, {Grady}, {Guyon},
  {Hashimoto}, {Hayano}, {Hayashi}, {Hayashi}, {Hodapp}, {Ishii}, {Iye},
  {Knapp}, {Kudo}, {Kusakabe}, {Matsuo}, {Miyama}, {Morino}, {Moro-Martin},
  {Nishimura}, {Pyo}, {Serabyn}, {Suto}, {Suzuki}, {Takami}, {Takato},
  {Terada}, {Turner}, {Watanabe}, {Wisniewski}, {Yamada}, {Takami}, {Usuda}, \&
  {Tamura}}]{carson12}
{Carson}, J., {Thalmann}, C., {Janson}, M., {et~al.} 2012, ArXiv e-prints

\bibitem[{{Carson} {et~al.}(2006){Carson}, {Eikenberry}, {Smith}, \&
  {Cordes}}]{carson06}
{Carson}, J.~C., {Eikenberry}, S.~S., {Smith}, J.~J., \& {Cordes}, J.~M. 2006,
  \aj, 132, 1146

\bibitem[{{Chauvin} {et~al.}(2012){Chauvin}, {Lagrange}, {Beust}, {Bonnefoy},
  {Boccaletti}, {Apai}, {Allard}, {Ehrenreich}, {Girard}, {Mouillet}, \&
  {Rouan}}]{chauvin12}
{Chauvin}, G., {Lagrange}, A.-M., {Beust}, H., {et~al.} 2012, ArXiv e-prints

\bibitem[{{Chauvin} {et~al.}(2010){Chauvin}, {Lagrange}, {Bonavita},
  {Zuckerman}, {Dumas}, {Bessell}, {Beuzit}, {Bonnefoy}, {Desidera}, {Farihi},
  {Lowrance}, {Mouillet}, \& {Song}}]{chauvin10}
{Chauvin}, G., {Lagrange}, A.-M., {Bonavita}, M., {et~al.} 2010, \aap, 509, A52

\bibitem[{{Chauvin} {et~al.}(2004){Chauvin}, {Lagrange}, {Dumas}, {Zuckerman},
  {Mouillet}, {Song}, {Beuzit}, \& {Lowrance}}]{chauvin04}
{Chauvin}, G., {Lagrange}, A.-M., {Dumas}, C., {et~al.} 2004, \aap, 425, L29

\bibitem[{{Chauvin} {et~al.}(2005){Chauvin}, {Lagrange}, {Zuckerman}, {Dumas},
  {Mouillet}, {Song}, {Beuzit}, {Lowrance}, \& {Bessell}}]{chauvin05}
{Chauvin}, G., {Lagrange}, A.-M., {Zuckerman}, B., {et~al.} 2005, \aap, 438,
  L29

\bibitem[{{Crepp} \& {Johnson}(2011)}]{crepp11}
{Crepp}, J.~R. \& {Johnson}, J.~A. 2011, \apj, 733, 126

\bibitem[{{Crida} {et~al.}(2009){Crida}, {Masset}, \& {Morbidelli}}]{crida09}
{Crida}, A., {Masset}, F., \& {Morbidelli}, A. 2009, \apjl, 705, L148

\bibitem[{{Cumming} {et~al.}(2008){Cumming}, {Butler}, {Marcy}, {Vogt},
  {Wright}, \& {Fischer}}]{cumming08}
{Cumming}, A., {Butler}, R.~P., {Marcy}, G.~W., {et~al.} 2008, \pasp, 120, 531

\bibitem[{{Cutri} {et~al.}(2003){Cutri}, {Skrutskie}, {van Dyk}, {Beichman},
  {Carpenter}, {Chester}, {Cambresy}, {Evans}, {Fowler}, {Gizis}, {Howard},
  {Huchra}, {Jarrett}, {Kopan}, {Kirkpatrick}, {Light}, {Marsh}, {McCallon},
  {Schneider}, {Stiening}, {Sykes}, {Weinberg}, {Wheaton}, {Wheelock}, \&
  {Zacarias}}]{cutri03}
{Cutri}, R.~M., {Skrutskie}, M.~F., {van Dyk}, S., {et~al.} 2003, VizieR Online
  Data Catalog, 2246, 0

\bibitem[{{De Rosa} {et~al.}(2011){De Rosa}, {Bulger}, {Patience}, {Leland},
  {Macintosh}, {Schneider}, {Song}, {Marois}, {Graham}, {Bessell}, \&
  {Doyon}}]{rosa11}
{De Rosa}, R.~J., {Bulger}, J., {Patience}, J., {et~al.} 2011, \mnras, 415, 854

\bibitem[{{Delorme} {et~al.}(2012){Delorme}, {Lagrange}, {Chauvin}, {Bonavita},
  {Lacour}, {Bonnefoy}, {Ehrenreich}, \& {Beust}}]{delorme12}
{Delorme}, P., {Lagrange}, A.~M., {Chauvin}, G., {et~al.} 2012, \aap, 539, A72

\bibitem[{{Devillar}(1997)}]{devillar97}
{Devillar}, N. 1997, The Messenger, 87

\bibitem[{{Dodson-Robinson} {et~al.}(2009){Dodson-Robinson}, {Veras}, {Ford},
  \& {Beichman}}]{dodson09}
{Dodson-Robinson}, S.~E., {Veras}, D., {Ford}, E.~B., \& {Beichman}, C.~A.
  2009, \apj, 707, 79

\bibitem[{{Dommanget} \& {Nys}(2002)}]{dommanget02}
{Dommanget}, J. \& {Nys}, O. 2002, VizieR Online Data Catalog, 1274, 0

\bibitem[{{Durisen} {et~al.}(2007){Durisen}, {Boss}, {Mayer}, {Nelson},
  {Quinn}, \& {Rice}}]{durisen07}
{Durisen}, R.~H., {Boss}, A.~P., {Mayer}, L., {et~al.} 2007, Protostars and
  Planets V, 607

\bibitem[{{Ehrenreich} {et~al.}(2010){Ehrenreich}, {Lagrange}, {Montagnier},
  {Chauvin}, {Galland}, {Beuzit}, \& {Rameau}}]{ehrenreich10}
{Ehrenreich}, D., {Lagrange}, A.-M., {Montagnier}, G., {et~al.} 2010, \aap,
  523, A73

\bibitem[{{Endl} {et~al.}(2006){Endl}, {Cochran}, {K{\"u}rster}, {Paulson},
  {Wittenmyer}, {MacQueen}, \& {Tull}}]{endl06}
{Endl}, M., {Cochran}, W.~D., {K{\"u}rster}, M., {et~al.} 2006, \apj, 649, 436

\bibitem[{{Esposito} {et~al.}(2012){Esposito}, {Mesa}, {Skemer}, {Arcidiacono},
  {Claudi}, {Desidera}, {Gratton}, {Mannucci}, {Marzari}, {Masciadri}, {Close},
  {Hinz}, {Kulesa}, {McCarthy}, {Males}, {Agapito}, {Argomedo}, {Boutsia},
  {Briguglio}, {Brusa}, {Busoni}, {Cresci}, {Fini}, {Fontana}, {Guerra},
  {Hill}, {Miller}, {Paris}, {Pinna}, {Puglisi}, {Quiros-Pacheco}, {Riccardi},
  {Stefanini}, {Testa}, {Xompero}, \& {Woodward}}]{esposito12}
{Esposito}, S., {Mesa}, D., {Skemer}, A., {et~al.} 2012, ArXiv e-prints

\bibitem[{{Fortney} {et~al.}(2008){Fortney}, {Lodders}, {Marley}, \&
  {Freedman}}]{fortney08}
{Fortney}, J.~J., {Lodders}, K., {Marley}, M.~S., \& {Freedman}, R.~S. 2008,
  \apj, 678, 1419

\bibitem[{{Fukagawa} {et~al.}(2006){Fukagawa}, {Tamura}, {Itoh}, {Kudo},
  {Imaeda}, {Oasa}, {Hayashi}, \& {Hayashi}}]{fukagawa06}
{Fukagawa}, M., {Tamura}, M., {Itoh}, Y., {et~al.} 2006, \apjl, 636, L153

\bibitem[{{Galvagni} {et~al.}(2012){Galvagni}, {Hayfield}, {Boley}, {Mayer},
  {Ro{\v s}kar}, \& {Saha}}]{galvagni12}
{Galvagni}, M., {Hayfield}, T., {Boley}, A., {et~al.} 2012, \mnras, 427, 1725

\bibitem[{{Gammie}(2001)}]{gammie01}
{Gammie}, C.~F. 2001, \apj, 553, 174

\bibitem[{{Girard} {et~al.}(2010){Girard}, {Kasper}, {Quanz}, {Kenworthy},
  {Rengaswamy}, {Sch{\"o}del}, {Gallenne}, {Gillessen}, {Huerta}, {Kervella},
  {Kornweibel}, {Lenzen}, {M{\'e}rand}, {Montagnier}, {O'Neal}, \&
  {Zins}}]{girard10}
{Girard}, J.~H.~V., {Kasper}, M., {Quanz}, S.~P., {et~al.} 2010, in Society of
  Photo-Optical Instrumentation Engineers (SPIE) Conference Series, Vol. 7736,
  Society of Photo-Optical Instrumentation Engineers (SPIE) Conference Series

\bibitem[{{Girard} {et~al.}(2012){Girard}, {O'Neal}, {Mawet}, {Kasper}, {Zins},
  {Neichel}, {Kolb}, {Christiaens}, \& {Tourneboeuf}}]{girard12}
{Girard}, J.~H.~V., {O'Neal}, J., {Mawet}, D., {et~al.} 2012, in Society of
  Photo-Optical Instrumentation Engineers (SPIE) Conference Series, Vol. 8447,
  Society of Photo-Optical Instrumentation Engineers (SPIE) Conference Series

\bibitem[{{Guillot} {et~al.}(2006){Guillot}, {Santos}, {Pont}, {Iro}, {Melo},
  \& {Ribas}}]{guillot06}
{Guillot}, T., {Santos}, N.~C., {Pont}, F., {et~al.} 2006, \aap, 453, L21

\bibitem[{{Haakonsen} \& {Rutledge}(2009)}]{haakonsen09}
{Haakonsen}, C.~B. \& {Rutledge}, R.~E. 2009, \apjs, 184, 138

\bibitem[{{Hillenbrand} {et~al.}(2008){Hillenbrand}, {Carpenter}, {Kim},
  {Meyer}, {Backman}, {Moro-Mart{\'{\i}}n}, {Hollenbach}, {Hines}, {Pascucci},
  \& {Bouwman}}]{hillenbrand08}
{Hillenbrand}, L.~A., {Carpenter}, J.~M., {Kim}, J.~S., {et~al.} 2008, \apj,
  677, 630

\bibitem[{{Holmberg} {et~al.}(2007){Holmberg}, {Nordstr{\"o}m}, \&
  {Andersen}}]{homlberg07}
{Holmberg}, J., {Nordstr{\"o}m}, B., \& {Andersen}, J. 2007, \aap, 475, 519

\bibitem[{{Janson} {et~al.}(2010){Janson}, {Bergfors}, {Goto}, {Brandner}, \&
  {Lafreni{\`e}re}}]{janson10}
{Janson}, M., {Bergfors}, C., {Goto}, M., {Brandner}, W., \& {Lafreni{\`e}re},
  D. 2010, \apjl, 710, L35

\bibitem[{{Janson} {et~al.}(2012){Janson}, {Bonavita}, {Klahr}, \&
  {Lafreni{\`e}re}}]{janson12}
{Janson}, M., {Bonavita}, M., {Klahr}, H., \& {Lafreni{\`e}re}, D. 2012, \apj,
  745, 4

\bibitem[{{Janson} {et~al.}(2011){Janson}, {Bonavita}, {Klahr},
  {Lafreni{\`e}re}, {Jayawardhana}, \& {Zinnecker}}]{janson11}
{Janson}, M., {Bonavita}, M., {Klahr}, H., {et~al.} 2011, \apj, 736, 89

\bibitem[{{Johnson}(2007)}]{johnson07}
{Johnson}, J.~A. 2007, in Bulletin of the American Astronomical Society,
  Vol.~39, American Astronomical Society Meeting Abstracts, 150.03

\bibitem[{{Johnson} {et~al.}(2010){Johnson}, {Aller}, {Howard}, \&
  {Crepp}}]{johnson10}
{Johnson}, J.~A., {Aller}, K.~M., {Howard}, A.~W., \& {Crepp}, J.~R. 2010,
  \pasp, 122, 905

\bibitem[{{Kains} {et~al.}(2011){Kains}, {Wyatt}, \& {Greaves}}]{kains11}
{Kains}, N., {Wyatt}, M.~C., \& {Greaves}, J.~S. 2011, \mnras, 414, 2486

\bibitem[{{Kalas} {et~al.}(2008){Kalas}, {Graham}, {Chiang}, {Fitzgerald},
  {Clampin}, {Kite}, {Stapelfeldt}, {Marois}, \& {Krist}}]{kalas08}
{Kalas}, P., {Graham}, J.~R., {Chiang}, E., {et~al.} 2008, Science, 322, 1345

\bibitem[{{Kalas} {et~al.}(2005){Kalas}, {Graham}, \& {Clampin}}]{kalas05}
{Kalas}, P., {Graham}, J.~R., \& {Clampin}, M. 2005, \nat, 435, 1067

\bibitem[{{Kasper} {et~al.}(2007){Kasper}, {Apai}, {Janson}, \&
  {Brandner}}]{kasper07}
{Kasper}, M., {Apai}, D., {Janson}, M., \& {Brandner}, W. 2007, \aap, 472, 321

\bibitem[{{Kennedy} \& {Kenyon}(2008)}]{kennedy08}
{Kennedy}, G.~M. \& {Kenyon}, S.~J. 2008, \apj, 682, 1264

\bibitem[{{Kenyon} \& {Hartmann}(1987)}]{kenyon87}
{Kenyon}, S.~J. \& {Hartmann}, L. 1987, \apj, 323, 714

\bibitem[{{Kley} \& {Dirksen}(2006)}]{kley06}
{Kley}, W. \& {Dirksen}, G. 2006, \aap, 447, 369

\bibitem[{{Kley} \& {Nelson}(2012)}]{kley12}
{Kley}, W. \& {Nelson}, R.~P. 2012, \araa, 50, 211

\bibitem[{{Kratter} \& {Murray-Clay}(2011)}]{kratter11}
{Kratter}, K.~M. \& {Murray-Clay}, R.~A. 2011, \apj, 740, 1

\bibitem[{{Kratter} {et~al.}(2010){Kratter}, {Murray-Clay}, \&
  {Youdin}}]{kratter10}
{Kratter}, K.~M., {Murray-Clay}, R.~A., \& {Youdin}, A.~N. 2010, \apj, 710,
  1375

\bibitem[{{Krivov}(2010)}]{krikov10}
{Krivov}, A.~V. 2010, Research in Astronomy and Astrophysics, 10, 383

\bibitem[{{Lafreni{\`e}re} {et~al.}(2007){Lafreni{\`e}re}, {Marois}, {Doyon},
  {Nadeau}, \& {Artigau}}]{lafreniere07}
{Lafreni{\`e}re}, D., {Marois}, C., {Doyon}, R., {Nadeau}, D., \& {Artigau},
  {\'E}. 2007, \apj, 660, 770

\bibitem[{{Lagrange} {et~al.}(2012{\natexlab{a}}){Lagrange}, {Boccaletti},
  {Milli}, {Chauvin}, {Bonnefoy}, {Mouillet}, {Augereau}, {Girard}, {Lacour},
  \& {Apai}}]{lagrange12b}
{Lagrange}, A.-M., {Boccaletti}, A., {Milli}, J., {et~al.} 2012{\natexlab{a}},
  \aap, 542, A40

\bibitem[{{Lagrange} {et~al.}(2010){Lagrange}, {Bonnefoy}, {Chauvin}, {Apai},
  {Ehrenreich}, {Boccaletti}, {Gratadour}, {Rouan}, {Mouillet}, {Lacour}, \&
  {Kasper}}]{lagrange10}
{Lagrange}, A.-M., {Bonnefoy}, M., {Chauvin}, G., {et~al.} 2010, Science, 329,
  57

\bibitem[{{Lagrange} {et~al.}(2012{\natexlab{b}}){Lagrange}, {De Bondt},
  {Meunier}, {Sterzik}, {Beust}, \& {Galland}}]{lagrange12a}
{Lagrange}, A.-M., {De Bondt}, K., {Meunier}, N., {et~al.} 2012{\natexlab{b}},
  \aap, 542, A18

\bibitem[{{Lagrange} {et~al.}(2009{\natexlab{a}}){Lagrange}, {Desort},
  {Galland}, {Udry}, \& {Mayor}}]{lagrange09a}
{Lagrange}, A.-M., {Desort}, M., {Galland}, F., {Udry}, S., \& {Mayor}, M.
  2009{\natexlab{a}}, \aap, 495, 335

\bibitem[{{Lagrange} {et~al.}(2009{\natexlab{b}}){Lagrange}, {Gratadour},
  {Chauvin}, {Fusco}, {Ehrenreich}, {Mouillet}, {Rousset}, {Rouan}, {Allard},
  {Gendron}, {Charton}, {Mugnier}, {Rabou}, {Montri}, \&
  {Lacombe}}]{lagrange09b}
{Lagrange}, A.-M., {Gratadour}, D., {Chauvin}, G., {et~al.} 2009{\natexlab{b}},
  \aap, 493, L21

\bibitem[{{Lagrange} {et~al.}(2012{\natexlab{c}}){Lagrange}, {Meunier},
  {Chauvin}, {Sterzik}, {Galland}, {Lo Curto}, {Rameau}, \&
  {Sosnowska}}]{lagrange12c}
{Lagrange}, A.-M., {Meunier}, N., {Chauvin}, G., {et~al.} 2012{\natexlab{c}},
  \aap

\bibitem[{{Lenzen} {et~al.}(2003){Lenzen}, {Hartung}, {Brandner}, {Finger},
  {Hubin}, {Lacombe}, {Lagrange}, {Lehnert}, {Moorwood}, \&
  {Mouillet}}]{lenzen03}
{Lenzen}, R., {Hartung}, M., {Brandner}, W., {et~al.} 2003, in SPIE, Vol. 4841,
  944--952

\bibitem[{{Lin} {et~al.}(1996){Lin}, {Bodenheimer}, \& {Richardson}}]{lin96}
{Lin}, D.~N.~C., {Bodenheimer}, P., \& {Richardson}, D.~C. 1996, \nat, 380, 606

\bibitem[{{L{\'o}pez-Santiago} {et~al.}(2006){L{\'o}pez-Santiago}, {Montes},
  {Crespo-Chac{\'o}n}, \& {Fern{\'a}ndez-Figueroa}}]{lopez06}
{L{\'o}pez-Santiago}, J., {Montes}, D., {Crespo-Chac{\'o}n}, I., \&
  {Fern{\'a}ndez-Figueroa}, M.~J. 2006, \apj, 643, 1160

\bibitem[{{Lovis} \& {Mayor}(2007)}]{lovis07}
{Lovis}, C. \& {Mayor}, M. 2007, \aap, 472, 657

\bibitem[{{Lowrance} {et~al.}(2000){Lowrance}, {Schneider}, {Kirkpatrick},
  {Becklin}, {Weinberger}, {Zuckerman}, {Plait}, {Malmuth}, {Heap}, {Schultz},
  {Smith}, {Terrile}, \& {Hines}}]{lowrance00}
{Lowrance}, P.~J., {Schneider}, G., {Kirkpatrick}, J.~D., {et~al.} 2000, \apj,
  541, 390

\bibitem[{{Lubow} {et~al.}(1999){Lubow}, {Seibert}, \& {Artymowicz}}]{lubow99}
{Lubow}, S.~H., {Seibert}, M., \& {Artymowicz}, P. 1999, \apj, 526, 1001

\bibitem[{{Macintosh} {et~al.}(2008){Macintosh}, {Graham}, {Palmer}, {Doyon},
  {Dunn}, {Gavel}, {Larkin}, {Oppenheimer}, {Saddlemyer}, {Sivaramakrishnan},
  {Wallace}, {Bauman}, {Erickson}, {Marois}, {Poyneer}, \&
  {Soummer}}]{macintosh08}
{Macintosh}, B.~A., {Graham}, J.~R., {Palmer}, D.~W., {et~al.} 2008, in Society
  of Photo-Optical Instrumentation Engineers (SPIE) Conference Series, Vol.
  7015, Society of Photo-Optical Instrumentation Engineers (SPIE) Conference
  Series

\bibitem[{{Makarov} \& {Kaplan}(2005)}]{makarov05}
{Makarov}, V.~V. \& {Kaplan}, G.~H. 2005, \aj, 129, 2420

\bibitem[{{Marley} {et~al.}(2007){Marley}, {Fortney}, {Hubickyj},
  {Bodenheimer}, \& {Lissauer}}]{marley07}
{Marley}, M.~S., {Fortney}, J.~J., {Hubickyj}, O., {Bodenheimer}, P., \&
  {Lissauer}, J.~J. 2007, \apj, 655, 541

\bibitem[{{Marois} {et~al.}(2006){Marois}, {Lafreni{\`e}re}, {Doyon},
  {Macintosh}, \& {Nadeau}}]{marois06}
{Marois}, C., {Lafreni{\`e}re}, D., {Doyon}, R., {Macintosh}, B., \& {Nadeau},
  D. 2006, \apj, 641, 556

\bibitem[{{Marois} {et~al.}(2008){Marois}, {Macintosh}, {Barman}, {Zuckerman},
  {Song}, {Patience}, {Lafreni{\`e}re}, \& {Doyon}}]{marois08}
{Marois}, C., {Macintosh}, B., {Barman}, T., {et~al.} 2008, Science, 322, 1348

\bibitem[{{Marois} {et~al.}(2010){Marois}, {Zuckerman}, {Konopacky},
  {Macintosh}, \& {Barman}}]{marois10}
{Marois}, C., {Zuckerman}, B., {Konopacky}, Q.~M., {Macintosh}, B., \&
  {Barman}, T. 2010, \nat, 468, 1080

\bibitem[{{Masciadri} {et~al.}(2005){Masciadri}, {Mundt}, {Henning}, {Alvarez},
  \& {Barrado y Navascu{\'e}s}}]{masciadri05}
{Masciadri}, E., {Mundt}, R., {Henning}, T., {Alvarez}, C., \& {Barrado y
  Navascu{\'e}s}, D. 2005, \apj, 625, 1004

\bibitem[{{Mason} {et~al.}(2001){Mason}, {Wycoff}, {Hartkopf}, {Douglass}, \&
  {Worley}}]{mason2001}
{Mason}, B.~D., {Wycoff}, G.~L., {Hartkopf}, W.~I., {Douglass}, G.~G., \&
  {Worley}, C.~E. 2001, \aj, 122, 3466

\bibitem[{{Mayor} {et~al.}(2011){Mayor}, {Marmier}, {Lovis}, {Udry},
  {S{\'e}gransan}, {Pepe}, {Benz}, {Bertaux}, {Bouchy}, {Dumusque}, {Lo Curto},
  {Mordasini}, {Queloz}, \& {Santos}}]{mayor11}
{Mayor}, M., {Marmier}, M., {Lovis}, C., {et~al.} 2011, ArXiv e-prints

\bibitem[{{McCaughrean} \& {Stauffer}(1994)}]{mccaughrean94}
{McCaughrean}, M.~J. \& {Stauffer}, J.~R. 1994, \aj, 108, 1382

\bibitem[{{Meru} \& {Bate}(2011)}]{meru11}
{Meru}, F. \& {Bate}, M.~R. 2011, in IAU Symposium, Vol. 276, IAU Symposium,
  ed. A.~{Sozzetti}, M.~G. {Lattanzi}, \& A.~P. {Boss}, 438--440

\bibitem[{{Mesa} {et~al.}(2011){Mesa}, {Gratton}, {Berton}, {Antichi},
  {Verinaud}, {Boccaletti}, {Kasper}, {Claudi}, {Desidera}, {Giro}, {Beuzit},
  {Dohlen}, {Feldt}, {Mouillet}, {Chauvin}, \& {Vigan}}]{mesa11}
{Mesa}, D., {Gratton}, R., {Berton}, A., {et~al.} 2011, \aap, 529, A131

\bibitem[{{Miller} \& {Fortney}(2011)}]{miller11}
{Miller}, N. \& {Fortney}, J.~J. 2011, \apjl, 736, L29

\bibitem[{{Mizuno}(1980)}]{mizuno80}
{Mizuno}, H. 1980, Progress of Theoretical Physics, 64, 544

\bibitem[{{Mizusawa} {et~al.}(2012){Mizusawa}, {Rebull}, {Stauffer}, {Bryden},
  {Meyer}, \& {Song}}]{mizusawa12}
{Mizusawa}, T.~F., {Rebull}, L.~M., {Stauffer}, J.~R., {et~al.} 2012, \aj, 144,
  135

\bibitem[{{Moffat}(1969)}]{moffat69}
{Moffat}, A.~F.~J. 1969, \aap, 3, 455

\bibitem[{{Morales} {et~al.}(2011){Morales}, {Rieke}, {Werner}, {Bryden},
  {Stapelfeldt}, \& {Su}}]{morales11}
{Morales}, F.~Y., {Rieke}, G.~H., {Werner}, M.~W., {et~al.} 2011, \apjl, 730,
  L29

\bibitem[{{Mordasini} {et~al.}(2009){Mordasini}, {Alibert}, \&
  {Benz}}]{mordasini09}
{Mordasini}, C., {Alibert}, Y., \& {Benz}, W. 2009, \aap, 501, 1139

\bibitem[{{Mordasini} {et~al.}(2012){Mordasini}, {Alibert}, {Benz}, {Klahr}, \&
  {Henning}}]{mordasini12}
{Mordasini}, C., {Alibert}, Y., {Benz}, W., {Klahr}, H., \& {Henning}, T. 2012,
  \aap, 541, A97

\bibitem[{{Mouillet} {et~al.}(1997){Mouillet}, {Larwood}, {Papaloizou}, \&
  {Lagrange}}]{mouillet97}
{Mouillet}, D., {Larwood}, J.~D., {Papaloizou}, J.~C.~B., \& {Lagrange}, A.~M.
  1997, \mnras, 292, 896

\bibitem[{{Nakajima} \& {Morino}(2012)}]{nakajima12}
{Nakajima}, T. \& {Morino}, J.-I. 2012, \aj, 143, 2

\bibitem[{{Neuh{\"a}user} {et~al.}(2011){Neuh{\"a}user}, {Ginski}, {Schmidt},
  \& {Mugrauer}}]{neuhauser11}
{Neuh{\"a}user}, R., {Ginski}, C., {Schmidt}, T.~O.~B., \& {Mugrauer}, M. 2011,
  \mnras, 416, 1430

\bibitem[{{Nielsen} \& {Close}(2010)}]{nielsen10}
{Nielsen}, E.~L. \& {Close}, L.~M. 2010, \apj, 717, 878

\bibitem[{{Papaloizou} {et~al.}(2007){Papaloizou}, {Nelson}, {Kley}, {Masset},
  \& {Artymowicz}}]{papaloizou07}
{Papaloizou}, J.~C.~B., {Nelson}, R.~P., {Kley}, W., {Masset}, F.~S., \&
  {Artymowicz}, P. 2007, Protostars and Planets V, 655

\bibitem[{{Patience} {et~al.}(2011){Patience}, {Bulger}, {King}, {Ayliffe},
  {Bate}, {Song}, {Pinte}, {Koda}, {Dowell}, \& {Kov{\'a}cs}}]{patience11}
{Patience}, J., {Bulger}, J., {King}, R.~R., {et~al.} 2011, \aap, 531, L17

\bibitem[{{Pecaut} \& {Mamajek}(2010)}]{pecaut10}
{Pecaut}, M. \& {Mamajek}, E. 2010, in Bulletin of the American Astronomical
  Society, Vol.~42, American Astronomical Society Meeting Abstracts \#215,
  455.30

\bibitem[{{Pollack} {et~al.}(1996){Pollack}, {Hubickyj}, {Bodenheimer},
  {Lissauer}, {Podolak}, \& {Greenzweig}}]{pollack96}
{Pollack}, J.~B., {Hubickyj}, O., {Bodenheimer}, P., {et~al.} 1996, \icarus,
  124, 62

\bibitem[{{Pourbaix} {et~al.}(2004){Pourbaix}, {Tokovinin}, {Batten}, {Fekel},
  {Hartkopf}, {Levato}, {Morrell}, {Torres}, \& {Udry}}]{pourbaix09}
{Pourbaix}, D., {Tokovinin}, A.~A., {Batten}, A.~H., {et~al.} 2004, \aap, 424,
  727

\bibitem[{{Rafikov} \& {Goldreich}(2005)}]{rafikov05}
{Rafikov}, R.~R. \& {Goldreich}, P. 2005, \apj, 631, 488

\bibitem[{{Rameau} {et~al.}(2012){Rameau}, {Chauvin}, {Lagrange}, {Thebault},
  {Milli}, \& {Bonnefoy}}]{rameau12}
{Rameau}, J., {Chauvin}, G., {Lagrange}, A.-M., {et~al.} 2012, \aap

\bibitem[{{Raymond} {et~al.}(2012){Raymond}, {Armitage}, {Moro-Mart{\'{\i}}n},
  {Booth}, {Wyatt}, {Armstrong}, {Mandell}, {Selsis}, \& {West}}]{raymond12}
{Raymond}, S.~N., {Armitage}, P.~J., {Moro-Mart{\'{\i}}n}, A., {et~al.} 2012,
  \aap, 541, A11

\bibitem[{{Rebull} {et~al.}(2008){Rebull}, {Stapelfeldt}, {Werner}, {Mannings},
  {Chen}, {Stauffer}, {Smith}, {Song}, {Hines}, \& {Low}}]{rebull08}
{Rebull}, L.~M., {Stapelfeldt}, K.~R., {Werner}, M.~W., {et~al.} 2008, \apj,
  681, 1484

\bibitem[{{Rhee} {et~al.}(2007){Rhee}, {Song}, {Zuckerman}, \&
  {McElwain}}]{rhee07}
{Rhee}, J.~H., {Song}, I., {Zuckerman}, B., \& {McElwain}, M. 2007, \apj, 660,
  1556

\bibitem[{{Roccatagliata} {et~al.}(2009){Roccatagliata}, {Henning}, {Wolf},
  {Rodmann}, {Corder}, {Carpenter}, {Meyer}, \& {Dowell}}]{rocca09}
{Roccatagliata}, V., {Henning}, T., {Wolf}, S., {et~al.} 2009, \aap, 497, 409

\bibitem[{{Rousset} {et~al.}(2003){Rousset}, {Lacombe}, {Puget}, {Hubin},
  {Gendron}, {Fusco}, {Arsenault}, {Charton}, {Feautrier}, {Gigan}, {Kern},
  {Lagrange}, {Madec}, {Mouillet}, {Rabaud}, {Rabou}, {Stadler}, \&
  {Zins}}]{rousset03}
{Rousset}, G., {Lacombe}, F., {Puget}, P., {et~al.} 2003, in SPIE, Vol. 4839,
  140--149

\bibitem[{{Schlieder} {et~al.}(2012){Schlieder}, {L{\'e}pine}, \&
  {Simon}}]{schlieder12}
{Schlieder}, J.~E., {L{\'e}pine}, S., \& {Simon}, M. 2012, \aj, 143, 80

\bibitem[{{Seager} \& {Rogers}(2011)}]{seager11}
{Seager}, S. \& {Rogers}, L.~A. 2011, in American Astronomical Society Meeting
  Abstracts \#218, 211.08

\bibitem[{{Siess} {et~al.}(2000){Siess}, {Dufour}, \& {Forestini}}]{siess00}
{Siess}, L., {Dufour}, E., \& {Forestini}, M. 2000, \aap, 358, 593

\bibitem[{{Smith} {et~al.}(2009){Smith}, {Churcher}, {Wyatt}, {Moerchen}, \&
  {Telesco}}]{smith09}
{Smith}, R., {Churcher}, L.~J., {Wyatt}, M.~C., {Moerchen}, M.~M., \&
  {Telesco}, C.~M. 2009, \aap, 493, 299

\bibitem[{{Song} {et~al.}(2001){Song}, {Caillault}, {Barrado y Navascu{\'e}s},
  \& {Stauffer}}]{song01}
{Song}, I., {Caillault}, J.-P., {Barrado y Navascu{\'e}s}, D., \& {Stauffer},
  J.~R. 2001, \apj, 546, 352

\bibitem[{{Soummer} {et~al.}(2011){Soummer}, {Brendan Hagan}, {Pueyo},
  {Thormann}, {Rajan}, \& {Marois}}]{soummer11}
{Soummer}, R., {Brendan Hagan}, J., {Pueyo}, L., {et~al.} 2011, \apj, 741, 55

\bibitem[{{Sousa} {et~al.}(2011){Sousa}, {Santos}, {Israelian}, {Mayor}, \&
  {Udry}}]{sousa11}
{Sousa}, S.~G., {Santos}, N.~C., {Israelian}, G., {Mayor}, M., \& {Udry}, S.
  2011, \aap, 533, A141

\bibitem[{{Stamatellos} \& {Whitworth}(2009)}]{stamatellos09}
{Stamatellos}, D. \& {Whitworth}, A.~P. 2009, \mnras, 392, 413

\bibitem[{{Su} {et~al.}(2006){Su}, {Rieke}, {Stansberry}, {Bryden},
  {Stapelfeldt}, {Trilling}, {Muzerolle}, {Beichman}, {Moro-Martin}, {Hines},
  \& {Werner}}]{su06}
{Su}, K.~Y.~L., {Rieke}, G.~H., {Stansberry}, J.~A., {et~al.} 2006, \apj, 653,
  675

\bibitem[{{Toomre}(1981)}]{toomre81}
{Toomre}, A. 1981, {Structure and Evolution of Normal Galaxies}, ed. J.~A.
  {Fall} \& D.~{Lynden-Bell}, 111

\bibitem[{{Torres} {et~al.}(2008){Torres}, {Quast}, {Melo}, \&
  {Sterzik}}]{torres08}
{Torres}, C.~A.~O., {Quast}, G.~R., {Melo}, C.~H.~F., \& {Sterzik}, M.~F. 2008,
  {Young Nearby Loose Associations}, ed. B.~{Reipurth}, 757

\bibitem[{{Valenti} \& {Fischer}(2005)}]{valenti05}
{Valenti}, J.~A. \& {Fischer}, D.~A. 2005, \apjs, 159, 141

\bibitem[{{van Leeuwen}(2007)}]{leeuwen07}
{van Leeuwen}, F. 2007, \aap, 474, 653

\bibitem[{{Veran} \& {Rigaut}(1998)}]{veran98}
{Veran}, J.-P. \& {Rigaut}, F.~J. 1998, in Society of Photo-Optical
  Instrumentation Engineers (SPIE) Conference Series, Vol. 3353, Society of
  Photo-Optical Instrumentation Engineers (SPIE) Conference Series, ed.
  D.~{Bonaccini} \& R.~K. {Tyson}, 426--437

\bibitem[{{Vigan} {et~al.}(2012){Vigan}, {Patience}, {Marois}, {Bonavita}, {De
  Rosa}, {Macintosh}, {Song}, {Doyon}, {Zuckerman}, {Lafreni{\`e}re}, \&
  {Barman}}]{vigan12}
{Vigan}, A., {Patience}, J., {Marois}, C., {et~al.} 2012, \aap, 544, A9

\bibitem[{{Whitworth} {et~al.}(2007){Whitworth}, {Bate}, {Nordlund},
  {Reipurth}, \& {Zinnecker}}]{whitworth07}
{Whitworth}, A., {Bate}, M.~R., {Nordlund}, {\AA}., {Reipurth}, B., \&
  {Zinnecker}, H. 2007, Protostars and Planets V, 459

\bibitem[{{Zuckerman} {et~al.}(2006){Zuckerman}, {Bessell}, {Song}, \&
  {Kim}}]{zuckerman06}
{Zuckerman}, B., {Bessell}, M.~S., {Song}, I., \& {Kim}, S. 2006, \apjl, 649,
  L115

\bibitem[{{Zuckerman} {et~al.}(2011){Zuckerman}, {Rhee}, {Song}, \&
  {Bessell}}]{zuckerman11}
{Zuckerman}, B., {Rhee}, J.~H., {Song}, I., \& {Bessell}, M.~S. 2011, \apj,
  732, 61

\bibitem[{{Zuckerman} {et~al.}(2004){Zuckerman}, {Song}, \&
  {Bessell}}]{zuckerman04}
{Zuckerman}, B., {Song}, I., \& {Bessell}, M.~S. 2004, \apjl, 613, L65

\bibitem[{{Zuckerman} {et~al.}(2001){Zuckerman}, {Song}, {Bessell}, \&
  {Webb}}]{zuckerman01}
{Zuckerman}, B., {Song}, I., {Bessell}, M.~S., \& {Webb}, R.~A. 2001, \apjl,
  562, L87

\end{thebibliography}

\newpage
\clearpage
\newpage
\begin{appendix} 
\section{Statistical formalism}
\label{annexe}
Our likelihood analysis approach follows the work done by \citet{carson06,lafreniere07,vigan12}. We nevertheless recall here the steps.

The principle of detecting a planet around a star is a Bernoulli event. We note, $p_j$ the probability to detect a giant planet around a star $j$ if it is indeed here. 
This probability depends on the distance, the luminosity, and the age of the star, and on the projected position, and the luminosity of a planet, and also on the
instrumental performances. We also note the fraction of stars $f$ harboring at least a planet in the interval $[m_\mathrm{min},m_\mathrm{max}]$ and $[a_\mathrm{min},a_\mathrm{max}]$.
We assume $f$ to be constant around the star sample. For a given $j$ star, the probability of detecting a giant planet companion is $fp_j$. Given the observational results of a survey,
one can let $d_j$ to represent the detection efficiency such that $d_j=1$ if a planet has been detecting around the star $j$ and $0$ otherwise.
Therefore, the likelihood function of the data for a set of $N$ star will be the product of each Bernoulli event since there are independent, so that :
\begin{equation}
\label{eq:likelihood}
\mathcal{L}(\{d_j\}|f)=\prod_{j=1}^N(fp_j)^{d_j}(1-fp_j)^{1-d_j}
\end{equation}

Then, we can apply the Bayes'rule which links the likelihood function of the data ($\{d_j\}$) given the model $f$ $\mathcal{L}(\{d_j\}|f)$ to the probability density of the model given the data, or posterior distribution 
$P(f|\{d_j\})$. We get :
\begin{equation}
\label{eq:proba_density}
P(f|\{d_j\})=\frac{\mathcal{L}(\{d_j\}|f)P(f)}{\int_0^1\mathcal{L}(\{d_j\}|f)P(f)df}
\end{equation}

The Bayes' rule also remains on the assumption on the initial probability of the model, or prior distribution $P(f)$, which can be the most controversial part. One can use the posterior distribution from previous studies or we can construct
priors by considering no previous knowledge on $f$ so that $P(f)=1$, excluding any bias on $f$. The later will be our assumption for a direct comparison between surveys.

As for any estimation of a random variable, here $f$, the confidence interval $[f_\mathrm{min},f_\mathrm{max}]$ in which the true $f$ is can be determined with the equation, considering a confidence level $CL$ :
\begin{equation}
\label{eq:CL_all}
CL = \int_{f_\mathrm{min}}^{f_\mathrm{max}}P(f|\{d_j\})df
\end{equation}

which can be split into implicit equations on $f_\mathrm{min}$ and $f_\mathrm{max}$ :
\begin{equation}
\label{eq:tail}
\frac{1-CL}{2}=\int_0^{f_\mathrm{min}}P(f|\{d_j\})df=\int_{f_\mathrm{max}}^1P(f|\{d_j\})df
\end{equation}

In case of a null detection, Poisson statistic dictates the probability of detecting a giant planet around a given star such that the likelihood equation \ref{eq:likelihood} becomes :
\begin{equation}
\label{eq:L_null}
\mathcal{L}(\{d_j\}|f)=\prod_{j=1}^Ne^{-fp_j}
\end{equation}

Null detection sets $f_\mathrm{min}=0$ and the equation \ref{eq:CL_all} becomes an explicit equation for $f_\mathrm{max}$ given CL :
\begin{equation}
\label{eq:CL_null}
f_\mathrm{max}=\frac{-ln(1-CL)}{N\langle p_j\rangle}
\end{equation}

\end{appendix}

\end{document}